\documentclass[a4paper, 11pt, oneside]{article}
\usepackage[a4paper,inner=2cm,outer=2cm,top=3cm,bottom=2.5cm]{geometry}

\usepackage[tbtags]{amsmath}
\usepackage{amssymb}
\usepackage{amsthm}
\usepackage{dsfont}
\usepackage{slashed}
\usepackage{mathrsfs}
\usepackage[tbtags]{mathtools}
\usepackage{color}

\usepackage[export]{adjustbox}

\usepackage[english]{babel}

\usepackage[utf8x]{inputenc}

\usepackage{graphicx,epsfig}

\usepackage{setspace}

\usepackage{booktabs}

\usepackage{float}

\usepackage{dcolumn}

\usepackage{nextpage}

\usepackage{newclude}

\usepackage[small,bf]{caption}
\usepackage{subcaption}

\usepackage{url}

\usepackage[colorlinks=true,
	linkcolor=black,
	citecolor=black,
	urlcolor=blue,
	filecolor=black
]{hyperref}

\usepackage[T1]{fontenc}

\usepackage{cancel}

\usepackage{pifont}

\usepackage{etex}

\usepackage{appendix}

\usepackage[numbers,sort&compress]{natbib}

\usepackage{placeins}

\usepackage{ragged2e}

\usepackage{ifmtarg}

\usepackage{tikz}
\usetikzlibrary{shapes,arrows}

\newcommand{\ChPT}{ChPT}

\renewcommand{\O}{\mathcal{O}}
\newcommand{\p}{\partial}

\renewcommand{\Re}{\text{Re}\,}
\renewcommand{\Im}{\text{Im}\,}
\newcommand{\Imspipi}{\text{Im}_s^{\pi\pi}}

\newcommand{\<}{\langle}
\renewcommand{\>}{\rangle}

\newcommand{\atan}{\mathrm{atan}}

\newcommand{\dprime}{{\prime\prime}}

\newcommand{\mpi}{M_{\pi}}

\newcommand{\mv}{M_V}
\newcommand{\BR}{\text{BR}}

\newcommand{\avec}{\boldsymbol{a}}
\newcommand{\FW}{\text{FW}}
\newcommand{\beq}{\begin{equation}}
\newcommand{\eeq}{\end{equation}}
\newcommand{\GeV}{\,\text{GeV}}
\newcommand{\Fpi}{F_\pi}

\newcommand{\remark}[1]{}

\newcommand{\mytag}{\\[-\baselineskip] \stepcounter{equation}\tag{\theequation}}

\makeatletter
\newcommand{\Cr}[2]{\@ifmtarg{#2}{\mathcal{C}_{#1}}{\mathcal{C}_{#1}\big[#2\big]}}
\makeatother

\newcommand{\norm}[1]{\left\lvert#1\right\rvert}

\newcolumntype{.}{D{.}{.}{2} } 
\newcolumntype{d}{D{.}{.}{2.2} }
\newcolumntype{L}[1]{>{\RaggedRight\hspace{0pt}}p{#1}}
\newcolumntype{R}[1]{>{\RaggedLeft\hspace{0pt}}p{#1}}

\makeatletter
  \def\my@tag@font{\normalsize}
  \def\maketag@@@#1{\hbox{\m@th\normalfont\my@tag@font#1}}
  \let\amsmath@eqref\eqref
  \renewcommand\eqref[1]{{\let\my@tag@font\relax\amsmath@eqref{#1}}}
\makeatother

\makeatletter
\renewcommand\paragraph{\@startsection{paragraph}{4}{\z@}%
  {-3.25ex \@plus -1ex \@minus -0.2ex}%
  {0.01pt}%
  {\bfseries}%
}
\makeatother

\makeatletter
\def\@xfootnote[#1]{%
  \protected@xdef\@thefnmark{#1}%
  \@footnotemark\@footnotetext}
\makeatother

\usepackage{abstract}

\allowdisplaybreaks[1]

\begin{document}

\mbox{}

\vspace{-1.75cm}
\hfill{}\begin{minipage}[t][0cm][t]{5cm}
\raggedleft
\footnotesize
INT-PUB-17-009 \\
CERN-TH-2017-041 \\
NSF-KITP-17-036
\end{minipage}
\vspace{1.25cm}

\bigskip

\begin{center}
{\LARGE{\bf \boldmath Dispersion relation for hadronic light-by-light scattering:\\[2mm] two-pion contributions}}

\vspace{0.5cm}

Gilberto Colangelo${}^a$, Martin Hoferichter${}^{b,c}$, Massimiliano Procura${}^{d}$\footnote[$\dagger$]{On leave from the University of Vienna.}, Peter Stoffer${}^{e,f}$

\vspace{1em}

\begin{center}
\it
${}^a$Albert Einstein Center for Fundamental Physics, Institute for Theoretical Physics, \\
University of Bern, Sidlerstrasse~5, 3012 Bern, Switzerland \\
\mbox{} \\
${}^b$Institute for Nuclear Theory, University of Washington, Seattle, WA 98195-1550, USA \\
\mbox{}\\
${}^c$Kavli Institute for Theoretical Physics, University of California, Santa Barbara, CA 93106, USA\\
\mbox{} \\
${}^d$Theoretical Physics Department, CERN, Geneva, Switzerland \\
\mbox{} \\
${}^e$Helmholtz-Institut f\"ur Strahlen- und Kernphysik (Theory) and Bethe Center for Theoretical Physics, University of Bonn, 53115 Bonn, Germany\\
\mbox{}\\
${}^f$Department of Physics, University of California at San Diego, La Jolla, CA 92093, USA
\end{center} 

\end{center}

\vspace{3em}

\hrule

\begin{abstract}
  In this third paper of a series dedicated to a dispersive treatment of
  the hadronic light-by-light (HLbL) tensor, we derive a partial-wave
  formulation for two-pion intermediate states in the HLbL contribution to
  the anomalous magnetic moment of the muon $(g-2)_\mu$, including a
  detailed discussion of the unitarity relation for arbitrary partial
  waves. We show that obtaining a final expression free from unphysical
  helicity partial waves is a subtle issue, which we thoroughly clarify. As
  a by-product, we obtain a set of sum rules that could be used to
  constrain future calculations of $\gamma^*\gamma^*\to\pi\pi$. We validate
  the formalism extensively using the pion-box contribution, defined by
  two-pion intermediate states with a pion-pole left-hand cut, and
  demonstrate how the full known result is reproduced when resumming the
  partial waves. Using dispersive fits to high-statistics data for the pion
  vector form factor, we provide an evaluation of the full pion box,
  $a_\mu^{\pi\text{-box}}=-15.9(2)\times 10^{-11}$. As an application of
  the partial-wave formalism, we present a first calculation of
  $\pi\pi$-rescattering effects in HLbL scattering, with
  $\gamma^*\gamma^*\to\pi\pi$ helicity partial waves constructed
  dispersively using $\pi\pi$ phase shifts derived from the
  inverse-amplitude method. In this way, the isospin-$0$ part of our
  calculation can be interpreted as the contribution of the $f_0(500)$ to
  HLbL scattering in $(g-2)_\mu$. We argue that the contribution due to
  charged-pion rescattering implements corrections related to the
  corresponding pion polarizability and show that these are moderate. Our
  final result for the sum of pion-box contribution and its $S$-wave rescattering
  corrections reads $a_\mu^{\pi\text{-box}} +
  a_{\mu,J=0}^{\pi\pi,\pi\text{-pole LHC}}=-24(1)\times 10^{-11}$.
\end{abstract}

\hrule

\newpage

\setcounter{tocdepth}{3}
\tableofcontents

\numberwithin{equation}{section}


\section{Introduction}

\begin{figure}[t]
	\begin{center}
		\includegraphics[width=4cm]{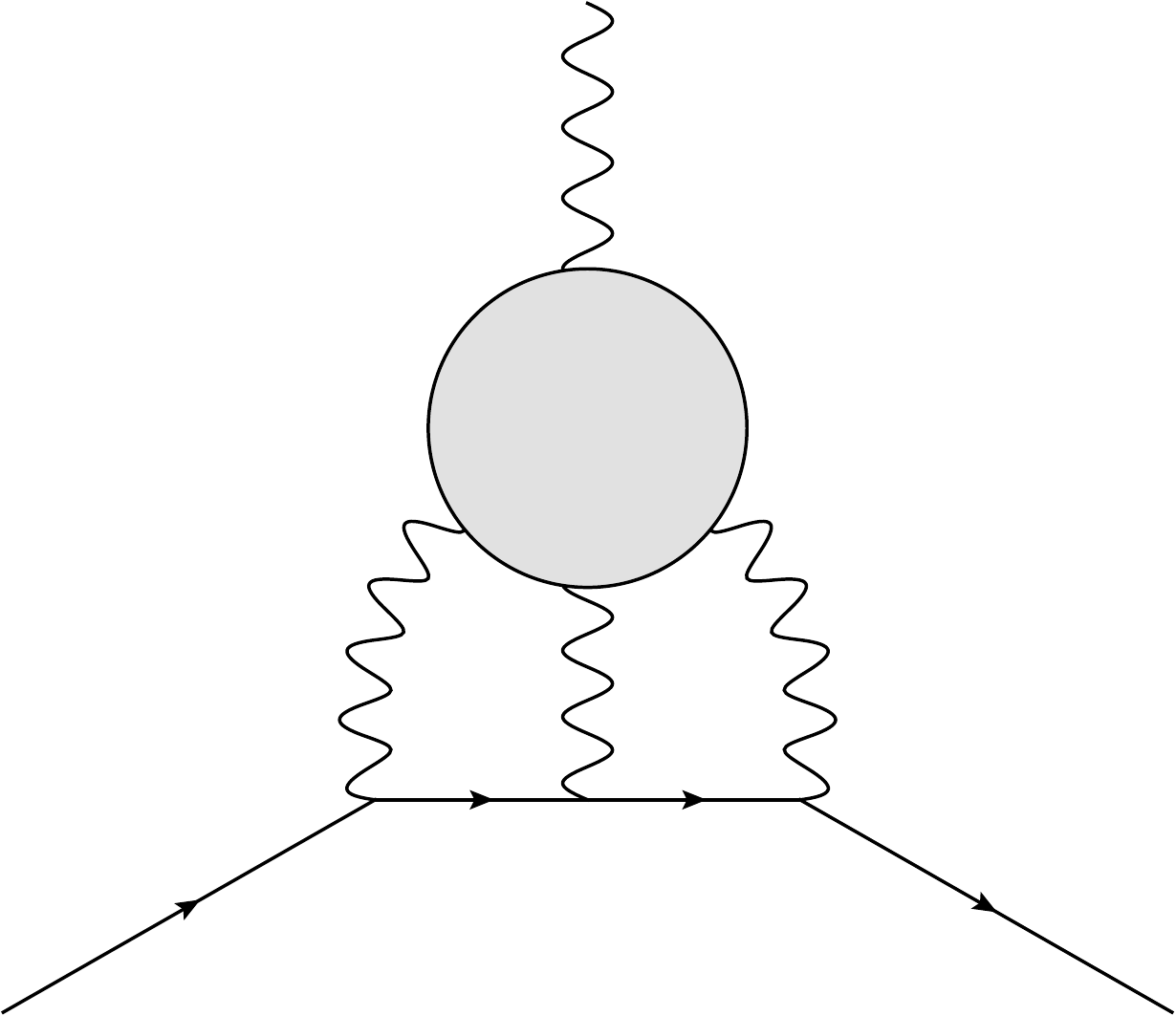}
		\caption{HLbL contribution to the anomalous magnetic moment of the muon $(g-2)_\mu$.}
		\label{img:HLbLinGminus2}
	\end{center}
\end{figure}

The long-standing discrepancy between the standard-model determination and the experimental measurement~\cite{Bennett:2006fi} (updated to the latest muon--proton magnetic moment ratio~\cite{Mohr:2015ccw})
\beq
a_\mu^\text{exp}=116\ 592\ 089(63)\times 10^{-11}
\eeq
of the anomalous magnetic moment of the muon $(g-2)_\mu$ has triggered substantial interest in the subject on both the theoretical and the experimental side. The ongoing E989 experiment at Fermilab~\cite{Grange:2015fou} as well as complementary efforts by J-PARC E34~\cite{Saito:2012zz}
aim at improving the precision by a factor of $4$, see~\cite{Gorringe:2015cma} for a detailed account of the experimental strategies in both cases. On the theory side, the uncertainty is dominated by hadronic effects~\cite{Jegerlehner:2009ry,Prades:2009tw,Benayoun:2014tra}, while QED~\cite{Aoyama:2012wk} and electroweak~\cite{Gnendiger:2013pva} contributions are under control at the level of at least $1\times10^{-11}$. Currently, the dominant source of hadronic uncertainties is hadronic vacuum polarization (HVP) at $\O(\alpha^2)$ in the fine-structure constant, closely followed by the $\O(\alpha^3)$
hadronic light-by-light (HLbL) contribution, depicted in Fig.~\ref{img:HLbLinGminus2}, and with higher-order insertions of the same hadronic amplitudes already under sufficient control~\cite{Calmet:1976kd,Hagiwara:2011af,Kurz:2014wya,Colangelo:2014qya}.
In view of improved data input for the dispersion relation for HVP~\cite{Blum:2013xva}, it is likely that the stumbling block will eventually become the sub-leading HLbL contribution.

Current estimates for HLbL scattering in $(g-2)_\mu$ are largely based on hadronic models~\cite{deRafael:1993za,Bijnens:1995cc,Bijnens:1995xf,Bijnens:2001cq,Hayakawa:1995ps,Hayakawa:1996ki,Hayakawa:1997rq,Knecht:2001qg,Knecht:2001qf,RamseyMusolf:2002cy,Melnikov:2003xd,Goecke:2010if}, which despite implementing different limits of QCD, such as large-$N_c$, chiral symmetry, or constraints from perturbative QCD, all involve a certain amount of uncontrollable uncertainties without offering a systematic path forward.  
In order to improve the determination of the HLbL contribution, we proposed a dispersive framework~\cite{Colangelo:2014dfa}, based on the fundamental principles of analyticity, unitarity, gauge invariance, and crossing symmetry, which opens up a path towards a data-driven evaluation~\cite{Colangelo:2014pva}. 
As the next step~\cite{Stoffer:2014rka,Colangelo:2015ama}, we presented 
a comprehensive solution to the task of constructing a basis for the HLbL tensor devoid of kinematic singularities, defining scalar functions that are amenable to a dispersive treatment.
In particular, we derived a Lorentz decomposition of the HLbL tensor that manifestly implements crossing symmetry and gauge invariance, with scalar coefficient functions free of kinematic singularities and zeros that fulfill the Mandelstam double-spectral representation. In this framework, we worked out how to define unambiguously and in a model-independent way both the pion-pole and the pion-box contribution.\footnote{For a dispersive approach not for the HLbL tensor, but for the Pauli form factor instead see~\cite{Pauk:2014rfa}. Complementary to the dispersive approach, 
a model-independent determination of the HLbL contribution could be achieved using lattice QCD, see~\cite{Blum:2014oka,Green:2015sra,Blum:2015gfa,Gerardin:2016cqj,Blum:2016lnc} for recent progress in this direction.}

With pion- as well as $\eta$-, $\eta'$-pole contributions determined by their doubly-virtual transition form factors, which by themselves are strongly constrained by unitarity, analyticity, and perturbative QCD in combination with experimental data~\cite{Stollenwerk:2011zz,Niecknig:2012sj,Schneider:2012ez,Hoferichter:2012pm,Hanhart:2013vba,Hoferichter:2014vra,Kubis:2015sga,Xiao:2015uva,Nyffeler:2016gnb}, 
we here apply our framework to extend the partial-wave formulation of two-pion rescattering effects for $S$-waves~\cite{Colangelo:2014dfa} to arbitrary partial waves. To this end, we identify a special set of (unambiguously defined) scalar functions that fulfill unsubtracted dispersion relations and can be expressed as linear combinations of helicity amplitudes. Their imaginary part, the input required in the dispersion relations, is provided in terms of helicity partial waves for $\gamma^*\gamma^*\to\pi\pi$ by means of unitarity. Working out explicitly the basis change to the helicity amplitudes, we generalize the unitarity relation derived in~\cite{Colangelo:2014dfa} up to $D$-waves only to arbitrary partial waves. We demonstrate that indeed the summation of the partial waves reproduces the known full result for the pion box, to which the $\pi\pi$-rescattering contribution is expected to produce the dominant correction. We provide the details of a first numerical analysis~\cite{Colangelo:2017qdm} of these rescattering effects based on helicity partial waves for $\gamma^*\gamma^*\to\pi\pi$ that we construct dispersively from a pion-pole left-hand cut (LHC) and $\pi\pi$ phase shifts from the inverse-amplitude method, an approach that isolates pure $\pi\pi$ contributions and thus, in the isospin-$0$ channel, provides an estimate for the impact of the $f_0(500)$ resonance on HLbL scattering.
In the same way, our $\gamma^*\gamma^*\to\pi\pi$ amplitudes reproduce the phenomenological value for the charged-pion polarizability, thereby clarifying the role of the associated corrections in $(g-2)_\mu$~\cite{Engel:2012xb,Engel:2013kda,Bijnens:2016hgx}. 
These results lay the groundwork for a future global analysis of two-meson intermediate states in the HLbL contribution.

The outline is as follows: Sect.~\ref{sec:HelicityFormalism} is devoted to a thorough derivation of partial-wave dispersion relations for the HLbL tensor, with tensor decomposition, dispersion relations, sum rules, and partial-wave expansion addressed in Sects.~\ref{sec:TensorDecomposition}--\ref{sec:HelAmpPWE}. A short summary of the strategy is provided at the beginning of Sect.~\ref{sec:HelicityFormalism}, complemented by a summary of the most important results in Sect.~\ref{sec:FormalismOverview}.
In Sect.~\ref{sec:PionBoxTests}, a numerical evaluation of the pion box is provided based on fits of the pion vector form factor to high-statistics time-like and space-like data. The pion box is further used to explicitly verify the general results derived in Sect.~\ref{sec:HelicityFormalism}, in particular to demonstrate the convergence of the partial-wave expansion for its contribution to $(g-2)_\mu$.  
Rescattering corrections to the pion box are discussed in Sect.~\ref{sec:Rescattering}, including a numerical analysis of the $S$-wave contribution, before we conclude in Sect.~\ref{sec:Conclusion}. 
Further details of the formalism are provided in the Appendices.


\section{Helicity formalism for HLbL}

\label{sec:HelicityFormalism}

In this section, we derive the formalism for the evaluation of the HLbL
two-pion contribution to $(g-2)_\mu$. The goal of our treatment is to
relate this contribution to helicity partial waves for the sub-process
$\gamma^*\gamma^*\to\pi\pi$, which in principle are measurable input
quantities or at least can be reconstructed dispersively. 

\begin{figure}[t]
	\centering
	\tikzstyle{block} = [rectangle, draw, fill=gray!20, align=center, rounded corners, minimum height=4em, inner sep=10pt]
	\tikzstyle{line} = [draw, -latex']
    
	\begin{tikzpicture}[node distance = 2cm, auto]
		\node[block] (BTT) {BTT tensor decomposition \\ Sects.~\ref{sec:TensorDecomposition} and \ref{sec:RelationToObservables}};
		\node[block, right of=BTT, node distance=8cm, double, thick] (mafo) {master formula~\eqref{eq:MasterFormulaPolarCoord}: \\[0.2cm] $\begin{aligned}a_\mu^\mathrm{HLbL} = \int d\tilde\Sigma\, dr\, d\phi \, (\ldots) \sum\limits_{i=1}^{12} T_i \; \bar\Pi_i\end{aligned}$};
		\node[block, below of=BTT, node distance=3cm] (mandel) {Mandelstam representation \\ Sect.~\ref{sec:DRforHLbL}};
		\node[block, right of=mandel, node distance=8cm, thick] (DR) {$\pi\pi$ dispersion relation~\eqref{eq:DRrescattering},~\eqref{eq:PicheckDispRel}: \\[0.3cm] $\begin{aligned}\check\Pi_i(s) = \int ds' \frac{\Im \check\Pi_i(s')}{s'-s} \end{aligned}$};
		\node[block, below of=mandel, node distance=3cm] (unitarity) {unitarity relation \\ Sect.~\ref{sec:HelAmpPWE}};
		\node[block, right of=unitarity, node distance=8cm, thick] (impart) {imaginary parts of scalar functions~\eqref{eq:HLbLScalarFunctionsPWImaginaryParts}: \\[0.3cm] $\begin{aligned}\Im\check\Pi_i(s) \propto \sum_j \check c_{ij} \sum_J  ( h_{J,\lambda_1\lambda_2} h^*_{J,\lambda_3\lambda_4} )_j \end{aligned}$};
		\node[block, below of=unitarity, node distance=3cm] (roysteiner) {Roy--Steiner equations for $\gamma^*\gamma^*\to\pi\pi$ \\ Sect.~\ref{sec:Rescattering}};
		\node[block, right of=roysteiner, node distance=8cm, thick] (helamps) {helicity partial waves for $\gamma^*\gamma^*\to\pi\pi$: \\[0.3cm] $\begin{aligned}h_{J,\lambda_1\lambda_2}(s,q_1^2,q_2^2) \end{aligned}$};

		\path [line, dashed] (BTT) -- (mafo);
		\path [line, dashed] (mandel) -- (DR);
		\path [line, dashed] (BTT) -- (DR);
		\path [line, dashed] (unitarity) -- (impart);
		\path [line, dashed] (roysteiner) -- (helamps);
		\path [line, double] (DR) -- (mafo);
		\path [line, double] (impart) -- (DR);
		\path [line, double] (helamps) -- (impart);
	\end{tikzpicture}
	\caption{Outline of the formalism for the HLbL two-pion contribution to $(g-2)_\mu$. The dashed lines denote a derivation or calculation, the double lines indicate the insertion of results.}
	\label{img:Gm2CalculationLogic}
\end{figure}
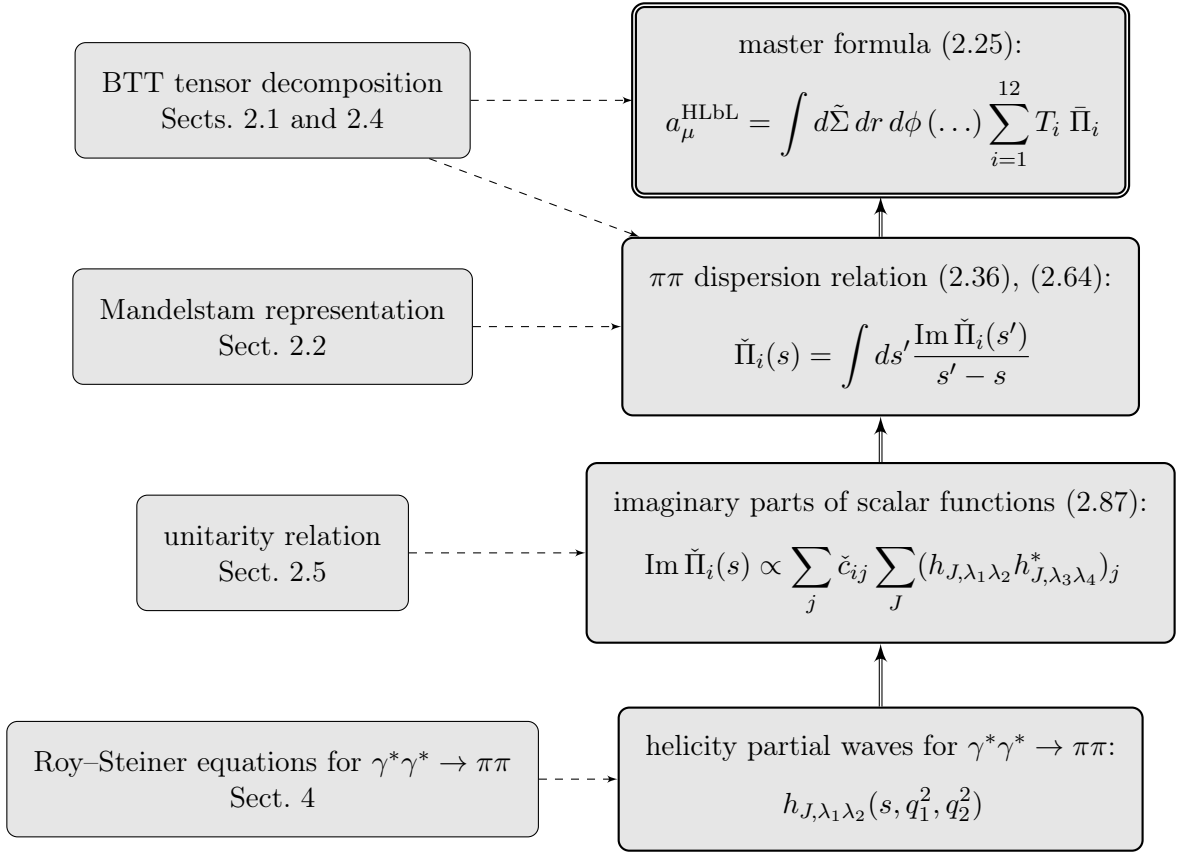

The outline of this derivation is illustrated as a flowchart in
Fig.~\ref{img:Gm2CalculationLogic}. The first step is the decomposition of
the HLbL tensor into Lorentz structures and scalar functions that are free
of kinematic singularities and zeros. We have solved this problem
in~\cite{Colangelo:2015ama} and recapitulate the results in
Sect.~\ref{sec:TensorDecomposition}. This representation, referred to as
BTT tensor decomposition~\cite{Bardeen:1969aw,Tarrach:1975tu} in
Fig.~\ref{img:Gm2CalculationLogic}, allows us to write the HLbL
contribution to $(g-2)_\mu$ in full generality as a master formula that
involves only three integrals. This master
formula~\eqref{eq:MasterFormulaPolarCoord} applies to any conceivable HLbL
tensor, as long as it is consistent with general properties that should be
fulfilled by any admissible HLbL amplitude: gauge invariance, crossing
symmetry, and the principle of maximal analyticity~\cite{Eden:1966}, i.e.\
the principle that the scattering amplitude can be represented by a complex
function that exhibits no further singularities except for those required
by unitarity and crossing symmetry. Any such singularities are of dynamical
origin, and thus have to be contained within the scalar functions $\bar
\Pi_i$ in the master formula. Phrased differently, if a given amplitude for
the HLbL tensor cannot be expressed in the BTT basis, e.g.\ due to the
appearance of kinematic singularities, this automatically implies that this
amplitude is at odds with said general properties.

The dynamics of HLbL scattering is thus contained in the scalar functions,
which are the objects that we describe
dispersively. In~\cite{Colangelo:2015ama}, we have used the Mandelstam
representation for the scalar functions to study the pion-box
contribution. In Sect.~\ref{sec:DRforHLbL}, we extend the dispersive
treatment and derive from the Mandelstam representation single-variable
dispersion relations for general two-pion contributions. Combining these
single-variable dispersion relations with unitarity constraints requires a
basis change to helicity amplitudes, since the partial-wave unitarity
relation becomes diagonal only for definite helicity amplitudes. However,
this basis change is complicated by the appearance of redundancies in the
representation which, together with the requirement that longitudinal
polarizations for on-shell photons not contribute in the final HLbL
representation, necessitates a more careful study of the BTT scalar
functions and their relation to helicity amplitudes. The solution to this
problem is the explicit derivation of a basis that removes all redundancies
and apparent contributions from unphysical polarizations, which is
presented in Sect.~\ref{sec:RelationToObservables}. As a by-product we
obtain a set of physical sum rules to be fulfilled by the scalar functions
and thereby the helicity amplitudes.  

After the basis change to helicity amplitudes, we can then employ the
unitarity relation to determine the imaginary parts in the dispersion
integrals in terms of helicity amplitudes for
$\gamma^*\gamma^*\to\pi\pi$. In particular, we perform a partial-wave
expansion of the helicity amplitudes and generalize the $S$-wave result
of~\cite{Colangelo:2014dfa} to arbitrary partial waves,
which is the main result of Sect.~\ref{sec:HelAmpPWE}. 
In performing this analysis the partial waves for
$\gamma^*\gamma^*\to\pi\pi$ are treated as known, given quantities, which
unfortunately they are not. The lack of experimental information can be
partly compensated by theory constraints, in particular by dispersion
relations in the form of Roy--Steiner
equations~\cite{GarciaMartin:2010cw,Hoferichter:2011wk,Moussallam:2013una,Hoferichter:2013ama}.
A simplified, $S$-wave variant of these will be solved in
Sect.~\ref{sec:Rescattering}.  

A summary of the main results is provided in
Sect.~\ref{sec:FormalismOverview}, including a glossary of the notation for
the scalar functions. The subtleties in the various basis changes
unfortunately require the introduction of different sets of scalar
functions, whose dimension, defining equation, and main properties are
summarized in Table~\ref{tab:FormalismOverview}.

\subsection[Tensor decomposition and master formula for $(g-2)_\mu$]{\boldmath Tensor decomposition and master formula for $(g-2)_\mu$}

\label{sec:TensorDecomposition}

In this subsection, we recapitulate the decomposition of the HLbL tensor
into a sum of gauge-invariant Lorentz structures times scalar functions
that are free of kinematic singularities. We slightly modify and improve
the master formula presented in~\cite{Stoffer:2014rka, Colangelo:2015ama}
in such a way that crossing symmetry between all three off-shell photons
remains manifest. The dynamical input in the master formula is encoded in
only six different scalar functions and their crossed versions. 

\subsubsection{BTT decomposition of the HLbL tensor}

The HLbL tensor is defined as the hadronic Green's function of four electromagnetic currents in pure QCD:
\begin{align}
	\label{eq:HLbLTensorDefinition}
	\Pi^{\mu\nu\lambda\sigma}(q_1,q_2,q_3) = -i \int d^4x \, d^4y \, d^4z \, e^{-i(q_1 \cdot x + q_2 \cdot y + q_3 \cdot z)} \< 0 | T \{ j_\mathrm{em}^\mu(x) j_\mathrm{em}^\nu(y) j_\mathrm{em}^\lambda(z) j_\mathrm{em}^\sigma(0) \} | 0 \> ,
\end{align}
where the electromagnetic current includes the three lightest quarks:
\begin{align}
	j_\mathrm{em}^\mu := \bar q Q \gamma^\mu q , \quad q = ( u , d, s )^T , \quad Q = \mathrm{diag}\left(\frac{2}{3}, -\frac{1}{3}, -\frac{1}{3}\right) .
\end{align}

The hadronic contribution to the helicity amplitudes for (off-shell) photon--photon scattering is given by the contraction of the HLbL tensor with polarization vectors:
\begin{align}
	\label{eq:HLbLHelicityAmplitudesDefinition}
	H_{\lambda_1\lambda_2,\lambda_3\lambda_4} = \epsilon_\mu^{\lambda_1}(q_1) \epsilon_\nu^{\lambda_2}(q_2) {\epsilon_\lambda^{\lambda_3}}^*(-q_3) {\epsilon_\sigma^{\lambda_4}}^*(q_4) \Pi^{\mu\nu\lambda\sigma}(q_1,q_2,q_3) ,
\end{align}
where $q_4 = q_1 + q_2 + q_3$.

The usual Mandelstam variables
\begin{align}
	s := (q_1+q_2)^2, \quad t := (q_1+q_3)^2, \quad u := (q_2 + q_3)^2
\end{align}
fulfill the linear relation
\begin{align}
	s + t + u = \sum_{i=1}^4 q_i^2 =: \Sigma .
\end{align}
Gauge invariance requires the HLbL tensor to satisfy the Ward--Takahashi identities
\begin{align}
	\label{eq:WardIdentitiesHLbLTensor}
	\{q_1^\mu, q_2^\nu, q_3^\lambda, q_4^\sigma\} \Pi_{\mu\nu\lambda\sigma}(q_1,q_2,q_3) = 0 .
\end{align}

Based on a recipe by Bardeen, Tung~\cite{Bardeen:1969aw}, and Tarrach~\cite{Tarrach:1975tu} (BTT), we have derived in~\cite{Stoffer:2014rka,Colangelo:2015ama} a decomposition of the HLbL tensor
\begin{align}
	\label{eqn:HLbLTensorKinematicFreeStructures}
	\Pi^{\mu\nu\lambda\sigma} &= \sum_{i=1}^{54} T_i^{\mu\nu\lambda\sigma} \Pi_i ,
\end{align}
with tensor structures reproduced here for completeness (all remaining ones follow from crossing symmetry~\cite{Colangelo:2015ama}) 
\begin{align*}
		\label{eq:HLbLBTTStructures}
		T_1^{\mu\nu\lambda\sigma} &= \epsilon^{\mu\nu\alpha\beta} \epsilon^{\lambda\sigma\gamma\delta} {q_1}_\alpha {q_2}_\beta {q_3}_\gamma {q_4}_\delta , \\
		T_4^{\mu\nu\lambda\sigma} &= \Big(q_2^\mu q_1^\nu - q_1 \cdot q_2 g^{\mu \nu} \Big) \Big( q_4^\lambda q_3^\sigma - q_3 \cdot q_4 g^{\lambda \sigma} \Big) , \\
		T_7^{\mu\nu\lambda\sigma} &= \Big(q_2^\mu q_1^\nu - q_1 \cdot q_2 g^{\mu \nu } \Big) \Big( q_1 \cdot q_4 \left(q_1^\lambda q_3^\sigma -q_1 \cdot q_3 g^{\lambda \sigma} \right) + q_4^\lambda q_1^\sigma q_1 \cdot q_3 - q_1^\lambda q_1^\sigma q_3 \cdot q_4 \Big) , \\
		 T_{19}^{\mu\nu\lambda\sigma} &= \Big( q_2^\mu q_1^\nu - q_1 \cdot q_2 g^{\mu \nu } \Big) \Big(q_2 \cdot q_4 \left(q_1^\lambda q_3^\sigma - q_1 \cdot q_3 g^{\lambda\sigma} \right)+q_4^\lambda q_2^\sigma q_1 \cdot q_3 - q_1^\lambda q_2^\sigma q_3 \cdot q_4 \Big) , \\
		T_{31}^{\mu\nu\lambda\sigma} &= \Big(q_2^\mu q_1^\nu - q_1\cdot q_2 g^{\mu\nu}\Big) \Big(q_2^\lambda q_1\cdot q_3 - q_1^\lambda q_2\cdot q_3\Big) \Big(q_2^\sigma q_1\cdot q_4 - q_1^\sigma q_2\cdot q_4\Big) , \\
		T_{37}^{\mu\nu\lambda\sigma} &= \Big( q_3^\mu q_1\cdot q_4 - q_4^\mu q_1\cdot q_3\Big) \begin{aligned}[t]
			& \Big( q_3^\nu q_4^\lambda q_2^\sigma - q_4^\nu q_2^\lambda q_3^\sigma + g^{\lambda\sigma} \left(q_4^\nu q_2\cdot q_3 - q_3^\nu q_2\cdot q_4\right) \\
			& + g^{\nu\sigma} \left( q_2^\lambda q_3\cdot q_4 - q_4^\lambda q_2\cdot q_3 \right) + g^{\lambda\nu} \left( q_3^\sigma q_2\cdot q_4 - q_2^\sigma q_3\cdot q_4 \right) \Big) , \end{aligned} \\
		T_{49}^{\mu\nu\lambda\sigma} &= q_3^\sigma  \begin{aligned}[t]
				& \Big( q_1\cdot q_3 q_2\cdot q_4 q_4^\mu g^{\lambda\nu} - q_2\cdot q_3 q_1\cdot q_4 q_4^\nu g^{\lambda\mu} + q_4^\mu q_4^\nu \left( q_1^\lambda q_2\cdot q_3 - q_2^\lambda q_1\cdot q_3 \right) \\
				& + q_1\cdot q_4 q_3^\mu q_4^\nu q_2^\lambda - q_2\cdot q_4 q_4^\mu q_3^\nu q_1^\lambda + q_1\cdot q_4 q_2\cdot q_4 \left(q_3^\nu g^{\lambda\mu} - q_3^\mu g^{\lambda\nu}\right) \Big) \end{aligned} \\
			& - q_4^\lambda \begin{aligned}[t]
				& \Big( q_1\cdot q_4 q_2\cdot q_3 q_3^\mu g^{\nu\sigma} - q_2\cdot q_4 q_1\cdot q_3 q_3^\nu g^{\mu\sigma} + q_3^\mu q_3^\nu \left(q_1^\sigma q_2\cdot q_4 - q_2^\sigma q_1\cdot q_4\right) \\
				& + q_1\cdot q_3 q_4^\mu q_3^\nu q_2^\sigma - q_2\cdot q_3 q_3^\mu q_4^\nu q_1^\sigma + q_1\cdot q_3 q_2\cdot q_3 \left( q_4^\nu g^{\mu\sigma} - q_4^\mu g^{\nu\sigma} \right) \Big) \end{aligned} \mytag\\
			& + q_3\cdot q_4 \Big(\left(q_1^\lambda q_4^\mu - q_1\cdot q_4 g^{\lambda\mu}\right) \left(q_3^\nu q_2^\sigma - q_2\cdot q_3 g^{\nu\sigma}\right) - \left(q_2^\lambda q_4^\nu - q_2\cdot q_4 g^{\lambda\nu}\right) \left(q_3^\mu q_1^\sigma - q_1\cdot q_3 g^{\mu\sigma}\right)\Big) .
\end{align*}
The BTT decomposition has the following properties:
\begin{itemize}
	\item all the Lorentz structures fulfill the Ward--Takahashi identities, i.e.
	\begin{align}
		\{q_1^\mu, q_2^\nu, q_3^\lambda, q_4^\sigma\} T^i_{\mu\nu\lambda\sigma}(q_1,q_2,q_3) = 0 , \quad \forall i \in \{1, \ldots, 54 \} ,
	\end{align}
	\item there are only seven distinct Lorentz structures, the remaining 47 ones are crossed versions thereof,
	\item the scalar functions $\Pi_i$ are free of kinematic singularities and zeros.
\end{itemize}
The first two properties make gauge invariance and crossing symmetry manifest, while the third property provides the foundation for writing dispersion relations: in a dispersive treatment, we exploit the analytic structure of the scalar functions dictated by unitarity and we have to make sure that the singularity structure due to the hadronic dynamics is not entangled with kinematic singularities.

Since the number of helicity amplitudes for fully off-shell photon--photon scattering is 41, the set of 54 structures $\{ T_i^{\mu\nu\lambda\sigma}\}$ does not form a basis, but exhibits a 13-fold redundancy, as we discussed in detail in~\cite{Colangelo:2015ama}. While 11 linear relations hold in general, two additional ones are present in four space-time dimensions~\cite{Eichmann:2014ooa}. Away from four space-time dimensions, a subset of 43 Lorentz structures forms a basis:
\begin{align}
	\label{eqn:HLbLTensorBasisDecomposition}
	\Pi^{\mu\nu\lambda\sigma} &= \sum_{i=1}^{43} \mathcal{B}_i^{\mu\nu\lambda\sigma} \tilde\Pi_i ,
\end{align}
where the basis-coefficient functions $\tilde\Pi_i$ are no longer free of kinematic singularities. However, the explicit structure of their kinematic singularities follows from the projection of the BTT decomposition onto this ``basis.''

\subsubsection[Master formula for the HLbL contribution to $(g-2)_\mu$]{\boldmath Master formula for the HLbL contribution to $(g-2)_\mu$}

\label{sec:MasterFormula}

Based on a projection technique in Dirac space, one can extract the HLbL contribution to $a_\mu := (g-2)_\mu/2$ from the following expression:
\begin{align}
	\begin{split}
		a_\mu^\mathrm{HLbL} &= - \frac{e^6}{48 m_\mu}  \int \frac{d^4q_1}{(2\pi)^4} \frac{d^4q_2}{(2\pi)^4} \frac{1}{q_1^2 q_2^2 (q_1+q_2)^2} \frac{1}{(p+q_1)^2 - m_\mu^2} \frac{1}{(p-q_2)^2 - m_\mu^2} \\
			& \quad \times \mathrm{Tr}\left( (\slashed p + m_\mu) [\gamma^\rho,\gamma^\sigma] (\slashed p + m_\mu) \gamma^\mu (\slashed p + \slashed q_1 + m_\mu) \gamma^\lambda (\slashed p - \slashed q_2 + m_\mu) \gamma^\nu \right)  \\
			& \quad \times  \sum_{i=1}^{54} \left( \frac{\p}{\p q_4^\rho} T^i_{\mu\nu\lambda\sigma}(q_1,q_2,q_4-q_1-q_2) \right) \bigg|_{q_4=0} \Pi_i(q_1,q_2,-q_1-q_2) .
	\end{split}
\end{align}
There are only 19 independent linear combinations of the structures $T_i^{\mu\nu\lambda\sigma}$ that contribute to $(g-2)_\mu$, hence we can make a basis change in the 54 structures
\begin{align}
	\label{eq:HLbLTensorPiHatDecomposition}
	\Pi^{\mu\nu\lambda\sigma} = \sum_{i=1}^{54} T_i^{\mu\nu\lambda\sigma} \Pi_i =  \sum_{i=1}^{54} \hat T_i^{\mu\nu\lambda\sigma} \hat \Pi_i ,
\end{align}
in such a way that in the limit $q_4\to0$ the derivative of 35 structures
$\hat T_i^{\mu\nu\lambda\sigma}$ vanishes. Since the loop integral and the
propagators are symmetric under $q_1 \leftrightarrow -q_2$,
in~\cite{Colangelo:2015ama} we made sure to preserve crossing symmetry
under exchange of $q_1$ and $q_2$, but did not yet exploit the fact that it
is even possible to preserve crossing symmetry between all three off-shell
photons---the limit $q_4\to0$ singles out one of the photons, but the
remaining three are completely equivalent. For the sake of simplifying
further calculations, we present here new structures $\hat
T_i^{\mu\nu\lambda\sigma}$ and the corresponding scalar functions $\hat 
\Pi_i$, superseding the ones given in~\cite{Colangelo:2015ama}. 

The 19 structures $\hat T_i^{\mu\nu\lambda\sigma}$ contributing to
$(g-2)_\mu$ can be chosen as follows: 
\begin{align}
	\begin{split}
		\label{eq:ThatStructures}
		\hat T_i^{\mu\nu\lambda\sigma} &=
                T_i^{\mu\nu\lambda\sigma}, \quad i=1,\ldots, 11, 13, 14,
                16, 17, 50, 51, 54, \\ 
		\hat T_{39}^{\mu\nu\lambda\sigma} &= \frac{1}{3} \left(
                  T_{39}^{\mu\nu\lambda\sigma} +
                  T_{40}^{\mu\nu\lambda\sigma} +
                  T_{46}^{\mu\nu\lambda\sigma} \right) . 
	\end{split}
\end{align}
The 35 structures
\begin{align}
	\big\{ \hat T_i^{\mu\nu\lambda\sigma} \big| i = 12, 15, 18, \ldots,
        38, 40, \ldots, 49, 52, 53 \big\} 
\end{align}
do not contribute to $(g-2)_\mu$ and are given in App.~\ref{sec:BasisChangeGm2}.

The set of 19 linear combinations of scalar functions that give a
contribution to $(g-2)_\mu$ is defined by (replacing Eq.~(D.1)
in~\cite{Colangelo:2015ama}) 
\begin{align*}
	\label{eq:PiHatFunctions}
	\hat\Pi_1 &= \Pi_1 + q_1 \cdot q_2 \Pi_{47} , \\
	\hat\Pi_4 &= \Pi_4 - q_1 \cdot q_3 \left( \Pi_{19} - \Pi_{42}
        \right) - q_2 \cdot q_3 \left( \Pi_{20} - \Pi_{43} \right) + q_1
        \cdot q_3  q_2 \cdot q_3 \Pi_{31} , \\ 
	\hat \Pi_7 &= \Pi_7 - \Pi_{19} + q_2 \cdot q_3 \Pi_{31} , \\
	\hat \Pi_{17} &= \Pi_{17} + \Pi_{42} + \Pi_{43} - \Pi_{47} , \\
	\hat \Pi_{39} &= \Pi_{39} + \Pi_{40} + \Pi_{46} , \\
	\hat \Pi_{54} &= \Pi_{42} - \Pi_{43} + \Pi_{54} ,
	\mytag
\end{align*}
together with the crossed versions thereof
\begin{align*}
	\label{eq:CrossingRelationsPiHat}
	\hat \Pi_2 &= \Cr{23}{\hat \Pi_1} , \quad \hat \Pi_3 = \Cr{13}{\hat
          \Pi_1} , \quad 
	\hat \Pi_5 = \Cr{23}{\hat \Pi_4} , \quad \hat \Pi_6 = \Cr{13}{\hat
          \Pi_4} , \\ 
	\hat \Pi_8 &= \Cr{12}{\hat \Pi_7} , \quad \hat \Pi_9 =
        \Cr{12}{\Cr{13}{\hat \Pi_7}} , \quad \hat \Pi_{10} = \Cr{23}{\hat
          \Pi_7} , \quad \hat \Pi_{13} = \Cr{13}{\hat \Pi_7} , \quad \hat
        \Pi_{14} = \Cr{12}{\Cr{23}{\hat \Pi_7}} , \\ 
	\hat \Pi_{11} &= \Cr{13}{\hat \Pi_{17}} , \quad \hat \Pi_{16} =
        \Cr{23}{\hat \Pi_{17}} , \quad 
	\hat \Pi_{50} = -\Cr{23}{\hat \Pi_{54}} , \quad \hat \Pi_{51} = 
        \Cr{13}{\hat \Pi_{54}} , 
	\mytag
\end{align*}
where the crossing operators $\mathcal{C}_{ij}$ exchange momenta and
Lorentz indices of the photons $i$ and $j$, e.g.\footnote{The composition
  of two crossing operators is understood to act e.g.\ in the following
  way: $\mathcal{C}_{12}[\mathcal{C}_{23}[f(q_1,q_2,q_3,q_4)]] =
  \mathcal{C}_{12}[f(q_1,q_3,q_2,q_4)] =
  f(q_2,q_3,q_1,q_4)$.\vspace{0.1cm}} 
\begin{align}
	\label{eq:DefCrossingOperator}
	\mathcal{C}_{12}[f] := f( \mu \leftrightarrow \nu, q_1
        \leftrightarrow q_2 ) , \quad \mathcal{C}_{14}[f] := f( \mu
        \leftrightarrow \sigma, q_1 \leftrightarrow -q_4 ) . 
\end{align}
The following intrinsic crossing symmetries are preserved (we do not list
the symmetries involving the fourth photon): 
\begin{align*}
	\label{eq:InternalCrossingSymmetriesPiHat}
	\hat \Pi_1 &= \Cr{12}{\hat \Pi_1} , \quad \hat \Pi_4 = \Cr{12}{\hat
          \Pi_4} , \quad \hat \Pi_{17} = \Cr{12}{\hat \Pi_{17}} , \\ 
	\hat \Pi_{39} &= \Cr{12}{\hat \Pi_{39}} = \Cr{13}{\hat \Pi_{39}} =
        \ldots , \quad \hat \Pi_{54} = - \Cr{12}{\hat \Pi_{54}}, 
	\mytag
\end{align*}
where the dots denote three more crossing relations that follow from the
given ones. Hence, the scalar functions $\hat \Pi_i$ contributing to
$(g-2)_\mu$ fall into only six distinct classes that are closed under
crossing symmetry of the off-shell photons 1, 2, and 3. Apart from
$\hat\Pi_{39}$, which is fully symmetric, the representatives
in~\eqref{eq:PiHatFunctions} are picked because they share a common
property: their $s$-channel is special as follows from the observation that
the corresponding Lorentz structures $\hat T_i^{\mu\nu\lambda\sigma}$ are
(anti-)symmetric under either $\Cr{12}{}$ or $\Cr{34}{}$ (or both). This is
reflected in the intrinsic crossing
symmetries~\eqref{eq:InternalCrossingSymmetriesPiHat}.\footnote{$\hat
  T_7^{\mu\nu\lambda\sigma}$ is symmetric under $\Cr{34}{}$, but not under 
  $\Cr{12}{}$. One could split the six elements in the crossing class of
  $\hat\Pi_7$ into two classes, one with an additional even, one with an
  odd intrinsic crossing symmetry.}

The HLbL contribution to $(g-2)_\mu$ can now be written as
\begin{align}
	\begin{split}
		a_\mu^\mathrm{HLbL} &= - e^6  \int \frac{d^4q_1}{(2\pi)^4} \frac{d^4q_2}{(2\pi)^4} \frac{1}{q_1^2 q_2^2 (q_1+q_2)^2} \frac{1}{(p+q_1)^2 - m_\mu^2} \frac{1}{(p-q_2)^2 - m_\mu^2} \\
			& \quad \times  \sum_{i\in G} \hat T_i(q_1,q_2;p) \hat\Pi_i(q_1,q_2,-q_1-q_2) ,
	\end{split}
\end{align}
where $G := \{ 1,\ldots, 11, 13, 14, 16, 17, 39, 50, 51, 54 \}$ and
\begin{align}
	\begin{split}
		\label{eq:DefinitionIntermediateKernels}
		\hat T_i(q_1,q_2;p) := \frac{1}{48 m_\mu} \mathrm{Tr}&\left( (\slashed p + m_\mu) [\gamma^\rho,\gamma^\sigma] (\slashed p + m_\mu) \gamma^\mu (\slashed p + \slashed q_1 + m_\mu) \gamma^\lambda (\slashed p - \slashed q_2 + m_\mu) \gamma^\nu \right)  \\
			& \times \left( \frac{\p}{\p q_4^\rho} \hat T^i_{\mu\nu\lambda\sigma}(q_1,q_2,q_4-q_1-q_2) \right) \bigg|_{q_4=0} .
	\end{split}
\end{align}
As in~\cite{Colangelo:2015ama}, we perform a Wick rotation, average the
result over the direction of the Euclidean four-momentum of the muon, and
use the Gegenbauer polynomial technique~\cite{Rosner:1967zz} to perform
five of the eight integrals in full generality, i.e.\ without prior
knowledge of the functions $\hat \Pi_i$. The symmetry properties of the
loop integral and the kernels $\hat T_i$ under $q_1 \leftrightarrow -q_2$
allow us to write the master formula for the HLbL contribution to
$(g-2)_\mu$ containing a sum of only 12 terms: 
\begin{align}
	\label{eq:MasterFormula3Dim}
	a_\mu^\mathrm{HLbL} &= \frac{2 \alpha^3}{3 \pi^2} \int_0^\infty dQ_1 \int_0^\infty dQ_2 \int_{-1}^1 d\tau \sqrt{1-\tau^2} Q_1^3 Q_2^3 \sum_{i=1}^{12} T_i(Q_1,Q_2,\tau) \bar \Pi_i(Q_1,Q_2,\tau) ,
\end{align}
where $Q_1 := |Q_1|$ and $Q_2 := |Q_2|$ denote the norm of the Euclidean four-vectors. The 12 scalar functions $\bar \Pi_i$ are a subset of the functions $\hat \Pi_i$:
\begin{align}
	\begin{aligned}
		\label{eq:PibarFunctions}
		\bar \Pi_1 &= \hat\Pi_1 , \quad &
		\bar \Pi_2 &= \hat\Pi_2 , \quad &
		\bar \Pi_3 &= \hat\Pi_4 , \quad &
		\bar \Pi_4 &= \hat\Pi_5 , \quad &
		\bar \Pi_5 &= \hat\Pi_7 , \quad &
		\bar \Pi_6 &= \hat\Pi_9 , \\
		\bar \Pi_7 &= \hat\Pi_{10} , \quad &
		\bar \Pi_8 &= \hat\Pi_{11} , \quad &
		\bar \Pi_9 &= \hat\Pi_{17} , \quad &
		\bar \Pi_{10} &= \hat\Pi_{39} , \quad &
		\bar \Pi_{11} &= \hat\Pi_{50} , \quad &
		\bar \Pi_{12} &= \hat\Pi_{54} .
	\end{aligned}
\end{align}
They have to be evaluated for the reduced $(g-2)_\mu$ kinematics
\begin{align}
	\label{eq:Gm2Kinematics}
	s &= q_3^2 = - Q_3^2 = - Q_1^2 - 2 Q_1 Q_2 \tau - Q_2^2 , \quad
	t = q_2^2 = -Q_2^2 , \quad
	u = q_1^2 = -Q_1^2 , \quad
	q_4^2 = 0.
\end{align}

Due to the basis change, the kernel functions $T_i$ differ slightly from the ones given in~\cite{Colangelo:2015ama}. We provide the explicit expressions in App.~\ref{sec:MaFoKernels}.

\begin{figure}[t]
	\centering
	\includegraphics{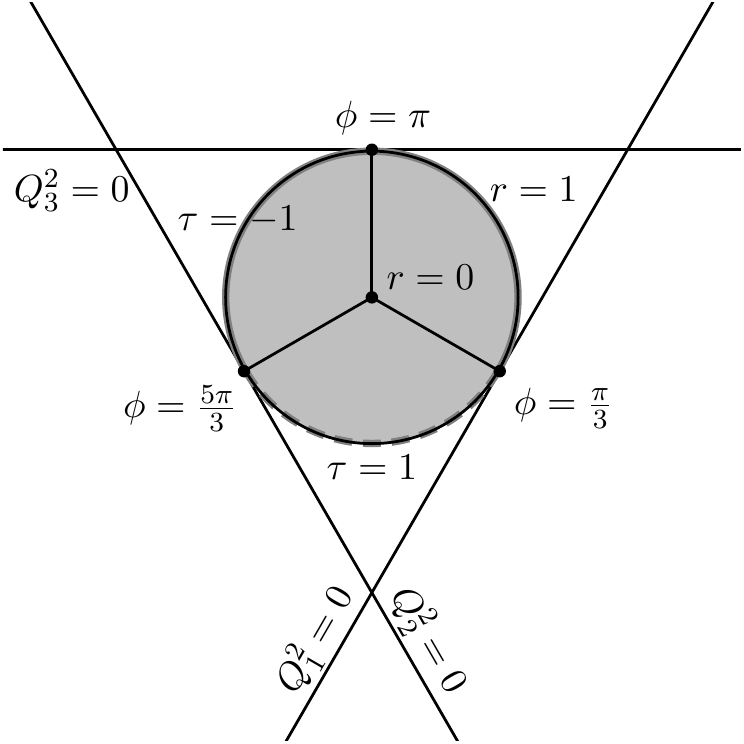}
	\caption{Integration region for $(g-2)_\mu$. The border of the integration region is at $r=1$ and corresponds to $\tau=-1$ for $\pi/3<\phi<5\pi/3$ (solid gray line) and $\tau=1$ otherwise (dashed gray line). The angles $\phi=\pi/3$, $\phi=\pi$, and $\phi=5\pi/3$ correspond to $Q_2^2=Q_3^2$, $Q_1^2=Q_2^2$, and $Q_1^2=Q_3^2$, respectively. The three points where one of the $Q_i^2$ is zero are singularities of the integration kernels. The height of the equilateral triangle is given by $\tilde\Sigma$.}
	\label{img:IntegrationRegionMasterFormula}
\end{figure}

In~\cite{Eichmann:2015nra} a different parametrization of the $(g-2)_\mu$ integration region has been proposed, which proved advantageous for the numerical implementation. We perform the following variable transformation in the master formula (note that $\tilde\Sigma = -\Sigma$ is the sum of the squared Euclidean virtualities, whereas $\Sigma$ denotes the sum of the squared Minkowskian virtualities):
\begin{align}
	\begin{split}
		Q_1^2 &= \frac{\tilde\Sigma}{3} \left( 1 - \frac{r}{2} \cos\phi - \frac{r}{2}\sqrt{3} \sin\phi \right) , \\
		Q_2^2 &= \frac{\tilde\Sigma}{3} \left( 1 - \frac{r}{2} \cos\phi + \frac{r}{2}\sqrt{3} \sin\phi \right) , \\
		Q_3^2 &= Q_1^2 + 2 Q_1 Q_2 \tau + Q_2^2 = \frac{\tilde\Sigma}{3} \left( 1 + r \cos\phi \right) .
	\end{split}
\end{align}
The range of integration is then $\tilde\Sigma \in [0, \infty)$, $r\in[0,1]$, and $\phi\in[0,2\pi]$. The integration region in the Mandelstam plane and the meaning of the variables is illustrated in Fig.~\ref{img:IntegrationRegionMasterFormula}.
After the variable transformation, the master formula becomes
\begin{align}
	\label{eq:MasterFormulaPolarCoord}
	a_\mu^\mathrm{HLbL} &= \frac{\alpha^3}{432\pi^2} \int_0^\infty d\tilde\Sigma\, \tilde\Sigma^3 \int_0^1 dr\, r\sqrt{1-r^2} \int_0^{2\pi} d\phi \,\sum_{i=1}^{12} T_i(Q_1,Q_2,\tau) \bar\Pi_i(Q_1,Q_2,\tau) ,
\end{align}
where $Q_1$, $Q_2$, and $\tau$ are understood as functions of $\tilde\Sigma$, $r$, and $\phi$.

The master formula for the HLbL contribution to $(g-2)_\mu$ is exact and
completely general: given any representation of the HLbL tensor, one can
project out the six scalar functions $\hat \Pi_i$
in~\eqref{eq:PiHatFunctions}. Using these and their crossed versions, one
can construct the 12 scalar functions $\bar \Pi_i$
in~\eqref{eq:PibarFunctions}, which encode the entire dynamical content of
HLbL scattering relevant for $(g-2)_\mu$. After their insertion into the
master formula~\eqref{eq:MasterFormulaPolarCoord}, only a three-dimensional
integral has to be carried out.

In a next step, we aim at reconstructing the scalar functions $\bar\Pi_i$
using dispersive methods, which will be the content of the remainder of
this section.


\subsection{Dispersion relations for the HLbL tensor}

\label{sec:DRforHLbL}

In this subsection, we discuss the dispersive framework that we employ for the reconstruction of the scalar functions. The starting point is the Mandelstam representation, which is a double-dispersion relation. Unitarity allows us to write the HLbL tensor as a sum of contributions from different intermediate states. After reviewing in Sect.~\ref{sec:MandelstamRepHLbL} the most important properties of the pion-pole and pion-box contributions, we continue by considering general two-pion intermediate states in Sect.~\ref{sec:TwoPionContributionsBeyondPionBox}.

In order to calculate the two-pion contributions beyond the pion box, input
on the sub-process $\gamma^*\gamma^*\to\pi\pi$ is needed. This input will
be in the form of helicity partial waves which, in principle, could be
measured or, in the absence of data on the doubly-virtual process, have to
be reconstructed
dispersively~\cite{GarciaMartin:2010cw,Hoferichter:2011wk,Moussallam:2013una,Hoferichter:2013ama}.
The partial-wave expansion turns, however, the amplitude into a polynomial in
the crossed-channel Mandelstam variables, i.e.\ the cut structure in the
crossed channel due to heavier (e.g.\ multi-pion) intermediate states gets
lost. Therefore, with $\gamma^*\gamma^*\to\pi\pi$ helicity partial waves as
input, one has to use a single-variable dispersion relation. We derive in
Sect.~\ref{sec:SingleVariableDR} a suitable form for such a dispersion
relation that follows from the Mandelstam representation.

\subsubsection{Mandelstam representation for HLbL}

\label{sec:MandelstamRepHLbL}

\begin{figure}[t]
	\centering
	\begin{align*}
		\includegraphics[width=2.5cm,valign=c]{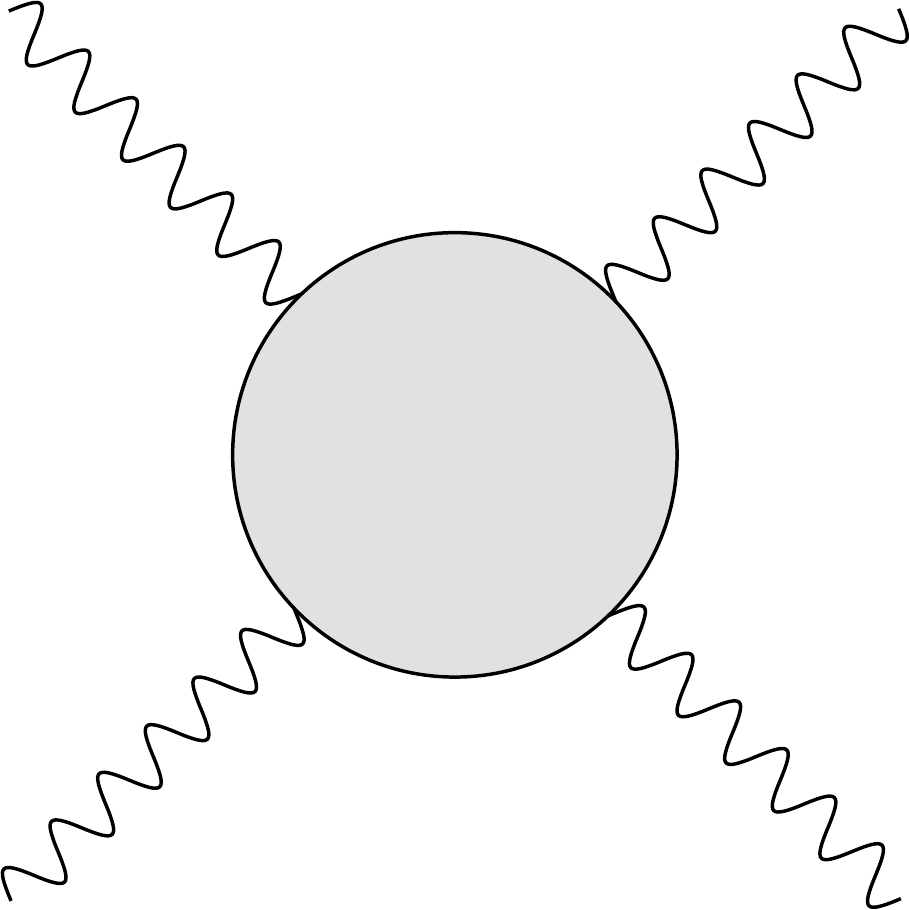}
		 \quad = \quad
		\includegraphics[width=2.5cm,valign=c]{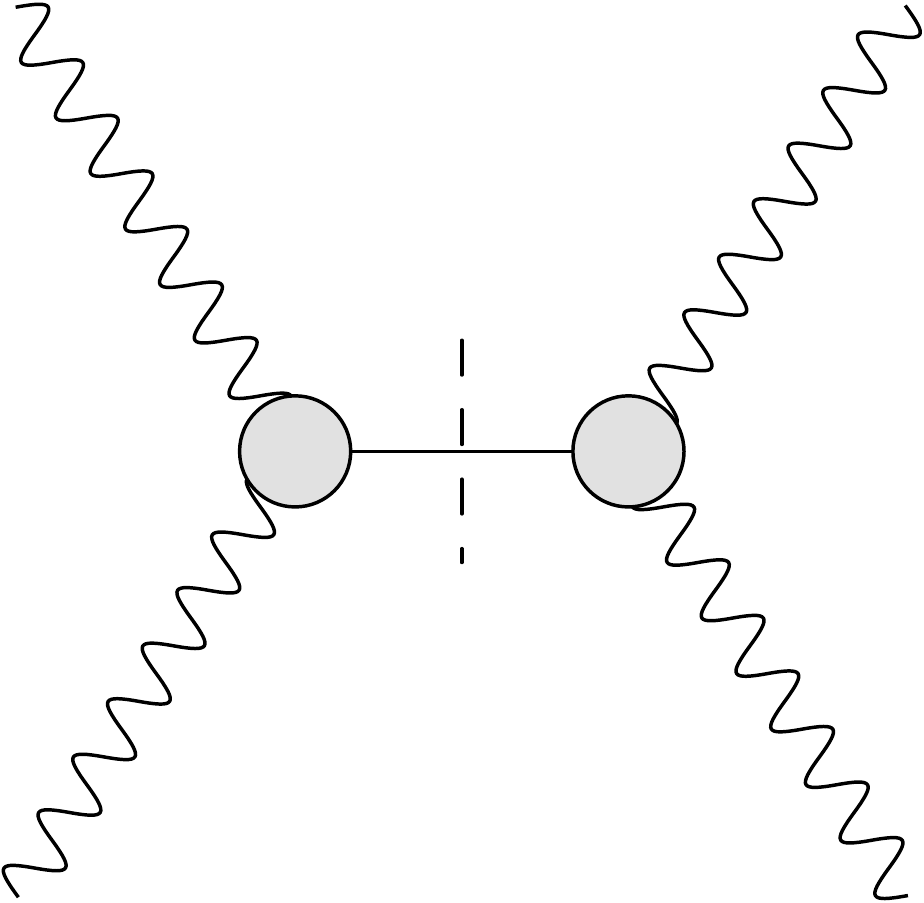}
		 \quad + \quad
		\includegraphics[width=2.5cm,valign=c]{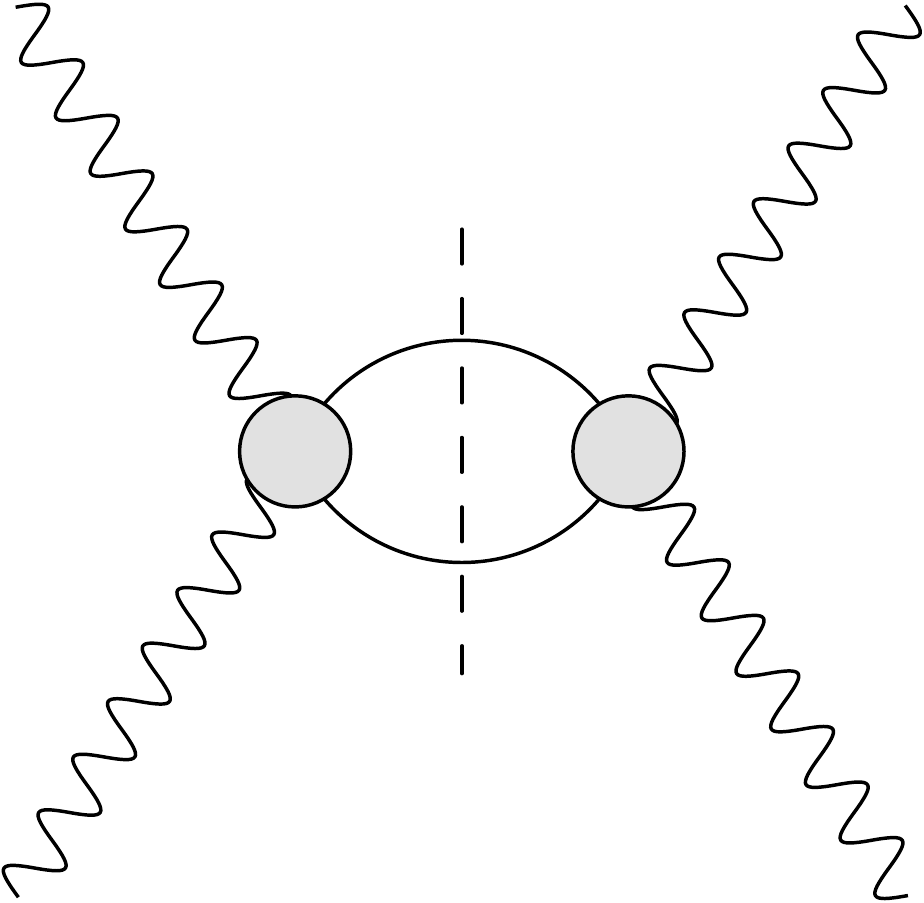}
		\quad + \quad \ldots
	\end{align*}
	\caption{Intermediate states in the direct channel: pion pole and two-pion cut.}
	\label{img:HLbLIntermediateStates}
\end{figure}

\begin{figure}[t]
	\centering
	\begin{align*}
		\includegraphics[width=2.5cm,valign=c]{images/TwoParticleCut}
		 =
		\includegraphics[width=2.5cm,valign=c]{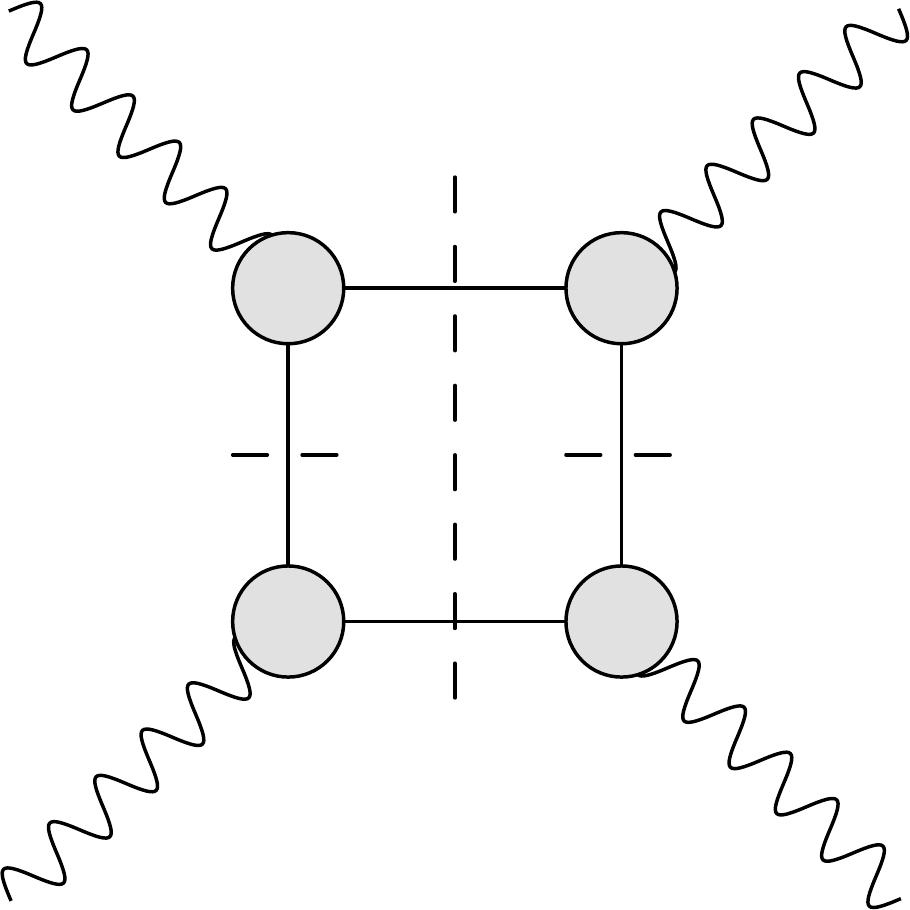}
		 +
		\includegraphics[width=2.5cm,valign=c,]{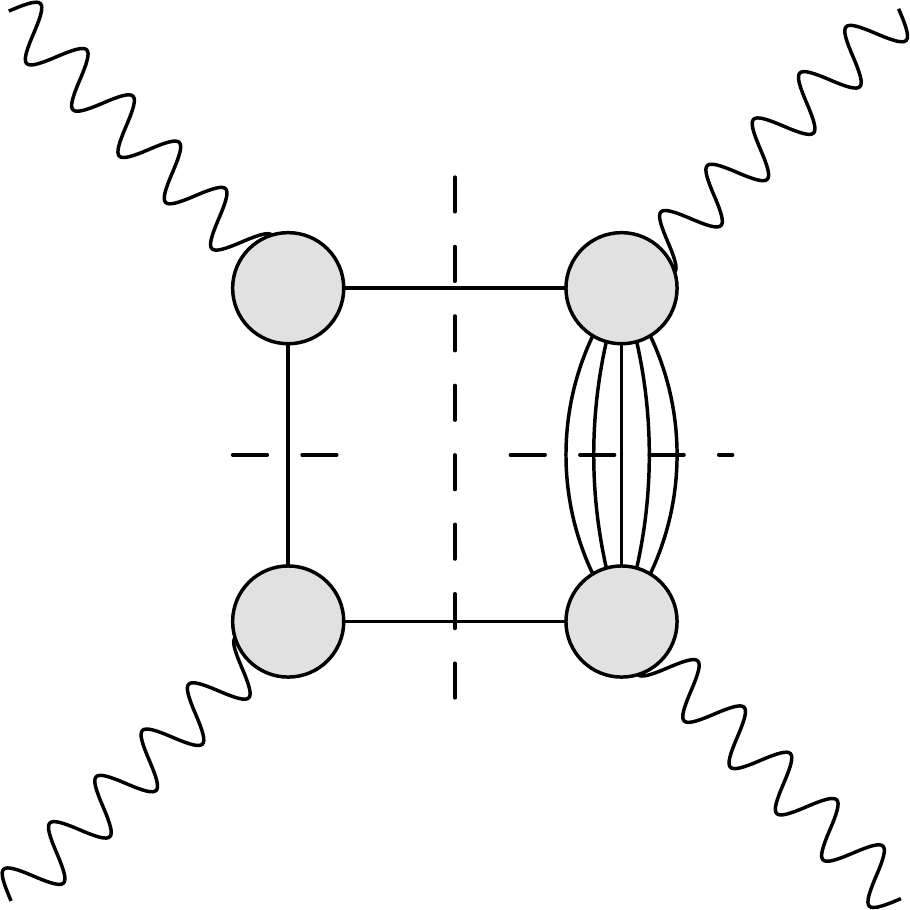}
		 +
		\includegraphics[width=2.5cm,valign=c]{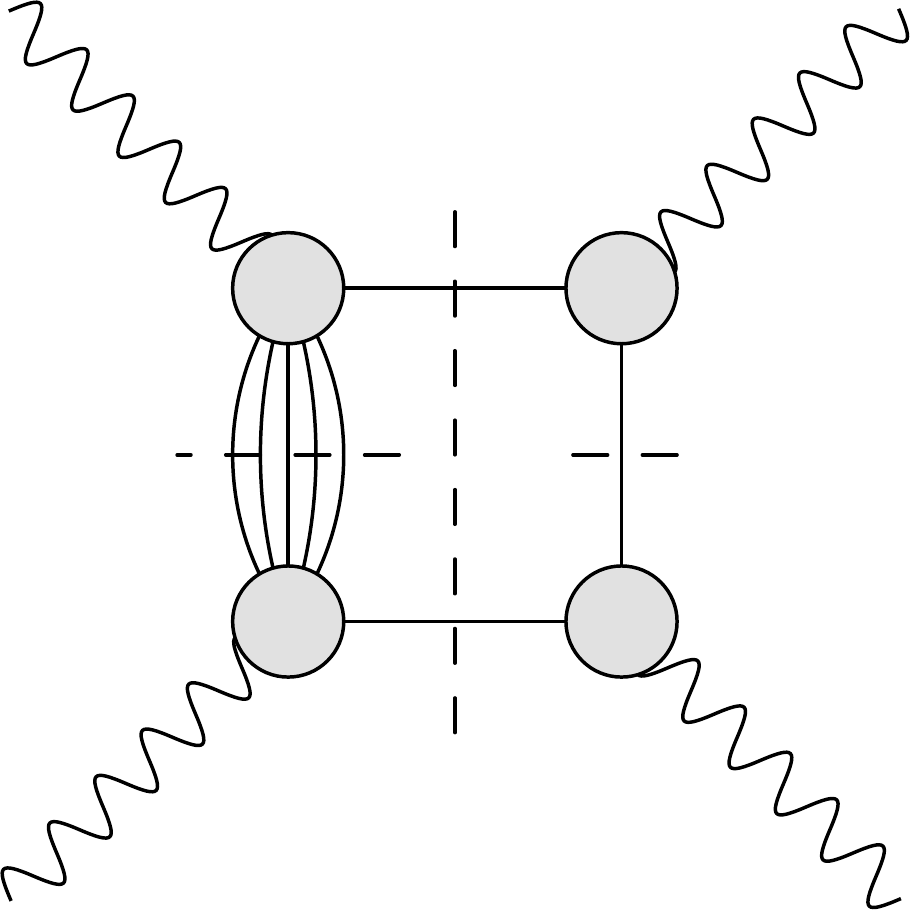}
		 +
		\includegraphics[width=2.5cm,valign=c]{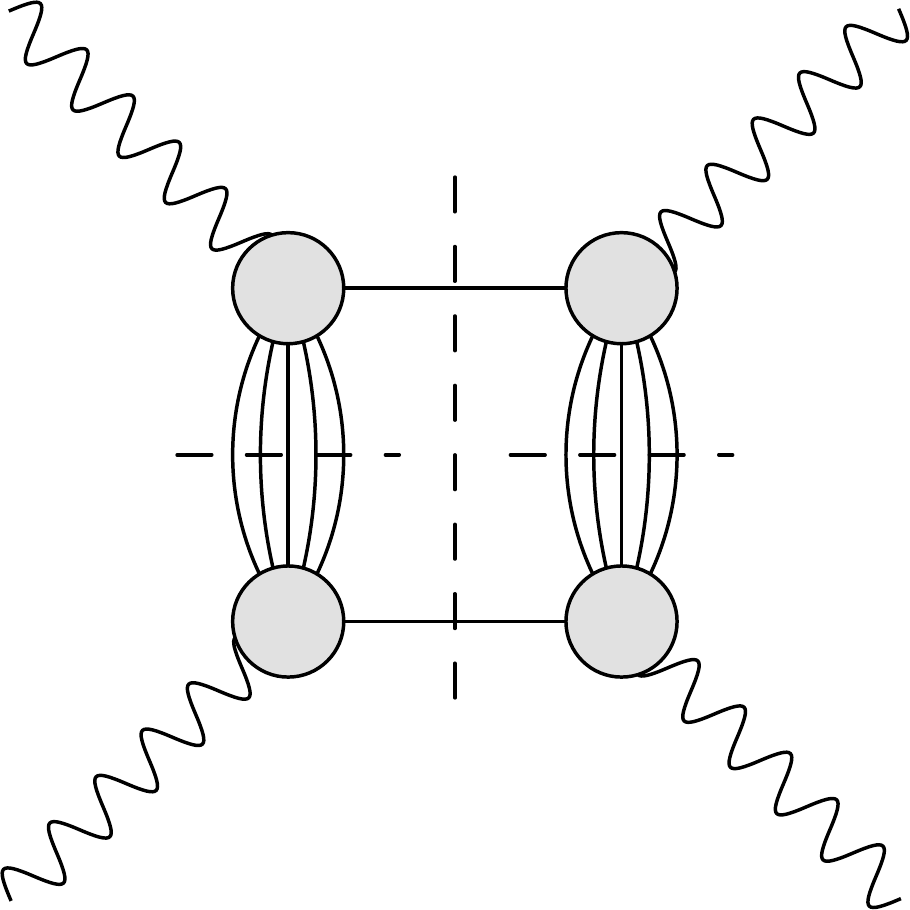}
		+ \ldots
	\end{align*}
	\caption{Two-pion contributions to HLbL. Further crossed diagrams are not shown explicitly.}
	\label{img:HLbLTwoPionContributions}
\end{figure}

In~\cite{Colangelo:2015ama}, we have used Mandelstam's double-spectral
representation~\cite{Mandelstam:1958xc} for the BTT scalar functions
$\Pi_i$ in order to split the HLbL contribution to $(g-2)_\mu$ into the following sum:
\begin{align}
	a_\mu^\mathrm{HLbL} = a_\mu^{\pi^0\text{-pole}} + a_\mu^{\pi\text{-box}} +  a_\mu^{\pi\pi} + \ldots
\end{align}
This sum directly reflects the sum over intermediate states in the
unitarity relation in which, by definition, all intermediate states enter
on-shell. While unitarity alone defines the imaginary parts, the real parts
are obtained from the dispersion integrals. In short, this amounts to the
following procedure: 
\begin{itemize}
	\item Write down the unitarity relation for the HLbL tensor.
	\item In the sum over intermediate (on-shell) states, the one-pion
          state contributes as a $\delta$-function to the imaginary part,
          which offsets the dispersion integral and defines the
          $\pi^0$-pole contribution. 
	\item The next-heavier intermediate state in the unitarity relation
          is a two-pion state. So far, we concentrate on one- and two-pion
          intermediate states, shown in
          Fig.~\ref{img:HLbLIntermediateStates}. 
	\item In the two-pion contribution, write down the crossed-channel
          unitarity relation for the sub-process
          $\gamma^*\gamma^*\to\pi\pi$. The one-pion contribution in this
          unitarity relation defines the $\pi$-pole contribution to
          $\gamma^*\gamma^*\to\pi\pi$. Separating this pole contribution
          corresponds to further splitting the two-pion contribution to
          HLbL into different box-type topologies, shown in Fig.~\ref{img:HLbLTwoPionContributions}.
	\item The two-pion phase-space integral in the HLbL unitarity
          relation can be converted into a second (crossed-channel)
          dispersion integral. This nontrivial but essential technical step
          is described in detail in App.~D of~\cite{Stoffer:2014rka}.
\item Finally, the symmetrization over the different channels produces the
  Mandelstam representation.
\end{itemize}

The double-spectral representation for the pion box has the following form:
\begin{align}
	\begin{split}
		\Pi_i^{\pi\text{-box}}(s,t,u;\{q_j^2\}) &= \frac{1}{\pi^2}
                \int_{4M_\pi^2}^\infty ds^\prime
                \int_{t^+(s^\prime;\{q_j^2\})}^\infty dt^\prime
                \frac{\rho_{i;st}^{\pi\text{-box}}(s^\prime,t^\prime;\{q_j^2\})}{(s^\prime - s)(t^\prime-t)} \\ 
			& +  \frac{1}{\pi^2} \int_{4M_\pi^2}^\infty
                        ds^\prime  \int_{u^+(s^\prime;\{q_j^2\})}^\infty
                        du^\prime
                        \frac{\rho_{i;su}^{\pi\text{-box}}(s^\prime,u^\prime;\{q_j^2\})}{(s^\prime - s)(u^\prime-u)} \\ 
			& +  \frac{1}{\pi^2} \int_{4M_\pi^2}^\infty
                        dt^\prime  \int_{u^+(t^\prime;\{q_j^2\})}^\infty
                        du^\prime
                        \frac{\rho_{i;tu}^{\pi\text{-box}}(t^\prime,u^\prime;\{q_j^2\})}{(t^\prime - t)(u^\prime-u)} , 
	\end{split}
\end{align}
where the functions $\rho_i^{\pi\text{-box}}$ denote the double-spectral
densities, which have been derived (though not given explicitly) in~\cite{Colangelo:2015ama}. The borders of the double-spectral regions $t^+$
and $u^+$ are defined in App.~G.3 of~\cite{Colangelo:2015ama}. 

In~\cite{Colangelo:2015ama}, we have explicitly shown that the Mandelstam
representation for the pion box is mathematically equivalent to a scalar
QED (sQED) one-loop calculation, multiplied by appropriate pion vector form
factors for the off-shell photons. First, the form factors only depend on
the virtualities $\{q_i^2\}$ and can be pulled out of the double-dispersion
integral. Second, triangle and bulb diagrams appear in the sQED calculation
only in order to ensure gauge invariance: indeed when projected onto our
gauge-invariant tensor structures, the analytic structure of sQED is 
the one of pure box topologies. In order to calculate the pion-box
contribution numerically, it is convenient to rather use a Feynman
parametrization instead of the dispersive representation. It turns out that
in the limit of $(g-2)_\mu$ kinematics, the Feynman parametrization of the
scalar functions $\hat \Pi_i$ defined in~\eqref{eq:PiHatFunctions} is very
compact. Due to the limit $q_4\to0$, only two-dimensional Feynman parameter
integrals appear: 
\begin{align}
	\hat \Pi_i^{\pi\text{-box}}(q_1^2,q_2^2,q_3^2) = F_\pi^V(q_1^2) F_\pi^V(q_2^2) F_\pi^V(q_3^2) \frac{1}{16\pi^2} \int_0^1 dx \int_0^{1-x} dy  I_i(x,y) ,
\end{align}
where $F_\pi^V$ is the electromagnetic pion vector form factor and the integrands $I_i$ can be found in App.~\ref{sec:FeynmanParametrizationPionBox}, written in a way that shows explicitly the absence of kinematic singularities.

The main goal of the present article is to describe two-pion contributions beyond the pion box, i.e.\ the topologies that involve a crossed-channel intermediate state heavier than one pion in one or both sub-processes.

\subsubsection{Two-pion contributions beyond the pion box}

\label{sec:TwoPionContributionsBeyondPionBox}

Let us examine in more detail the form of the Mandelstam representation as sketched in the previous subsection. The starting point is a fixed-$t$ dispersion relation with a discontinuity given by the two-pion contribution to the unitarity relation for the HLbL tensor:
\begin{align*}
	\Imspipi \Pi^{\mu\nu\lambda\sigma} &= \frac{1}{32\pi^2} \frac{\sigma_\pi(s)}{2} \int d\Omega_s^\dprime \begin{aligned}[t]
		& \bigg( W_{+-}^{\mu\nu}(p_1,p_2,q_1) {W_{+-}^{\lambda\sigma}}^*(p_1,p_2,-q_3) \\
		& + \frac{1}{2} W_{00}^{\mu\nu}(p_1,p_2,q_1) {W_{00}^{\lambda\sigma}}^*(p_1,p_2,-q_3) \bigg) , \end{aligned} \mytag
\end{align*}
where $W^{\mu\nu}$ are the matrix elements for $\gamma^*\gamma^*\to\pi\pi$. The subscripts $\{+-, 00\}$ denote the charges and $p_{1,2}$ the momenta of the intermediate pions. The phase-space factor is 
\beq
\sigma_\pi(s) := \sqrt{1 - \frac{4M_\pi^2}{s}}.
\eeq
In order to analytically continue the unitarity relation, these matrix elements have to be expressed in terms of fixed-$s$ dispersion relations for the scalar functions in a proper tensor decomposition, see~\cite{Colangelo:2015ama}:
\begin{align}
	\begin{split}
		W_{+-}^{\mu\nu} &= \sum_{i=1}^5 T^{\mu\nu}_i \begin{aligned}[t]
			& \Bigg( \frac{\rho_{i;t}^{s;+-}(s)}{t-M_\pi^2} + \frac{\rho_{i;u}^{s;+-}(s)}{u-M_\pi^2} + \frac{1}{\pi} \int_{4M_\pi^2}^\infty dt_1 \frac{ D_{i;t}^{s;+-}(t_1;s)}{t_1 - t} + \frac{1}{\pi} \int_{4M_\pi^2}^\infty du_1 \frac{ D_{i;u}^{s;+-}(u_1;s)}{u_1 - u} \Bigg) , \end{aligned} \\
		W_{00}^{\mu\nu} &= \sum_{i=1}^5 T^{\mu\nu}_i \begin{aligned}[t]
			& \Bigg( \frac{1}{\pi} \int_{4M_\pi^2}^\infty dt_1 \frac{ D_{i;t}^{s;00}(t_1;s)}{t_1 - t} + \frac{1}{\pi} \int_{4M_\pi^2}^\infty du_1 \frac{ D_{i;u}^{s;00}(u_1;s)}{u_1 - u} \Bigg) . \end{aligned}
	\end{split}
\end{align}
$W_{00}^{\mu\nu}$ does not contain any pole terms because the photon does not couple to two neutral pions due to angular momentum conservation and Bose symmetry.

If we pick the contribution of the pole terms on both sides of the cut, we single out box topologies:
{\small
\begin{align}
	\Imspipi \Pi^{\mu\nu\lambda\sigma} \Big|_\mathrm{box} &= \frac{1}{32\pi^2} \frac{\sigma_\pi(s)}{2} \int d\Omega_s^\dprime  \sum_{i,j=1,4} T^{\mu\nu}_i T^{\lambda\sigma}_j 
		\Bigg( \frac{\rho_{i;t}^{s;+-}(s)}{t^\prime-M_\pi^2} + \frac{\rho_{i;u}^{s;+-}(s)}{u^\prime-M_\pi^2} \Bigg) \Bigg( \frac{\rho_{j;t}^{s;+-}(s)}{t^\dprime-M_\pi^2} + \frac{\rho_{j;u}^{s;+-}(s)}{u^\dprime-M_\pi^2} \Bigg)^* ,
\end{align}}%
where the primed variables belong to the sub-process on the left-hand side and the double-primed variables to the sub-process on the right-hand side of the cut. This contribution was the subject of study in~\cite{Colangelo:2015ama}. We consider now the contributions with discontinuities either in one or both of the sub-processes:
{\small
\begin{align*}
	\Imspipi \Pi^{\mu\nu\lambda\sigma} \Big|_\mathrm{1disc} &= \frac{1}{32\pi^2} \frac{\sigma_\pi(s)}{2} \int d\Omega_s^\dprime  \sum_{i,j=1}^5 T^{\mu\nu}_i T^{\lambda\sigma}_j \\
			& \times \begin{aligned}[t]
			& \Bigg[ \bigg( \frac{\rho_{i;t}^{s;+-}(s)}{t^\prime-M_\pi^2} + \frac{\rho_{i;u}^{s;+-}(s)}{u^\prime-M_\pi^2} \bigg) \bigg( \frac{1}{\pi} \int_{4M_\pi^2}^\infty dt_2 \frac{D_{j;t}^{s;+-}(t_2;s)}{t_2 - t^\dprime} + \frac{1}{\pi} \int_{4M_\pi^2}^\infty du_2 \frac{D_{j;u}^{s;+-}(u_2;s)}{u_2 - u^\dprime} \bigg)^* \\
			& + \bigg(\frac{1}{\pi} \int_{4M_\pi^2}^\infty dt_1 \frac{D_{i;t}^{s;+-}(t_1;s)}{t_1 - t^\prime} + \frac{1}{\pi} \int_{4M_\pi^2}^\infty du_1 \frac{D_{i;u}^{s;+-}(u_1;s)}{u_1 - u^\prime} \bigg) \bigg( \frac{\rho_{j;t}^{s;+-}(s)}{t^\dprime-M_\pi^2} + \frac{\rho_{j;u}^{s;+-}(s)}{u^\dprime-M_\pi^2} \bigg)^* \Bigg] , \end{aligned} \\
		\Imspipi \Pi^{\mu\nu\lambda\sigma} \Big|_\mathrm{2disc} &= \frac{1}{32\pi^2} \frac{\sigma_\pi(s)}{2} \int d\Omega_s^\dprime \sum_{i,j=1}^5 T^{\mu\nu}_i T^{\lambda\sigma}_j \\*
			&\times \Bigg[ \begin{aligned}[t]
				& \bigg( \frac{1}{\pi} \int_{4M_\pi^2}^\infty dt_1 \frac{D_{i;t}^{s;+-}(t_1;s)}{t_1 - t^\prime} + \frac{1}{\pi} \int_{4M_\pi^2}^\infty du_1 \frac{D_{i;u}^{s;+-}(u_1;s)}{u_1 - u^\prime} \bigg) \\
				& \times \bigg( \frac{1}{\pi} \int_{4M_\pi^2}^\infty dt_2 \frac{D_{j;t}^{s;+-}(t_2;s)}{t_2 - t^\dprime} + \frac{1}{\pi} \int_{4M_\pi^2}^\infty du_2 \frac{D_{j;u}^{s;+-}(u_2;s)}{u_2 - u^\dprime} \bigg)^* \\
				& + \frac{1}{2} \bigg( \frac{1}{\pi} \int_{4M_\pi^2}^\infty dt_1 \frac{D_{i;t}^{s;00}(t_1;s)}{t_1 - t^\prime} + \frac{1}{\pi} \int_{4M_\pi^2}^\infty du_1 \frac{D_{i;u}^{s;00}(u_1;s)}{u_1 - u^\prime} \bigg) \\
				& \times \bigg( \frac{1}{\pi} \int_{4M_\pi^2}^\infty dt_2 \frac{D_{j;t}^{s;00}(t_2;s)}{t_2 - t^\dprime} + \frac{1}{\pi} \int_{4M_\pi^2}^\infty du_2 \frac{D_{j;u}^{s;00}(u_2;s)}{u_2 - u^\dprime} \bigg)^* \Bigg] . \end{aligned} \mytag
\end{align*}}%
If the order of phase-space and dispersive integrals are exchanged, the phase-space integrals can be performed by applying a tensor reduction to the quantities
\begin{align}
	\int d\Omega_s^\dprime \sum_{i,j=1}^5 T_i^{\mu\nu} T_j^{\lambda\sigma} \frac{1}{t_1-t^\prime}\frac{1}{t_2-t^\dprime} .
\end{align}
The reduced scalar phase-space integrals can then be transformed into another dispersive integral. Together with the dispersion integral $ds^\prime$ of the primary cut, this produces a double-dispersion relation. The case of the simplest scalar phase-space integral is explained in~\cite{Stoffer:2014rka}. Here, we do not try to calculate explicitly the tensor phase-space integrals, because we are interested just in the analytic structure of the ``1disc'' and ``2disc'' contributions, i.e.\ the boxes with heavier intermediate states in one or both of the sub-processes.

In order to obtain the full double-spectral representation, one has to
consider not only a fixed-$t$ dispersion relation as a starting point but
also the crossed versions, i.e.\ fixed-$s$ and fixed-$u$ dispersion
relations. The symmetrization leads to the Mandelstam representation. For a
more detailed discussion in the case of the pion box, see
again~\cite{Colangelo:2015ama}. We consider now the ``1disc'' and ``2disc''
contributions, where the pole in one or both of the sub-processes is
replaced by a discontinuity. As the symmetrization procedure is identical
in both cases, we only discuss the case of a discontinuity in both
sub-processes. 

\begin{figure}[t]
	\centering
	\begin{subfigure}[b]{0.24\textwidth}
		\centering
		\includegraphics[width=3cm]{images/Box_2Disc}
		\caption{$\rho_{st}$}
		\label{img:HLbLBox2DiscA}
	\end{subfigure}
	\begin{subfigure}[b]{0.24\textwidth}
		\centering
		\includegraphics[width=3cm]{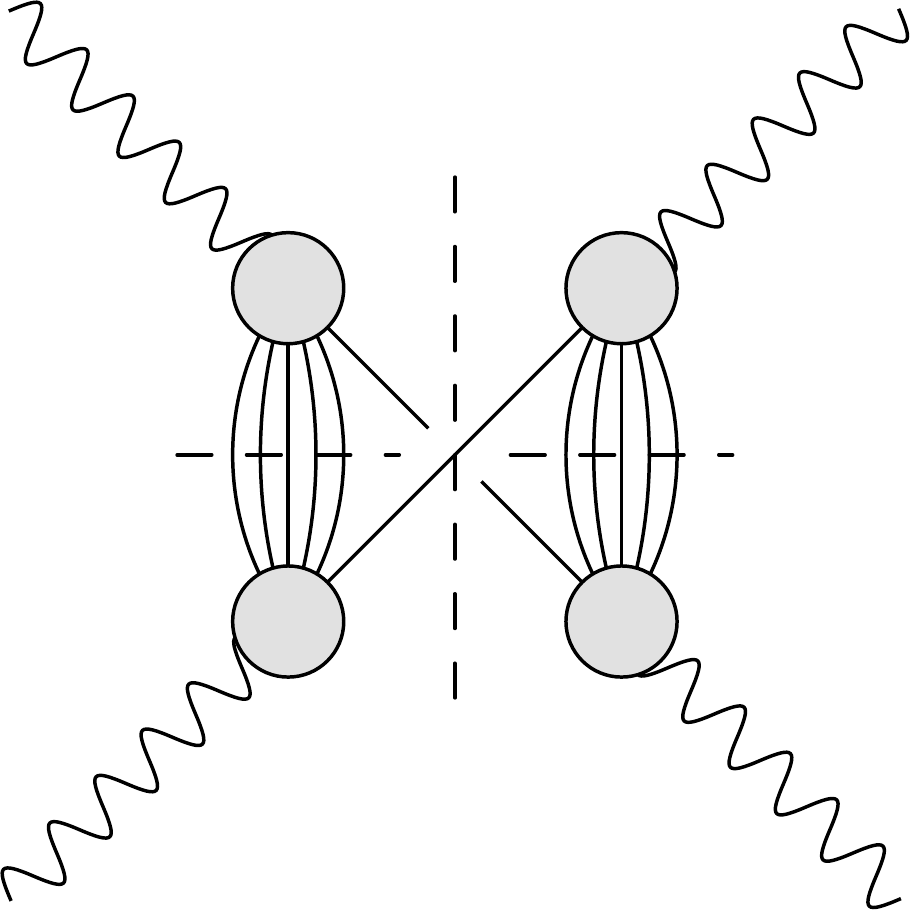}
		\caption{$\rho_{su}$}
		\label{img:HLbLBox2DiscB}
	\end{subfigure}
	\begin{subfigure}[b]{0.24\textwidth}
		\centering
		\includegraphics[width=3cm]{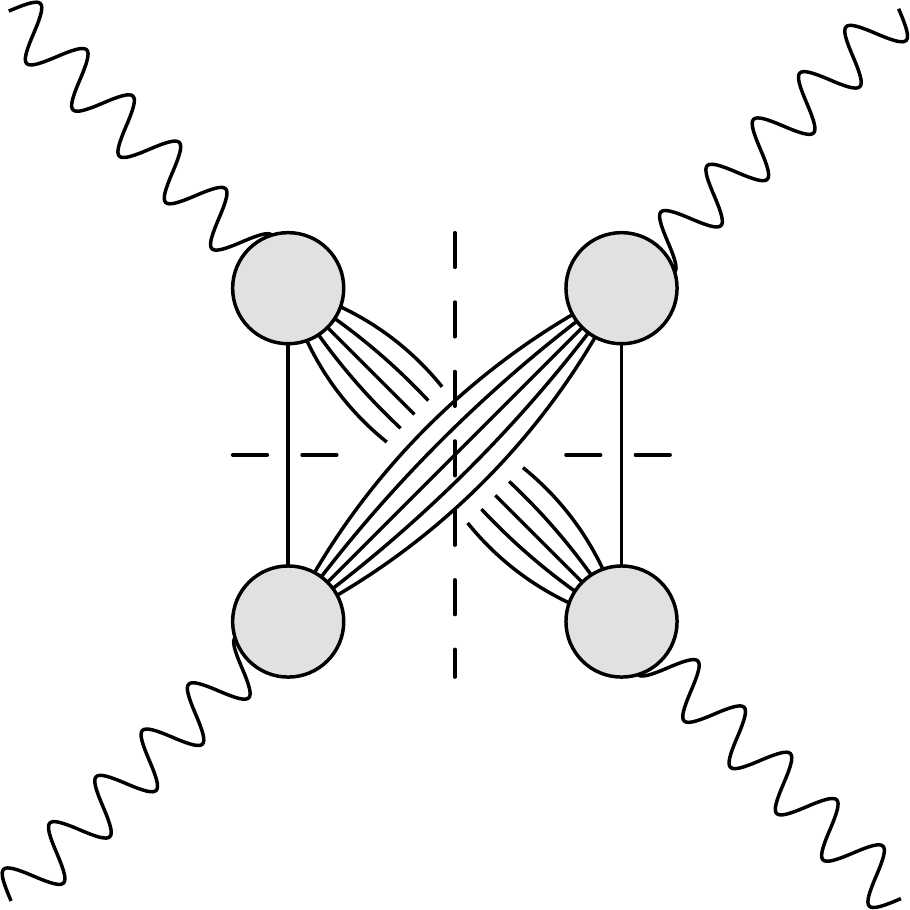}
		\caption{$\rho_{us}$}
		\label{img:HLbLBox2DiscC}
	\end{subfigure}
	\begin{subfigure}[b]{0.24\textwidth}
		\centering
		\includegraphics[width=3cm]{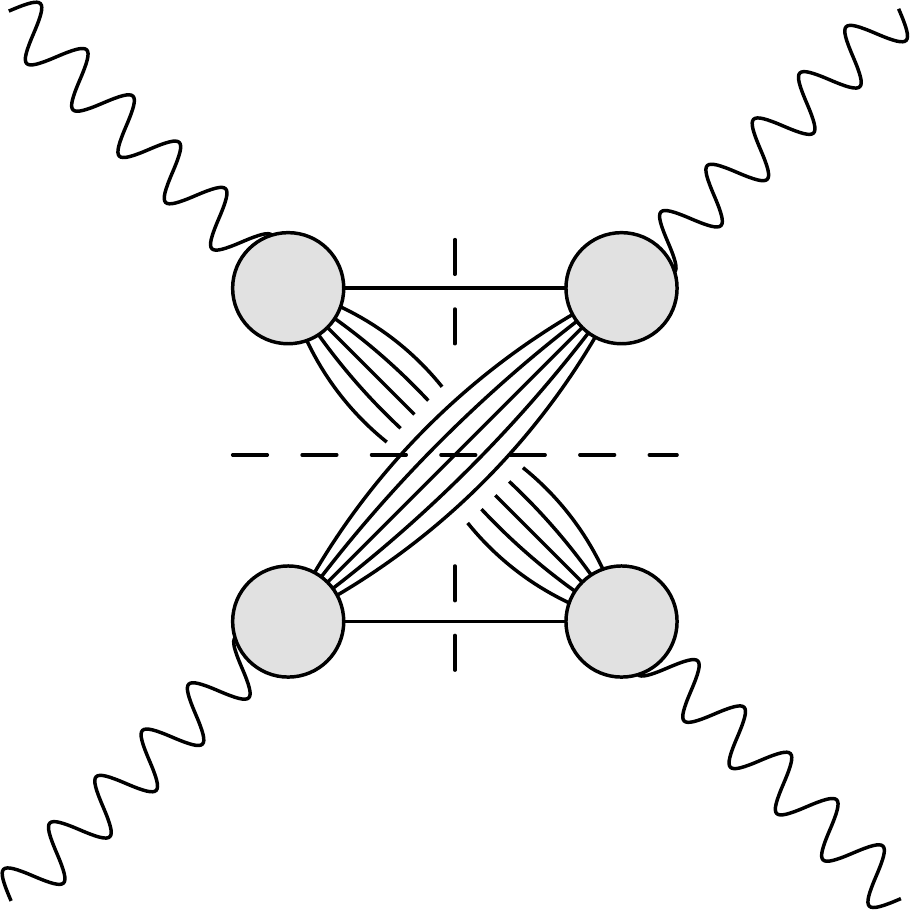}
		\caption{$\rho_{ut}$}
		\label{img:HLbLBox2DiscD}
	\end{subfigure}
	\caption{Unitarity diagrams representing the ``2disc''-box contributions that are (partially) accessible through a fixed-$t$ dispersion relation.}
	\label{img:HLbLBox2Disc}
\end{figure}

Fig.~\ref{img:HLbLBox2Disc} shows the unitarity diagrams corresponding to the double-spectral representations that are generated if we start in our derivation from the fixed-$t$ dispersion relation: the diagrams~\ref{img:HLbLBox2DiscA} and \ref{img:HLbLBox2DiscB} generate a cut for $s>4M_\pi^2$, which is the right-hand cut in the fixed-$t$ dispersion relation. The diagrams~\ref{img:HLbLBox2DiscC} and \ref{img:HLbLBox2DiscD} are responsible for the left-hand cut for $u>4M_\pi^2$. In all cases the first cut is always the one through the two-pion intermediate state.

As discussed in~\cite{Stoffer:2014rka, Colangelo:2015ama}, an $(st)$-box diagram can be represented either by a fixed-$s$, fixed-$t$, or fixed-$u$ dispersion relation: in the case of a fixed-$t$ representation, there appears only one dispersion integral along the right-hand $s$-channel cut. Likewise, in a fixed-$s$ representation, only one dispersion integral along the $t$-channel cut is present. In the case of a fixed-$u$ representation, however, an $(st)$-box generates two integrals along both the $s$- and the $t$-channel cut. This particularity translates directly into the double-spectral representation: the $(st)$-box can be written as only one double-dispersion integral if one starts from a fixed-$s$ or fixed-$t$ representation. If one starts from the fixed-$u$ representation, one obtains a sum of two double-dispersion integrals, see App.~G.3 of~\cite{Colangelo:2015ama}.

\begin{figure}[t]
	\centering
	\includegraphics{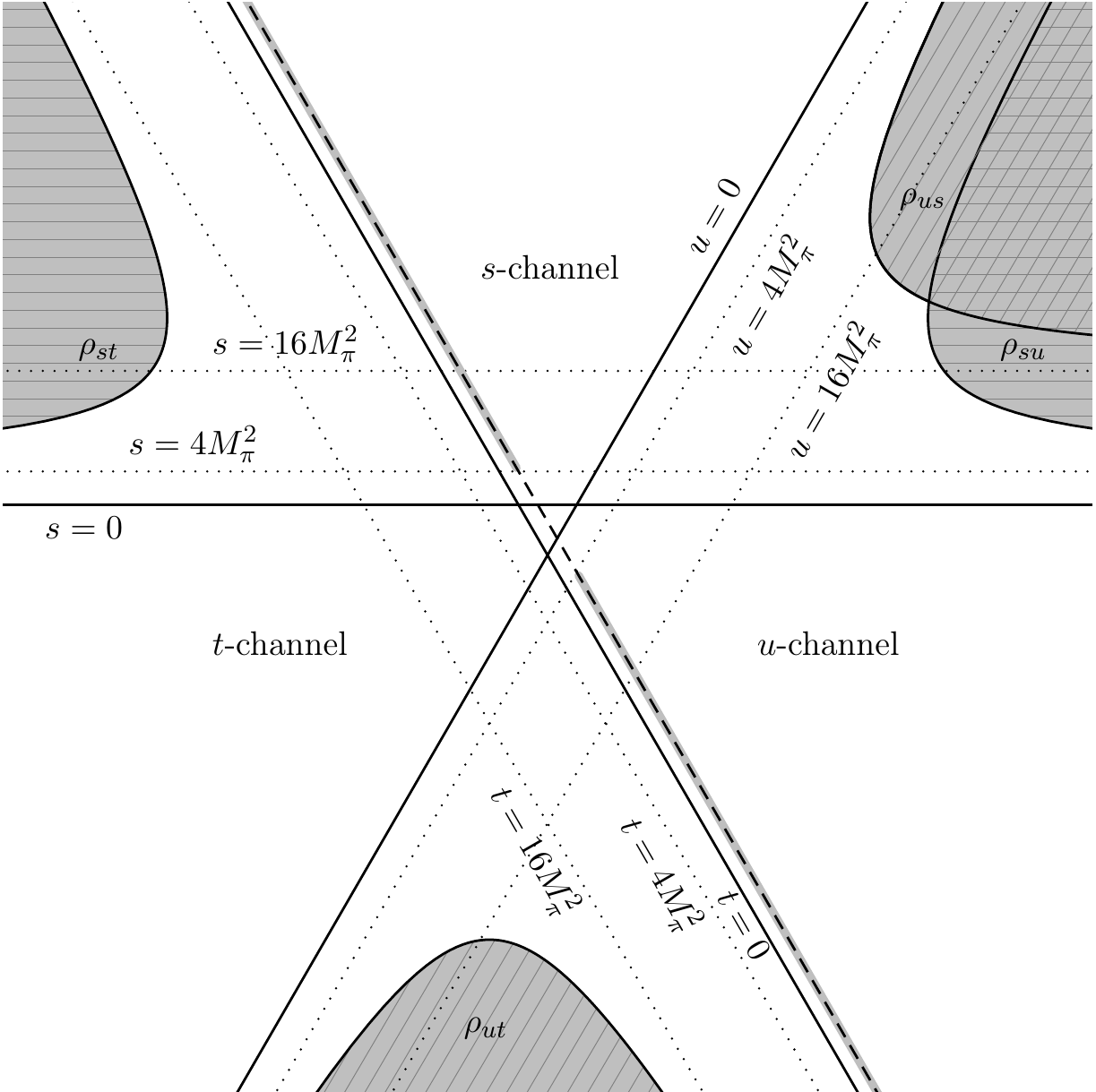}
	\caption{Mandelstam diagram for HLbL scattering for the case $q_1^2 = q_2^2 = q_3^2 = -2 M_\pi^2$, $q_4^2 = 0$. Only those double-spectral regions for ``2disc''-box topologies are shown that are reconstructed from the fixed-$t$ dispersion relation. The dashed line marks a line of fixed $t$ with its $s$- and $u$-channel cuts highlighted in gray.}
	\label{img:HLbLMandelstamDiagramBoxDiscFixedT}
\end{figure}

Consider now the Mandelstam diagram in Fig.~\ref{img:HLbLMandelstamDiagramBoxDiscFixedT}, which shows the double-spectral regions that we generate if we start from a fixed-$t$ dispersion relation. Because we consider in the primary cut only two-pion intermediate states, not all the contributions from the displayed double-spectral regions are generated. We understand from the above discussion of the $(st)$-box that $\rho_{st}$ and $\rho_{ut}$ are complete, but that the contributions from $\rho_{us}$ and $\rho_{su}$ are not, because only one double-spectral integral for each of these contributions is obtained. However, two double-spectral integrals would be needed to generate the full contribution of these regions: one of the two integrals has a primary cut at the higher threshold $16M_\pi^2$ and is neglected in the fixed-$t$ representation. Of course, two more double-spectral regions $\rho_{ts}$ and $\rho_{tu}$, which correspond to crossed boxes, are completely missing in the fixed-$t$ representation.

\begin{figure}[t]
	\centering
	\includegraphics{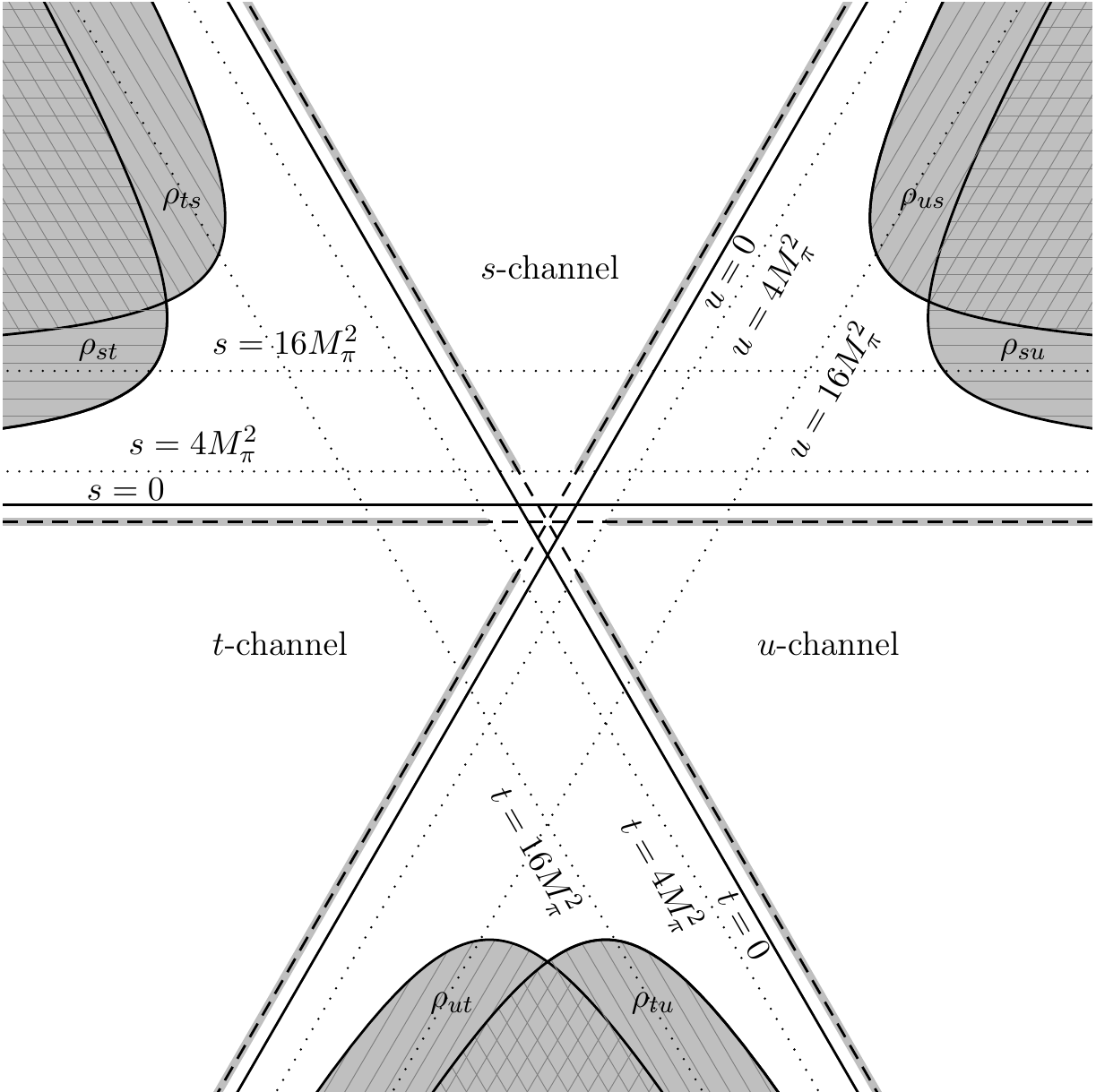}
	\caption{Mandelstam diagram for HLbL scattering for the case $q_1^2 = q_2^2 = q_3^2 = -2 M_\pi^2$, $q_4^2 = 0$ with all the double-spectral regions for ``2disc''-box topologies.}
	\label{img:HLbLMandelstamDiagramBoxDiscSymm}
\end{figure}

The complete set of double-spectral regions, which is obtained after symmetrization, is shown in Fig.~\ref{img:HLbLMandelstamDiagramBoxDiscSymm}. In the symmetric version, the double-spectral integrals over $\rho_{st}$ and $\rho_{ut}$ are taken from the fixed-$t$ representation, $\rho_{ts}$ and $\rho_{us}$ come from the fixed-$s$ representation, and finally $\rho_{su}$ and $\rho_{tu}$ stem from the fixed-$u$ dispersion relation.

In summary, we can write the contribution of higher intermediate states in the secondary channel as a double-spectral representation (we suppress the explicit dependence on the virtualities):
{\small
\begin{align}
	\begin{split}
		\label{eq:HLbLDoubleSpectralDisc}
		\Pi_i^{\pi\pi}(s,t,u) &= \frac{1}{\pi^2} \int_{4M_\pi^2}^\infty ds^\prime \int_{t^+(s^\prime)}^\infty dt^\prime \frac{\rho^{\pi\pi}_{i;st}(s^\prime,t^\prime)}{(s^\prime - s)(t^\prime - t)}  + \frac{1}{\pi^2} \int_{4M_\pi^2}^\infty ds^\prime \int_{u^+(s^\prime)}^\infty du^\prime \frac{\rho^{\pi\pi}_{i;su}(s^\prime,u^\prime)}{(s^\prime - s)(u^\prime - u)} \\
			&\quad + \frac{1}{\pi^2} \int_{4M_\pi^2}^\infty dt^\prime \int_{s^+(t^\prime)}^\infty ds^\prime \frac{\rho^{\pi\pi}_{i;ts}(t^\prime,s^\prime)}{(t^\prime - t)(s^\prime - s)} + \frac{1}{\pi^2} \int_{4M_\pi^2}^\infty dt^\prime \int_{u^+(t^\prime)}^\infty du^\prime \frac{\rho^{\pi\pi}_{i;tu}(t^\prime,u^\prime)}{(t^\prime - t)(u^\prime - u)} \\
			&\quad + \frac{1}{\pi^2} \int_{4M_\pi^2}^\infty du^\prime \int_{s^+(u^\prime)}^\infty ds^\prime \frac{\rho^{\pi\pi}_{i;us}(u^\prime,s^\prime)}{(u^\prime - u)(s^\prime - s)} + \frac{1}{\pi^2} \int_{4M_\pi^2}^\infty du^\prime \int_{t^+(u^\prime)}^\infty dt^\prime \frac{\rho^{\pi\pi}_{i;ut}(u^\prime,t^\prime)}{(u^\prime - u)(t^\prime - t)} .
	\end{split}
\end{align}}%
The border functions of the double-spectral regions approach asymptotically $t^+(s)\stackrel{s\to\infty}{\longrightarrow}9M_\pi^2$ for the ``1disc'' contribution or $16M_\pi^2$ for the ``2disc'' contribution.

\FloatBarrier

\subsubsection{Single-variable dispersion relation for two-pion contributions}

\label{sec:SingleVariableDR}

\begin{figure}[H]
	\centering
	\begin{subfigure}[b]{0.15\textwidth}
		\centering
		\includegraphics[width=2cm]{images/Box_2Disc}
		\caption{$\rho_{st}$}
		\label{img:HLbLBox2DiscApproxA}
	\end{subfigure}
	\begin{subfigure}[b]{0.15\textwidth}
		\centering
		\includegraphics[width=2cm]{images/Box_2Disc_Crossed1}
		\caption{$\rho_{su}$}
		\label{img:HLbLBox2DiscApproxB}
	\end{subfigure}
	\begin{subfigure}[b]{0.15\textwidth}
		\centering
		\includegraphics[width=2cm,angle=90,origin=c]{images/Box_2Disc}
		\caption{$\rho_{ts}$}
		\label{img:HLbLBox2DiscApproxC}
	\end{subfigure}
	\begin{subfigure}[b]{0.15\textwidth}
		\centering
		\reflectbox{\includegraphics[width=2cm,angle=90,origin=c]{images/Box_2Disc_Crossed1}}
		\caption{$\rho_{tu}$}
		\label{img:HLbLBox2DiscApproxD}
	\end{subfigure}
	\begin{subfigure}[b]{0.15\textwidth}
		\centering
		\includegraphics[width=2cm]{images/Box_2Disc_Crossed2}
		\caption{$\rho_{us}$}
		\label{img:HLbLBox2DiscApproxE}
	\end{subfigure}
	\begin{subfigure}[b]{0.15\textwidth}
		\centering
		\includegraphics[width=2cm]{images/Box_2Disc_Crossed3}
		\caption{$\rho_{ut}$}
		\label{img:HLbLBox2DiscApproxF}
	\end{subfigure}
	
	\vspace{0.25cm}
	
	$\stackrel{\underbrace{\hspace{4cm}}}{\stackrel{}{\approx}}$ \hspace{1.1cm} $\stackrel{\underbrace{\hspace{4cm}}}{\stackrel{}{\approx}}$  \hspace{1.1cm} $\stackrel{\underbrace{\hspace{4cm}}}{\stackrel{}{\approx}}$
	
	\vspace{0.25cm}
	
	\begin{subfigure}[b]{0.3\textwidth}
		\centering
		\includegraphics[width=2cm]{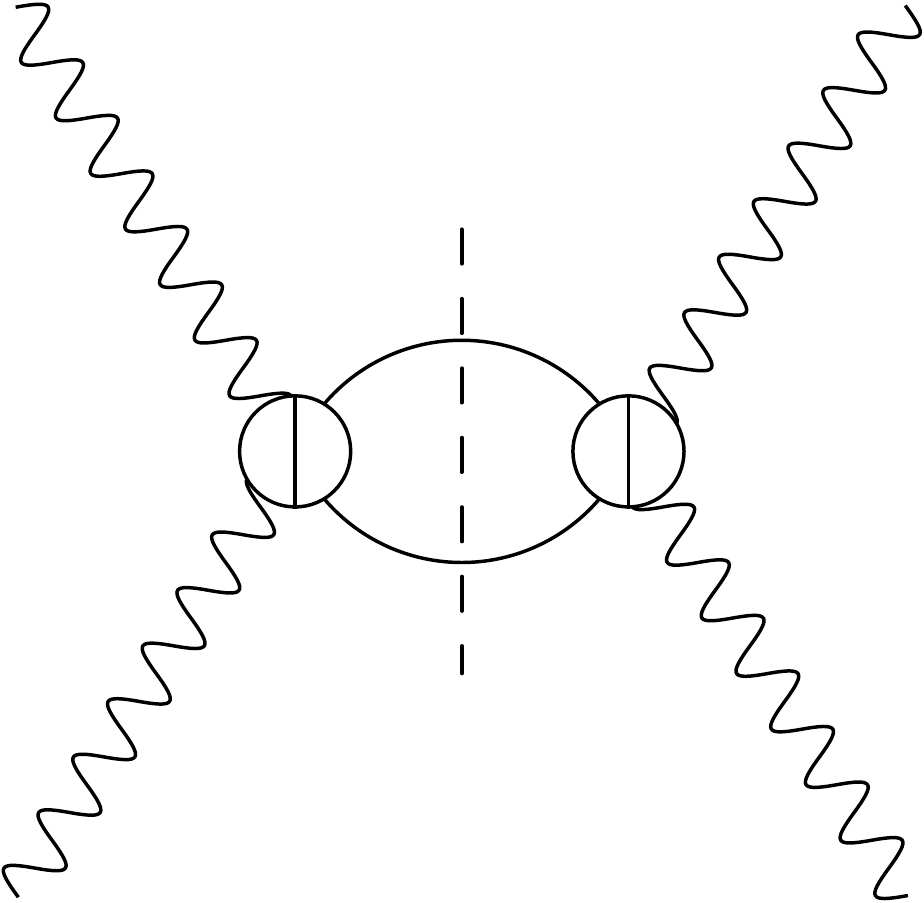}
		\caption{}
		\label{img:HLbLBox2DiscApproxG}
	\end{subfigure}
	\begin{subfigure}[b]{0.315\textwidth}
		\centering
		\includegraphics[width=2cm,angle=90,origin=c]{images/Bulb}
		\caption{}
		\label{img:HLbLBox2DiscApproxH}
	\end{subfigure}
	\begin{subfigure}[b]{0.3\textwidth}
		\centering
		\includegraphics[width=2cm]{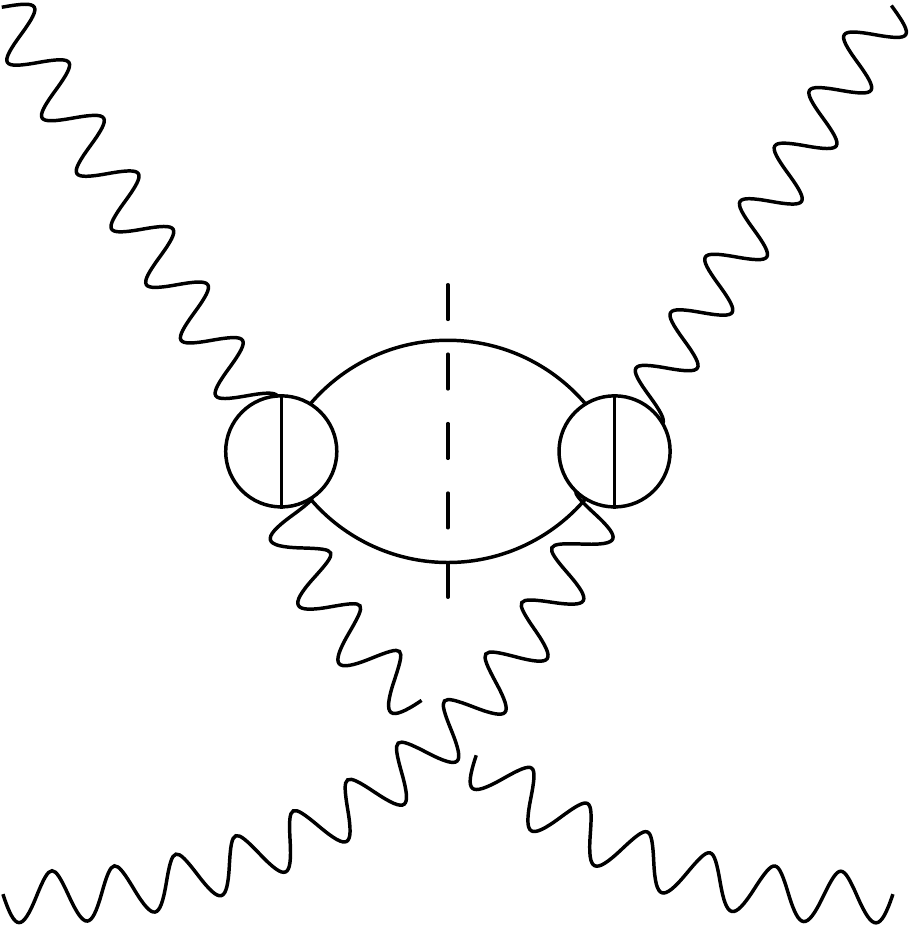}
		\caption{}
		\label{img:HLbLBox2DiscApproxI}
	\end{subfigure}
	
	\caption{{\bf (a)--(f)} Unitarity diagrams representing the complete set of ``2disc''-box contributions. \newline {\bf (g)--(i)} Partial-wave approximation: the sub-process becomes a polynomial in the crossed variable.}
	\label{img:HLbLBox2DiscApprox}
\end{figure}

When we expand the sub-process $\gamma^*\gamma^*\to\pi\pi$ into partial waves, we obtain a polynomial in the crossed-channel Mandelstam variables. This means that we neglect the crossed channel cut of the ``1disc'' or ``2disc'' boxes, reducing them effectively to triangle (in the case of ``1disc'' boxes) and bulb topologies (in the case of ``2disc'' boxes), as illustrated in Fig.~\ref{img:HLbLBox2DiscApprox}. After having applied the approximation, there is no way to distinguish e.g.~in Fig.~\ref{img:HLbLBox2DiscApproxG} between contributions coming originally from $\rho_{st}$ or $\rho_{su}$. Therefore, we discuss in the following what kind of single-variable dispersion relation is appropriate in the case of a partial-wave expanded input for the sub-process.

Consider again the situation for a fixed-$t$ dispersion relation with the corresponding Mandelstam diagram in~Fig.~\ref{img:HLbLMandelstamDiagramBoxDiscFixedT}. When constructing the Mandelstam representation, we selected from this representation only the contributions from $\rho_{st}$ and $\rho_{ut}$. After the partial-wave expansion, however, we are no longer able to drop the incomplete contributions from $\rho_{us}$ and $\rho_{su}$. Instead, let us assume that the neglected contributions from these two double-spectral regions are small: they are only due to the higher thresholds $9M_\pi^2$ (in the case of ``1disc'') or $16M_\pi^2$ (in the case of ``2disc''). Furthermore, their discontinuities, being generated by multi-particle intermediate states, are phase-space suppressed. Instead of combining the completely reconstructed double-spectral regions from fixed-$s$, fixed-$t$, and fixed-$u$ representations, we can simply sum all contributions from all three fixed-$(s,t,u)$ representations. Apart from the neglected higher cuts, each double-spectral contribution appears twice in this sum. The appropriate representation is therefore one half the sum of fixed-$(s,t,u)$ representations:
\begin{align*}
\label{eq:DRrescattering}
	\Pi_i^{\pi\pi}(s,t,u) \approx \frac{1}{2} \begin{aligned}[t]
		& \bigg( \frac{1}{\pi} \int_{4M_\pi^2}^\infty dt' \frac{\Im \Pi_i^{\pi\pi}(s,t',u')}{t'-t} + \frac{1}{\pi} \int_{4M_\pi^2}^\infty du' \frac{\Im \Pi_i^{\pi\pi}(s,t',u')}{u'-u} \\
		&+  \frac{1}{\pi} \int_{4M_\pi^2}^\infty ds' \frac{\Im \Pi_i^{\pi\pi}(s',t,u')}{s'-s} + \frac{1}{\pi} \int_{4M_\pi^2}^\infty du' \frac{\Im \Pi_i^{\pi\pi}(s',t,u')}{u'-u} \\
		&+  \frac{1}{\pi} \int_{4M_\pi^2}^\infty ds' \frac{\Im \Pi_i^{\pi\pi}(s',t',u)}{s'-s} + \frac{1}{\pi} \int_{4M_\pi^2}^\infty dt' \frac{\Im \Pi_i^{\pi\pi}(s',t',u)}{t'-t} \bigg) . \end{aligned} \mytag
\end{align*}
In the limit of infinitely heavy intermediate states in the crossed channel this relation is exact. In particular, the dominant $\pi\pi$-rescattering contributions that we consider in this paper can be understood as a unitarization of the pion pole in the crossed channel on a partial-wave basis. In this case, the dispersion relation~\eqref{eq:DRrescattering} provides a model-independent representation of the contribution of resonant effects in the $\pi\pi$ spectrum.

\begin{figure}[t]
	\centering
	\begin{subfigure}[b]{\textwidth}
		\centering
		\includegraphics{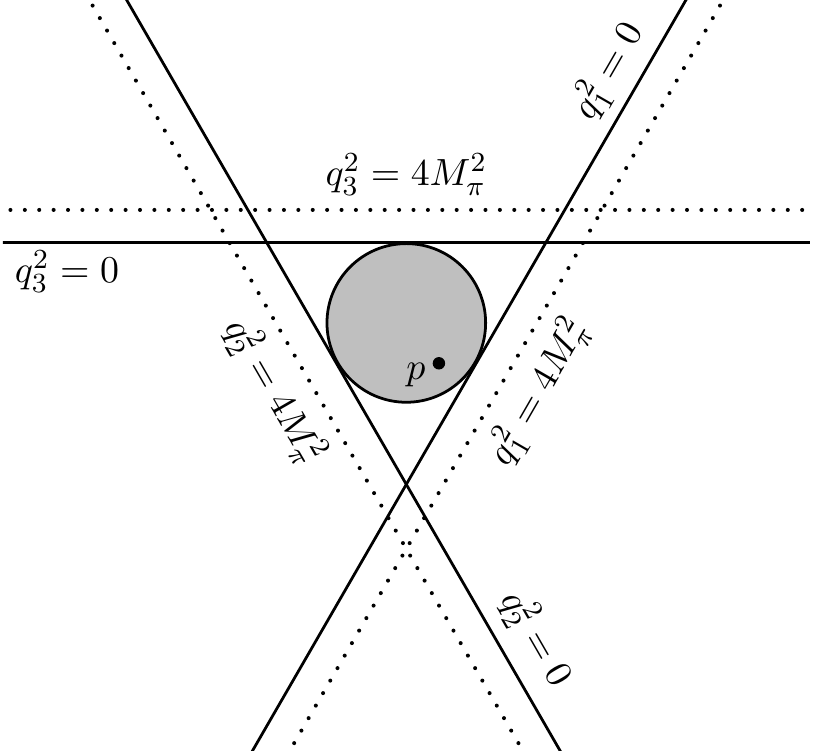}
		\caption{A point $p$ inside the $(g-2)_\mu$ integration region is selected and defines the external kinematics.}
		\label{img:HLbLMandelstamDiagramPWEA}
	\end{subfigure}
	\\[0.5cm]
	\begin{subfigure}[b]{\textwidth}
		\centering
		\includegraphics{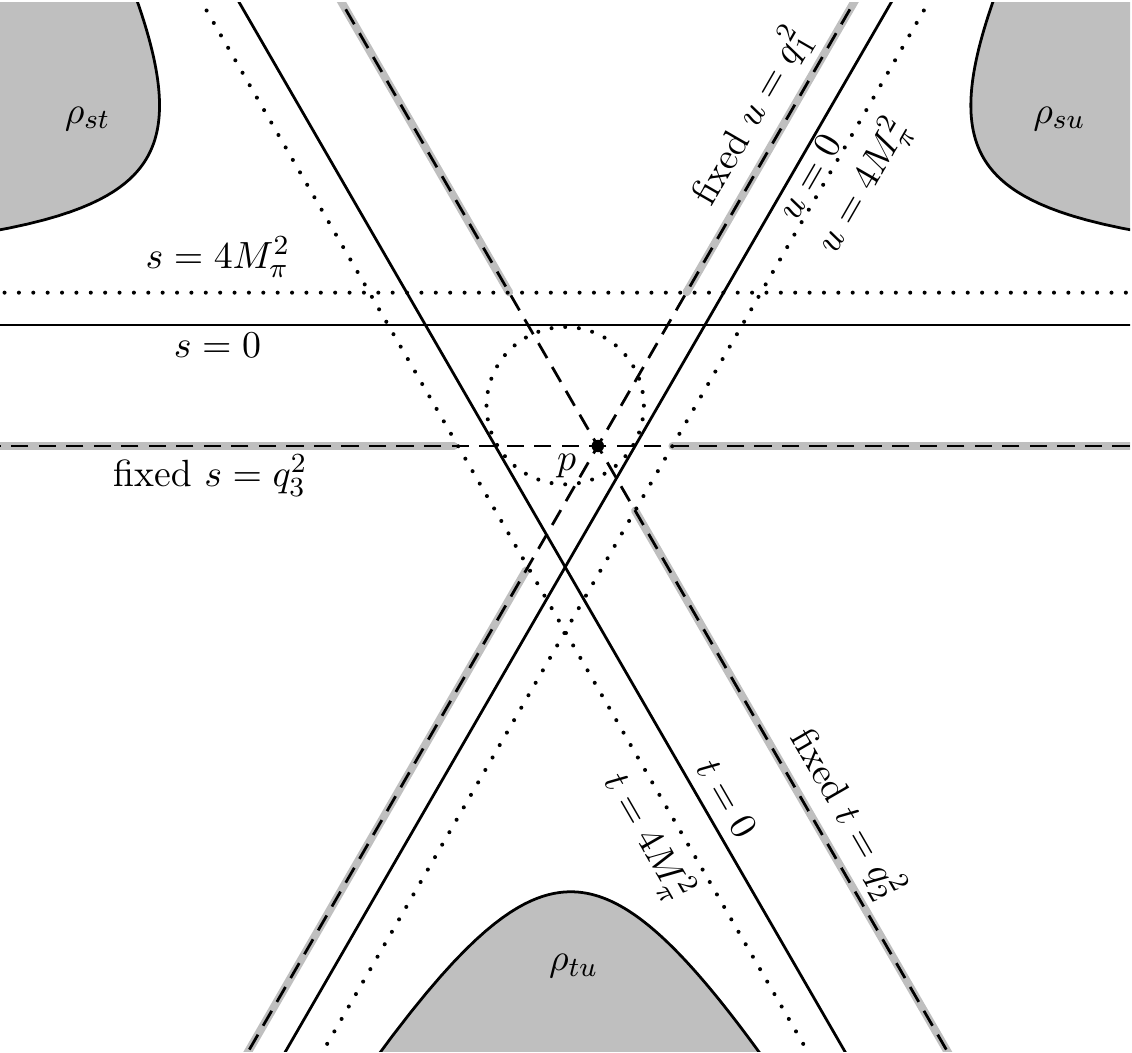}
		\caption{Mandelstam diagram for the selected kinematics of point $p$. The double-spectral regions for the pion box are shown. Lines of fixed $s$, $t$, and $u$ running through the point $p$ with $(s,t,u)=(q_3^2,q_2^2,q_1^2)$ are shown. They do not intersect any double-spectral region.}
		\label{img:HLbLMandelstamDiagramPWEB}
	\end{subfigure}

	\caption{For $(g-2)_\mu$ kinematics, the paths of the single-variable dispersion integrals never enter any double-spectral region, which enables a partial-wave expansion.}
	\label{img:HLbLMandelstamDiagramPWE}
\end{figure}

\FloatBarrier

In the case of $(g-2)_\mu$ kinematics, we are interested only in space-like momenta of the virtual photons. The lines of fixed-$(s,t,u)$ therefore never enter the double-spectral regions, see Fig.~\ref{img:HLbLMandelstamDiagramPWE}. This implies that a partial-wave expansion is valid without restrictions. This is true even in the case of the pion box, which provides the opportunity to check the partial-wave formalism in a case where we know the full result. However, one has to bear in mind that the double-spectral representation for the pion box differs from the ``1disc'' and ``2disc'' boxes: in the case of the pion box, there are only two-pion intermediate states, hence only three box topologies exist and there are only three double-spectral regions. Each fixed-$(s,t,u)$ representation reconstructs already all three double-spectral contributions, so that the full result can be obtained from a fixed-$(s,t,u)$ dispersion relation separately. Hence, in a symmetrized version for the pion-box one has to take one third of the sum of fixed-$(s,t,u)$ representations:
\begin{align*}
	\label{eq:FixedstuDRforPionBox}
	\Pi_i^{\pi\text{-box}}(s,t,u) = \frac{1}{3} \begin{aligned}[t]
		& \bigg( \frac{1}{\pi} \int_{4M_\pi^2}^\infty dt' \frac{\Im \Pi_i^{\pi\text{-box}}(s,t',u')}{t'-t} + \frac{1}{\pi} \int_{4M_\pi^2}^\infty du' \frac{\Im \Pi_i^{\pi\text{-box}}(s,t',u')}{u'-u} \\
		&+  \frac{1}{\pi} \int_{4M_\pi^2}^\infty ds' \frac{\Im \Pi_i^{\pi\text{-box}}(s',t,u')}{s'-s} + \frac{1}{\pi} \int_{4M_\pi^2}^\infty du' \frac{\Im \Pi_i^{\pi\text{-box}}(s',t,u')}{u'-u} \\
		&+  \frac{1}{\pi} \int_{4M_\pi^2}^\infty ds' \frac{\Im \Pi_i^{\pi\text{-box}}(s',t',u)}{s'-s} + \frac{1}{\pi} \int_{4M_\pi^2}^\infty dt' \frac{\Im \Pi_i^{\pi\text{-box}}(s',t',u)}{t'-t} \bigg) , \end{aligned} \mytag
\end{align*}
and the relation is exact.


\subsection{Sum rules for the BTT scalar functions}

\label{sec:SumRulesBTTFunctions}

The Lorentz decomposition of the HLbL tensor is only unique up to
transformations that do not introduce kinematic singularities, hence there
is a fair amount of freedom in choosing a particular representation. One
important aspect of such transformations concerns the fact that the
different mass dimensions of the Lorentz structures imply different mass
dimensions of the scalar functions $\Pi_i$, which must be reflected in a
different asymptotic behavior. Indeed if we assume, as it is natural, a
uniform asymptotic behavior of the whole HLbL tensor, i.e.\ in all
Mandelstam variables and for all tensor components, this implies that
functions multiplying Lorentz structures of higher mass dimension should
fall down even faster for asymptotic values of the Mandelstam variables. In
order to have a predictive framework, we require, that all BTT scalar
functions satisfy unsubtracted (i.e.~parameter-free) dispersion relations,
and in particular that those multiplying the Lorentz structures with lowest
mass dimensions fall down like the inverse of the Mandelstam variables at
infinity. This hypothesis, which will be tested later on, implies that the
HLbL tensor behaves asymptotically as 
\begin{align}
	\Pi^{\mu\nu\lambda\sigma} \asymp s, t, u ,
\end{align}
and that the BTT scalar functions behave (up to logarithmic corrections)
according to: 
\begin{align}
	\begin{split}
		\label{eq:BTTAsymptoticBehaviour}
		\Pi_1, \Pi_4 &\asymp \frac{1}{s}, \frac{1}{t}, \frac{1}{u} , \\
		\Pi_7, \Pi_{19}, \Pi_{37}, \Pi_{49} &\asymp \frac{1}{s^2}, \frac{1}{t^2}, \frac{1}{u^2} , \\
		\Pi_{31} &\asymp \frac{1}{s^3}, \frac{1}{t^3}, \frac{1}{u^3} ,
	\end{split}
\end{align}
with analogous asymptotics for the functions related by crossing
symmetry. Under this assumption, the functions $\Pi_1, \ldots, \Pi_6$
fulfill an unsubtracted dispersion relation. However, as they fall down to
zero even faster, the functions $\Pi_7$, $\ldots$ fulfill not only
unsubtracted dispersion relations, but even a set of sum rules. These sum
rules ensure that the result for the HLbL tensor is independent of the
choice of the tensor decomposition: the difference between the Mandelstam
representations for one set of scalar coefficient functions and a second,
equally valid set of functions will vanish as a consequence of the sum
rules (also known as ``superconvergence relations'' \cite{Martin:1970}). 

Consider for example $\Pi_7$ for fixed $t=t_b=q_2^2 + q_4^2$. At this
kinematic point, the Tarrach singularity is absent and $\Pi_7 = \tilde
\Pi_7$ is unambiguously defined (up to the redundancy in 4 space-time
dimensions), see~\cite{Colangelo:2015ama}. It fulfills an unsubtracted
fixed-$t$ dispersion relation:\footnote{All imaginary parts are understood
  to be evaluated on the upper rim of the cut in the respective channel.} 
\begin{align}
	\label{eq:Pi7UnsubtractedDR}
	\Pi_7\big|_{t=t_b} = \frac{1}{\pi} \int_{s_0}^\infty ds' \frac{\Im
          \Pi_7(s',t_b,\Sigma-t_b-s')}{s'-s} + \frac{1}{\pi}
        \int_{u_0}^\infty du' \frac{\Im\Pi_7(\Sigma-t_b-u',t_b,u')}{u'-u}, 
\end{align}
where $s_0$ and $u_0$ denote the threshold in the respective channel.
Due to the asymptotic behavior, $s\; \Pi_7$ fulfills an unsubtracted
dispersion relation as well: 
{\small
\begin{align}
	s \; \Pi_7\big|_{t=t_b} = \frac{1}{\pi} \int_{s_0}^\infty ds'
        \frac{s' \Im\Pi_7(s',t_b,\Sigma-t_b-s')}{s'-s} + \frac{1}{\pi}
        \int_{u_0}^\infty du' \frac{(\Sigma-t_b-u')
          \Im\Pi_7(\Sigma-t_b-u',t_b,u')}{u'-u} . 
\end{align}}%
We subtract this equation once using
\begin{align}
	\frac{1}{s'-s} = \frac{1}{s'} + \frac{s}{s'(s'-s)}
\end{align}
as well as
\begin{align}
	\frac{1}{s'-s} = \frac{1}{\Sigma-t_b-u' - (\Sigma-t_b-u)} = -\frac{1}{u'-u}
\end{align}
in the $u$-channel integral to obtain:
\begin{align}
	\begin{split}
		\Pi_7\big|_{t=t_b} &= \frac{1}{s} \left( \frac{1}{\pi}
                  \int_{s_0}^\infty ds' \Im\Pi_7(s',t_b,\Sigma-t_b-s') -
                  \frac{1}{\pi} \int_{u_0}^\infty du'
                  \Im\Pi_7(\Sigma-t_b-u',t_b,u') \right) \\ 
			&\quad + \frac{1}{\pi} \int_{s_0}^\infty ds'
                        \frac{\Im\Pi_7(s',t_b,\Sigma-t_b-s')}{s'-s} +
                        \frac{1}{\pi} \int_{u_0}^\infty du'
                        \frac{\Im\Pi_7(\Sigma-t_b-u',t_b,u')}{u'-u} . 
	\end{split}
\end{align}
The comparison with~\eqref{eq:Pi7UnsubtractedDR} gives the following sum rule:
\begin{align}
	\frac{1}{\pi} \int_{s_0}^\infty ds' \Im\Pi_7(s',t_b,\Sigma-t_b-s')
        - \frac{1}{\pi} \int_{u_0}^\infty du'
        \Im\Pi_7(\Sigma-t_b-u',t_b,u') = 0. 
\end{align}

In the case of $\Pi_{31}$, an even higher-degree sum rule is
fulfilled. Starting from the unsubtracted fixed-$s$ dispersion relation 
{\small
\begin{align}
	t^2 \; \Pi_{31}\big|_{s=s_b} = \frac{1}{\pi} \int_{t_0}^\infty dt' \frac{t'^2 \Im\Pi_{31}(s_b,t',\Sigma-s_b-t')}{t'-t} + \frac{1}{\pi} \int_{u_0}^\infty du' \frac{(\Sigma-s_b-u')^2 \Im\Pi_{31}(s_b,\Sigma-s_b-u',u')}{u'-u},
\end{align}}%
with $s_b=q_3^2+q_4^2$, two subtractions lead to
{\small
\begin{align}
	\begin{split}
		\Pi_{31}\big|_{s=s_b} &= \frac{1}{t^2} \left( \frac{1}{\pi} \int_{t_0}^\infty dt' t' \Im\Pi_{31}(s_b,t',\Sigma-s_b-t') - \frac{1}{\pi} \int_{u_0}^\infty du' (\Sigma-s_b-u') \Im\Pi_{31}(s_b,\Sigma-s_b-u',u') \right) \\
			&\quad + \frac{1}{t} \left( \frac{1}{\pi} \int_{t_0}^\infty dt' \Im\Pi_{31}(s_b,t',\Sigma-s_b-t') - \frac{1}{\pi} \int_{u_0}^\infty du' \Im\Pi_{31}(s_b,\Sigma-s_b-u',u') \right) \\
			&\quad +  \frac{1}{\pi} \int_{t_0}^\infty dt' \frac{\Im\Pi_{31}(s_b,t',\Sigma-s_b-t')}{t'-t} + \frac{1}{\pi} \int_{u_0}^\infty du' \frac{\Im\Pi_{31}(s_b,\Sigma-s_b-u',u')}{u'-u}.
	\end{split}
\end{align}}%
Both large brackets have to vanish, producing two independent sum rules for $\Pi_{31}$. We have verified these sum rules explicitly in the case of sQED, see Sect.~\ref{sec:sQEDSumRules}.

\subsection{Relation to observables}

\label{sec:RelationToObservables}

In Sect.~\ref{sec:DRforHLbL}, we have derived the form of the dispersion
relation for general two-pion contributions to $(g-2)_\mu$, writing the
results~\eqref{eq:DRrescattering} and~\eqref{eq:FixedstuDRforPionBox} for a
generic BTT function $\Pi_i$. In a next step, we want to use this
dispersion relation for the actual input in the $(g-2)_\mu$ master
formula~\eqref{eq:MasterFormulaPolarCoord}. Our goal is to establish via
unitarity a relation between the two-pion contribution to $(g-2)_\mu$ and
helicity amplitudes for the sub-process $\gamma^*\gamma^{(*)}\to\pi\pi$. 

While the BTT decomposition solves the problem of kinematic singularities,
the 54 scalar functions $\Pi_i$ have the disadvantage to form a redundant
set: there are 11 Tarrach redundancies~\cite{Colangelo:2015ama} and two
further ambiguities in four space-time dimensions~\cite{Eichmann:2014ooa},
as the number of helicity amplitudes for HLbL scattering is
41. Furthermore, in the on-shell limit of the external photon contributions
from its longitudinal polarization must not survive. This reduces the
number of helicity amplitudes to $\frac{3^3\times2}{2} = 27$. In the limit
of $(g-2)_\mu$ kinematics, the number of independent amplitudes is further
reduced to 19, see Sect.~\ref{sec:MasterFormula}. Note, however, that this
limit applies to the {\em outer} kinematics in the master formula and not
to the imaginary parts inside the dispersion integrals, where one
Mandelstam variable is integrated over and thus not fixed to $(g-2)_\mu$
kinematics~\eqref{eq:Gm2Kinematics}. 

While working with a redundant set of functions may, at first sight, seem
as a minor nuisance which should in the final result take care of itself
and lead to a unique and correct answer, this is not the case in our
context. The origin of the problem is that (i) establishing the relation
between the physical observables (i.e.\ the helicity amplitudes) and the BTT
functions, (ii) projecting on partial waves, and (iii) writing down
dispersion relations, are not necessarily commuting operations.
In~\cite{Colangelo:2015ama}, we constructed single-variable dispersion
relations that are free of the Tarrach redundancies. However, for most of
the scalar functions, we only found a dispersion relation in one of the
three channels, which was sufficient to obtain the dispersive reconstruction of
the pion box. For general two-pion contributions~\eqref{eq:DRrescattering}
we need all three fixed-$(s,t,u)$ dispersion relations. Furthermore, the
fact that longitudinal polarizations of the external photon do not
contribute is not immediately manifest in dispersion relations for the BTT
functions $\Pi_i$. In order to solve all these problems we must construct
another basis that is appropriate for the kinematics in the 
imaginary parts of the dispersion integrals. The scalar functions of this
basis, which we will call $\check\Pi_i$, are in one-to-one correspondence
with the 27 singly-on-shell helicity amplitudes.

In Sect.~\ref{sec:SinglyOnShellBasis}, we explain how to derive these
singly-on-shell basis functions $\check\Pi_i$. In the construction, we make
use of the sum rules for the BTT scalar functions $\Pi_i$, derived in
Sect.~\ref{sec:SumRulesBTTFunctions}. Readers who are not interested in the
technical details of the derivation may skip the following subsection
and jump directly to Sect.~\ref{sec:Pichecksolution}, where we present the
solution for the $\check\Pi_i$ functions. As a by-product in the derivation
of the singly-on-shell basis, we find a set of 15 sum rules for fixed-$t$
kinematics, presented in Sect.~\ref{sec:PhysicalSumRules}. These physical
sum rules are of relevance for the construction of the input on
$\gamma^*\gamma^{(*)}\to\pi\pi$ and can be considered a generalization of
certain sum rules for forward HLbL scattering
from~\cite{Pascalutsa:2012pr}.

\subsubsection{Construction of the singly-on-shell basis}

\label{sec:SinglyOnShellBasis}

The most efficient way to obtain a representation for the two-pion HLbL
contribution to $(g-2)_\mu$ involving only physical helicity amplitudes is
the construction of a basis $\{\check\Pi_i\}$ for singly-on-shell
kinematics that can be used together with unsubtracted single-variable
dispersion relations. In such a basis, contributions from longitudinal
polarizations of the external photon are manifestly absent. As we will see,
this construction is possible due to the presence of the sum rules for the
BTT scalar functions derived in Sect.~\ref{sec:SumRulesBTTFunctions}. The
rather surprising fact that contributions from unphysical polarizations are
not trivially absent in a representation involving redundancies is
explained in App.~\ref{sec:UnphysicalPolarizations}. 

Let us define the transformation from the BTT functions to $\hat\Pi_i$ as
the $54\times54$ matrix $t$: 
\begin{align}
	\hat\Pi_i = \sum_{j=1}^{54} t_{ij} \Pi_j .
\end{align}
In terms of the Lorentz structures $\hat T_i^{\mu\nu\lambda\sigma}$, both
the 11 Tarrach redundancies ($r$) and the two ambiguities in four
space-time dimensions ($a$) can be written as linear relations: 
\begin{align}
	\sum_{i=1}^{54} \hat T_i^{\mu\nu\lambda\sigma} r_{ij} = 0 , \quad j
        = 1, \ldots, 11 , \quad \sum_{i=1}^{54} \hat
        T_i^{\mu\nu\lambda\sigma} a_{ij} = 0 , \quad j = 1, 2 . 
\end{align}

Next, we study the unphysical polarizations: they multiply structures that
are proportional to $q_4^2$ or $q_4^\sigma$. Hence, we determine all linear
dependencies of the tensor structures in the limit $q_4^2, q_4^\sigma \to
0$, which leads to a matrix $u$ of rank 25: 
\begin{align}
	\sum_{i=1}^{54} \bigg( \lim_{q_4^2,q_4^\sigma\to0} \hat
        T_i^{\mu\nu\lambda\sigma} \bigg) u_{ij} = 0 , \quad j = 1, \ldots,
        25 , \quad \mathrm{rank}(u) = 25. 
\end{align}
If we join $u$ with the two 4d ambiguities, the rank is 27:
\begin{align}
	\mathrm{rank}(u,a) = \mathrm{rank}(u,a,r) = 27 .
\end{align}
Since $54 - 27 = 27$, this is consistent with the fact that in the
singly-on-shell limit there are 27 independent helicity amplitudes. In this
limit, the 11 Tarrach redundancies $r$ are linearly dependent on $a$ and
$u$. Moreover, in the singly-on-shell limit the transformations $u$ and $a$
can be interpreted as an ambiguity in the scalar functions: 
\begin{align}
	\label{eq:AmbiguitySinglyOnShell}
	\hat\Pi_i \mapsto \hat\Pi_i + \sum_{j=1}^{27} \bar u_{ij} \Delta_j ,
\end{align}
where we denote by $\bar u$ the $54\times27$ matrix $(u,a)$. 

We consider now the limit $q_4^2\to0$ and $t\to q_2^2$, which is relevant
for the fixed-$t$ dispersion relation. For a suitable choice of $u$ and
$a$, we still have $\mathrm{rank}(\bar u) = 27$ in this kinematic
limit. The goal is now to find all linear combinations of scalar functions
$\hat\Pi_i$ that are invariant under the
transformation~\eqref{eq:AmbiguitySinglyOnShell} and satisfy an
unsubtracted dispersion relation. Hence, we have to determine the matrix
$\hat p$, such that 
\begin{align}
	\hat p_{ki} \left( \hat\Pi_i + \sum_{j=1}^{27} \bar u_{ij} \Delta_j \right) = \hat p_{ki} \hat\Pi_i ,
\end{align}
for arbitrary $\Delta_j$, which corresponds to the null-space of $\bar u$:
\begin{align}
	\sum_{i=1}^{54} \hat p_{ki} \bar u_{ij} = 0 .
\end{align}
However, the requirement that $\hat p_{ki} \hat\Pi_i$ satisfy an
unsubtracted dispersion relation does not allow arbitrary $\hat p$. Here,
the sum rules for the BTT scalar functions $\Pi_i$, derived in
Sect.~\ref{sec:SumRulesBTTFunctions}, are employed in an essential way: the
linear combinations 
\begin{align}
	\sum_{i=1}^{54} \hat p_{ki} \hat\Pi_i\Bigg|{}_{\substack{q_4^2=0 \\ t=q_2^2}} = \sum_{i,j=1}^{54} \hat p_{ki} t_{ij} \Pi_j \Bigg|{}_{\substack{q_4^2=0 \\ t=q_2^2}} =: \sum_{j=1}^{54} p_{kj} \Pi_j \Bigg|{}_{\substack{q_4^2=0 \\ t=q_2^2}}
\end{align}
must only involve coefficients $p_{kj}$ that depend linearly on $s$ for
$j\ge7$ or at most quadratically for $j\in\{31, \ldots, 36\}$, because the
scalar functions $\Pi_j$ satisfy the linear sum rule for $j\ge7$ and the
quadratic sum rule for $j\in\{31, \ldots, 36\}$. Meanwhile, the
coefficients $p_{kj}$ can have an arbitrary dependence on the virtualities
$q_i^2$, which in the dispersion relation are fixed external
quantities. Hence, we write 
\begin{align}
	p_{kj}(s) = \sum_{l=0}^2 p_{kjl} s^l
\end{align}
and bear in mind the mentioned restrictions for $p_{kj1}$ and
$p_{kj2}$. Solving this linear algebra exercise is the major problem of the
calculation. With the help of computer algebra, we obtain a $42\times162$
matrix $(p_{kj0},p_{kj1},p_{kj2})$, whose contraction $p_{kj}(s)$ has again
rank 27 and is in one-to-one correspondence with the 27 singly-on-shell
helicity amplitudes. 

In a last step, we consider the limit $s\to q_3^2$ (which is now equivalent
to $q_4\to0$) and search for linear relations 
\begin{align}
	\hat\Pi_i \Big|_{q_4=0} = \sum_{k=1}^{42} \sum_{j=1}^{54} b_{ik} p_{kj}(q_3^2) \Pi_j \Big|_{q_4=0} , \quad i=1,\ldots, 11, 13, 14, 16, 17, 39, 50, 51, 54,
\end{align}
for all the functions contributing to $(g-2)_\mu$, where the coefficients
$b_{ik}$ are functions of the virtualities $q_1^2$, $q_2^2$, and
$q_3^2$. The solution of this system is not unique: $p_{kj}$ is a
$42\times54$ matrix of rank 27, hence there exist 15 null relations 
\begin{align}
	\label{eq:NullRelationsSumRules}
	0 = \sum_{k=1}^{42} n_{ik} p_{kj}(q_3^2) , \quad i=1,\ldots, 15,
\end{align}
again with coefficients $n_{ik}$ depending only on $q_1^2$, $q_2^2$, and
$q_3^2$. 

With the constructed solution for $p_{kj}$, we can build a singly-on-shell
basis by selecting a convenient set of 27 independent linear
combinations. We choose the basis functions $\check\Pi_i$ in such a way
that only the first 19 contribute to $(g-2)_\mu$: 
\begin{align}
	\check\Pi_i(s;q_i^2) := \sum_{k=1}^{42} \sum_{j=1}^{54} b_{g_i k} p_{kj}(s) \Pi_j \Bigg|{}_{\substack{q_4^2=0\\t=q_2^2}} , \quad i=1,\ldots, 19 ,
\end{align}
where $\{ g_i \} := G = \{1,\ldots, 11, 13, 14, 16, 17, 39, 50, 51, 54\}$.

\subsubsection{Scalar functions for the two-pion dispersion relations}

\label{sec:Pichecksolution}

In Sect.~\ref{sec:SinglyOnShellBasis}, we have detailed the derivation of
the 27 singly-on-shell basis functions $\check\Pi_i$. These functions have
the following four important properties.  
\begin{enumerate}
	\item They are linear combinations of the BTT functions $\Pi_i$ for
          fixed-$t$ with coefficients depending on $s$ only in such a way 
          that the sum rules in Sect.~\ref{sec:SumRulesBTTFunctions} allow
          an unsubtracted dispersion relation for the $\check\Pi_i$: 
		\begin{align}
			\label{eq:PicheckFixedTDR}
			\check\Pi_i\Bigg|{}_{\substack{q_4^2=0 \\ t=q_2^2}}
                        = \frac{1}{\pi} \int_{4M_\pi^2}^\infty ds'
                        \frac{\Im \check\Pi_i(s',q_2^2,u')}{s'-s} +
                        \frac{1}{\pi} \int_{4M_\pi^2}^\infty du' \frac{\Im
                          \check\Pi_i(s',q_2^2,u')}{u'-u} . 
		\end{align}
	\item In the limit $q_4\to0$, a subset of 19 functions reproduces
          the input for the master
          formula~\eqref{eq:MasterFormulaPolarCoord} for $(g-2)_\mu$: 
	\begin{align}
		\check\Pi_i \Big|_{s=q_3^2} = \hat\Pi_{g_i} \Big|_{q_4=0} , \quad i = 1, \ldots, 19.
	\end{align}
	\item They are free from Tarrach
          redundancies~\cite{Colangelo:2015ama} and the ambiguity in four
          space-time dimensions~\cite{Eichmann:2014ooa}. 
	\item A basis change relates them to the 27 singly-on-shell
          helicity amplitudes, hence the imaginary parts in the dispersion
          integrals~\eqref{eq:PicheckFixedTDR} can be expressed in terms of
          physical helicity amplitudes for $\gamma^*\gamma^*\to\pi\pi$ and
          $\gamma^*\gamma\to\pi\pi$. 
\end{enumerate}
The first point reflects the need to obtain a parameter-free prediction for
the two-pion contribution to $(g-2)_\mu$. The second point implies that we
can construct a dispersive representation for $\hat\Pi_i$ of the
form~\eqref{eq:DRrescattering} (or~\eqref{eq:FixedstuDRforPionBox} for the
pion box), by summing fixed-$(s,t,u)$ representations. The fixed-$t$
representation is given directly by~\eqref{eq:PicheckFixedTDR}, while
fixed-$s$ and fixed-$u$ representations follow from the crossing
relations~\eqref{eq:CrossingRelationsPiHat}.
The last two properties mean that we can relate the two-pion contribution
to $(g-2)_\mu$ to observable quantities. In particular, longitudinal
polarizations for the external photon must drop out in the limit
$q_4^2\to0$. 

The 19 functions contributing to $(g-2)_\mu$ can be written as (for
$q_4^2=0$ and $t=q_2^2$) 
\begin{align}
	\label{eq:PicheckFunctions}
	\check\Pi_i &= \hat\Pi_{g_i} + (s-q_3^2) \bar\Delta_i + (s-q_3^2)^2 \bar{\bar{\Delta}}_i ,
\end{align}
where $\{ g_i \} = G = \{1,\ldots, 11, 13, 14, 16, 17, 39, 50, 51, 54\}$ and where
\begin{align}
	\label{eq:PicheckPihatDiff}
	\bar\Delta_i &= \sum_{j=7}^{54} \bar d_{ij} \Pi_j , \quad
        \bar{\bar{\Delta}}_i = \sum_{j=31}^{36} \bar{\bar d}_{ij} \Pi_j 
\end{align}
are given explicitly in App.~\ref{sec:PiCheckFunctions}. The coefficients
$\bar d_{ij}$ and $\bar{\bar d}_{ij}$ depend only on $q_1^2$, $q_2^2$, and
$q_3^2$. To verify that the functions $\check\Pi_i$ fulfill unsubtracted
fixed-$t$ dispersion relations, we observe 
\begin{align}
	\begin{split}
		\label{eq:PicheckDispRel}
		\frac{1}{\pi} \int ds' \frac{\Im \check\Pi_i(s')}{s'-s} &=
                \frac{1}{\pi} \int ds' \frac{\Im \hat\Pi_{g_i}(s')}{s'-s}
                \\ 
			&\quad + \frac{1}{\pi} \int ds' \frac{(s'-q_3^2)
                          \Im \bar\Delta_i(s')}{s'-s} + \frac{1}{\pi} \int
                        ds' \frac{(s'-q_3^2)^2 \Im
                          \bar{\bar\Delta}_i(s')}{s'-s} \\ 
			&= \frac{1}{\pi} \int ds' \frac{\Im
                          \hat\Pi_{g_i}(s')}{s'-s} \\ 
			&\quad + (s-q_3^2) \frac{1}{\pi} \int ds' \frac{\Im
                          \bar\Delta_i(s')}{s'-s} + (s-q_3^2)^2
                        \frac{1}{\pi} \int ds' \frac{\Im
                          \bar{\bar\Delta}_i(s')}{s'-s} \\ 
			&\quad + \frac{1}{\pi} \int ds' \Im
                        \bar\Delta_i(s') + \frac{1}{\pi} \int ds'
                        (s+s'-2q_3^2) \Im \bar{\bar\Delta}_i(s') \\ 
			&= \hat\Pi_{g_i}(s) + (s-q_3^2) \bar\Delta_i(s) +
                        (s-q_3^2)^2 \bar{\bar\Delta}_i(s) \\ 
			&= \check\Pi_i(s) ,
	\end{split}
\end{align}
where we have used the sum rules for the BTT functions:
\begin{align}
	\begin{split}
		\label{eq:BTTsumrules}
		\int ds' \Im\Pi_i(s') &= 0 , \quad i \in\{7, \ldots, 54\} , \\
		\int ds' s' \Im\Pi_i(s') &= 0 , \quad i \in\{31, \ldots, 36\} ,
	\end{split}
\end{align}
and written both channels schematically as one integral.
This proves that the dispersion relation for $\check\Pi_i$ is indeed
fulfilled. In particular, the limit $s\to q_3^2$ provides a fixed-$t$
representation for $\hat\Pi_{g_i}$, the input for the $(g-2)_\mu$ master
formula. The solutions for fixed-$s$ and fixed-$u$ follow immediately from
the crossing relations~\eqref{eq:CrossingRelationsPiHat}
and~\eqref{eq:InternalCrossingSymmetriesPiHat}. 

Unfortunately, it turns out that it is not possible to find a
representation for the functions $\check\Pi_i$ with coefficients $\bar
d_{ij}$ and $\bar{\bar d}_{ij}$ in~\eqref{eq:PicheckPihatDiff} free of all
kinematic singularities. This is a final relic of the redundancy in the
tensor decomposition which is, however, not a real problem at all. Indeed
the contribution of $\bar\Delta_i$ and $\bar{\bar\Delta}_i$ in the
dispersion relation for $\hat\Pi_{g_i}$ vanishes due to the sum rules, and
the same is true for the residue of any kind of kinematic singularity in
the coefficients $\bar d_{ij}$ and $\bar{\bar d}_{ij}$. The residue is
defined in terms of physical quantities only and can thus be subtracted
explicitly, to obtain a representation that is manifestly free of kinematic
singularities.

Using the above sum rules, we can optimize the representation to a certain
degree. We have chosen our preferred representation in
App.~\ref{sec:PiCheckFunctions} according to the following criteria: 
\begin{itemize}
	\item We have avoided for scalar functions $\Pi_i$ that receive
          $S$-wave contributions to mix into other functions
          in~\eqref{eq:PicheckFunctions}. 
	\item We have made the singularity structure of the coefficients
          $\bar d_{ij}$ and $\bar{\bar d}_{ij}$ as simple as possible.
	\item We have optimized the convergence of the partial-wave representation
          of $(g-2)_\mu$ for the pion box.
\end{itemize}
The minimal singularity structure for the coefficients $\bar d_{ij}$ and
$\bar{\bar d}_{ij}$ consists of simple poles in $1/(q_1^2+q_3^2)$ and
singularities of the type $1/\lambda(q_1^2,q_2^2,q_3^2)$, where
$\lambda(a,b,c):=a^2+b^2+c^2-2(ab+bc+ca)$ is the K\"all\'en triangle
function. The first singularity lies on a straight line outside the
$(g-2)_\mu$ integration region, see
Fig.~\ref{img:IntegrationRegionMasterFormula}. Writing 
\begin{align} 
	\lambda_{123} := \lambda(q_1^2,q_2^2,q_3^2) = -\frac{1}{3} \tilde\Sigma^2 (1-r^2) ,
\end{align}
we see that the second type of singularity lies on the border of the
$(g-2)_\mu$ integration region. In the $(g-2)_\mu$ master
formula~\eqref{eq:MasterFormulaPolarCoord}, we subtract the residue of this
singularity at $r=1$, which vanishes due to the sum rules, to obtain a
representation without any kinematic singularities in the $(g-2)_\mu$
integration region.

\subsubsection{Physical sum rules}
\label{sec:PhysicalSumRules}

In the derivation of the singly-on-shell basis functions $\check\Pi_i$, we
have encountered the 15 null relations~\eqref{eq:NullRelationsSumRules},
which lead to sum rules involving only physical (singly-on-shell)
quantities. We build the 15 functions 
\begin{align}
	N_i(s;q_i^2) := \sum_{k=1}^{42} \sum_{j=1}^{54} n_{ik} p_{kj}(s) \Pi_j \Bigg|{}_{\substack{q_4^2=0\\t=q_2^2}} , \quad i=1,\ldots, 15 .
\end{align}
By using the null relations~\eqref{eq:NullRelationsSumRules}, we subtract
zero on the right-hand side and obtain 
\begin{align}
	N_i(s;q_i^2) &= \sum_{k=1}^{42} \sum_{j=1}^{54} n_{ik} \left( p_{kj}(s) - p_{kj}(q_3^2) \right) \Pi_j \Bigg|{}_{\substack{q_4^2=0\\t=q_2^2}} =  \sum_{k=1}^{42} \sum_{j=7}^{54} n_{ik} \left( p_{kj}(s) - p_{kj}(q_3^2) \right) \Pi_j \Bigg|{}_{\substack{q_4^2=0\\t=q_2^2}} ,
\end{align}
where the second equality follows from the fact that $p_{kj}(s) = p_{kj}(q_3^2)$ is constant for $j<7$. For $j\ge7$, $p_{kj}(s)$ is linear in $s$ or quadratic for $j\in\{31,\ldots,36\}$. Hence, we can write
\begin{align}
	p_{kj}(s) - p_{kj}(q_3^2) = (s-q_3^2) \tilde p_{kj}(s) , \quad j \ge 7 ,
\end{align}
where $\tilde p_{kj}$ is either constant or linear in $s$ for $j\in\{31, \ldots, 36\}$. Inserting $N_i$ into a dispersion integral leads to 15 linear combinations of the sum rules for the scalar functions, discussed in Sect.~\ref{sec:SumRulesBTTFunctions}:
\begin{align}
	\label{eq:PhysicalSumRules}
	\frac{1}{\pi} \int ds' \frac{\Im N_i(s')}{s'-q_3^2} &= \sum_{k=1}^{42} n_{ik} \sum_{j=7}^{54} \frac{1}{\pi} \int ds' \tilde p_{kj}(s') \Im \Pi_j(s') \Bigg|{}_{\substack{q_4^2=0\\t=q_2^2}} = 0.
\end{align}
These 15 sum rules are special: they are free of any ambiguity and only involve physical helicity amplitudes, i.e.~amplitudes with a transversely polarized external photon. They can be used to modify the fixed-$t$ representations~\eqref{eq:PicheckFunctions} of the 19 $\check\Pi_i$ functions contributing to $(g-2)_\mu$. The 15 sum rules can be written in very compact form in terms of the singly-on-shell basis functions $\check\Pi_i$, defined in App.~\ref{sec:PiCheckFunctions}:
\begin{align}
	\begin{split}
		\label{eq:SumRulesPicheck}
		0 &= \int ds' \Im \check\Pi_i(s') , \quad i = 7, 8, 9, 10, 12, 13, 16, 20, 21, 22, 23, 24 , \\
		0 &= \int ds' \Im\Big( \check\Pi_{11}(s') + \check\Pi_{18}(s') - \check\Pi_{19}(s') \Big) , \\
		0 &= \int ds' \Im\Big( \check\Pi_{15}(s') - \check\Pi_{18}(s') + \check\Pi_{19}(s') \Big) , \\
		0 &= \int ds' \Im\Big( \check\Pi_{17}(s') - \check\Pi_{18}(s') + \check\Pi_{19}(s') \Big) ,
	\end{split}
\end{align}
where fixed-$t$ kinematics is implicit. These sum rules are related to certain sum rules for forward HLbL scattering derived in~\cite{Pascalutsa:2012pr}, although we have derived them for a different kinematic situation (non-forward scattering but $q_4^2=0$). A detailed comparison is provided in App.~\ref{eq:ForwardScattering}.


\subsection{Helicity amplitudes and partial-wave expansion}
\label{sec:HelAmpPWE}

In order to determine the two-pion contribution to the scalar functions in
the master formula~\eqref{eq:MasterFormulaPolarCoord}, we write
fixed-$(s,t,u)$ dispersion relations of the form~\eqref{eq:DRrescattering},
where we take only the contribution of the two-pion intermediate state to
the imaginary parts into account. The scalar functions that fulfill
single-variable dispersion relations and reproduce the scalar functions in
the master formula are given in~\eqref{eq:PicheckFunctions}. The last
missing piece in the formalism for two-pion contributions to $(g-2)_\mu$ is
thus the link with helicity amplitudes and partial waves for
$\gamma^*\gamma^{(*)}\to\pi\pi$. 

Unitarity determines the imaginary part of the scalar functions, which is
the input in the dispersion relations, and is most conveniently expressed
in the basis of helicity amplitudes, expanded into partial waves: for
helicity partial waves the unitarity relation is diagonal. Furthermore, the
input on $\gamma^*\gamma^{(*)}\to\pi\pi$ is available in the form of
helicity partial waves: these are in principle observable quantities, even
though given the absence of double-virtual data they will have to be
reconstructed dispersively by means of the solution of a system of
Roy--Steiner
equations~\cite{Hoferichter:2011wk,Colangelo:2014dfa,Colangelo:2015ama}. In
Sect.~\ref{sec:Rescattering}, we will provide a first estimate of the
two-pion rescattering contribution by solving the Roy--Steiner equations
for $S$-waves, using a pion-pole LHC and $\pi\pi$ phase shifts based on the
inverse-amplitude
method~\cite{Dobado:1989qm,Dobado:1992ha,Dobado:1996ps,Guerrero:1998ei,GomezNicola:2001as,Nieves:2001de}.

The step from the singly-on-shell basis to the basis of
helicity amplitudes for HLbL is again rather tedious. The helicity
amplitudes can be easily expressed in terms of BTT scalar functions or the
singly-on-shell basis by contracting the HLbL with appropriate polarization
vectors, but expressing the scalar functions in terms of helicity
amplitudes requires the analytic inversion of a $27\times27$ matrix, which
is a formidable task. Here, we present the solution to this problem and
discuss the subtleties of the partial-wave expansion in connection with
$(g-2)_\mu$. In Sect.~\ref{sec:UnitarityHelAmps}, we recall the definitions
for the helicity amplitudes from~\cite{Colangelo:2015ama}. In
Sect.~\ref{sec:PWSumRules}, we comment on the implication of the sum rules
for the partial waves. In Sect.~\ref{sec:ResultsPWDispRel}, we discuss the
result for the dispersion relation in terms of helicity partial waves,
generalizing the $S$-wave result of~\cite{Colangelo:2014dfa} to arbitrary
partial waves. Some technical parts of the calculation are relegated to
App.~\ref{sec:BasisChangeHelAmps}. 

\subsubsection{Unitarity relation in the partial-wave picture}
\label{sec:UnitarityHelAmps}

Although the direct inversion of the $27\times27$ matrix is feasible, see
App.~\ref{sec:BasisChangeHelAmpsDirectInversion} for a summary of how to
achieve this task, there is a more elegant way to derive the partial-wave
unitarity relation without the need for a full inversion. We checked that
both derivations lead to identical results, but pursue the latter, more
physical approach in the main part of the paper. 

The strategy that avoids the inversion of the matrix describing the basis change relies on the following idea: by expanding the sub-process $\gamma^*\gamma^*\to\pi\pi$ into helicity partial waves, we can explicitly calculate the phase-space integral in the unitarity relation and determine the imaginary part as a sum of products of helicity partial waves. The phase-space integrals become more and more complicated for higher partial waves, but due to the fact that unitarity is diagonal for helicity partial waves, the contribution of arbitrary partial waves is determined as soon as the $S$-, $D$-, and $G$-wave discontinuities are calculated.

In phenomenological applications, we expect the contribution of partial waves beyond $D$-waves to be negligible. However, the calculation of higher partial waves allows us to check the convergence of the partial-wave series to the full result in the test case of the pion box and provides a very strong test of the formalism for the single-variable partial-wave dispersion relations. The numerical checks of the convergence will be discussed in Sect.~\ref{sec:PionBoxPWConvergence}.

In the following, we define the helicity amplitudes for HLbL and the sub-process $\gamma^*\gamma^{(*)}\to\pi\pi$. The definitions of angles and polarization vectors can be found in~\cite{Colangelo:2015ama}.

The helicity amplitudes of $\gamma^*\gamma^*\to\pi\pi$ are defined as
\begin{align}
	H_{\lambda_1\lambda_2} = e^{i(\lambda_2-\lambda_1)\phi} \epsilon_\mu^{\lambda_1}(q_1) \epsilon_\nu^{\lambda_2}(q_2) W^{\mu\nu}(p_1,p_2,q_1) .
\end{align}
For two off-shell photons, there are in principle $3^2 = 9$ helicity combinations. However, due to parity conservation and with our convention for the polarization vectors, we have the relation
\begin{align}
	\label{eq:ggpipiFlippedHelicities}
	H_{-\lambda_1-\lambda_2} = (-1)^{\lambda_1+\lambda_2} H_{\lambda_1\lambda_2} ,
\end{align}
which implies that only $\frac{3^2-1}{2}+1 = 5$ amplitudes are independent:
\begin{align}
	H_{++} = H_{--} , \quad H_{+-} = H_{-+} , \quad H_{+0} = -H_{-0} , \quad H_{0+} = - H_{0-} , \quad H_{00} .
\end{align}
Similarly, for the HLbL helicity amplitudes, defined by
\begin{align}
	H_{\lambda_1\lambda_2,\lambda_3\lambda_4} = \epsilon_\mu^{\lambda_1}(q_1) \epsilon_\nu^{\lambda_2}(q_2) {\epsilon_\lambda^{\lambda_3}}^*(-q_3) {\epsilon_\sigma^{\lambda_4}}^*(q_4) \Pi^{\mu\nu\lambda\sigma}(q_1,q_2,q_3) ,
\end{align}
there are $3^4$ helicity amplitudes, but only $\frac{3^4-1}{2}+1 = 41$ independent ones.

We introduce rescaled helicity amplitudes that remain finite in the limit $q_i^2\to0$:
\begin{align}
	\label{eq:HLbLFiniteHelicityAmplitudes}
	H_{\lambda_1\lambda_2} =: \kappa^1_{\lambda_1} \kappa^2_{\lambda_2} \bar H_{\lambda_1\lambda_2}, \quad H_{\lambda_1\lambda_2,\lambda_3\lambda_4} =: \kappa^1_{\lambda_1} \kappa^2_{\lambda_2} \kappa^3_{\lambda_3} \kappa^4_{\lambda_4} \bar H_{\lambda_1\lambda_2,\lambda_3\lambda_4} ,
\end{align}
where
\begin{align}
	\kappa^i_{\pm} = 1, \quad \kappa^i_0 = \frac{q_i^2}{\xi_i} ,
\end{align}
and $\xi_i$ refers to the normalization of the longitudinal polarization vectors. Since only the $\bar H$ amplitudes appear in the final results, this procedure avoids any confusion that might originate from a particular choice of normalization.

We define the helicity partial-wave expansion for $\gamma^*\gamma^*\to\pi\pi$ by
\begin{align}
	\label{eq:ggpipiHelicityPW}
	\bar H_{\lambda_1\lambda_2}(s,t,u) = \sum_J (2J+1) d^J_{m0}(z) h_{J,\lambda_1\lambda_2}(s) ,
\end{align}
where $m=|\lambda_1-\lambda_2|$, $z$ is the cosine of the scattering angle, and $d^J_{m_1m_2}(z)$ denotes the Wigner $d$-function.
For HLbL, we expand the helicity amplitudes as follows into partial waves:
\begin{align}
	\bar H_{\lambda_1\lambda_2,\lambda_3\lambda_4}(s,t,u) = \sum_J (2J+1) d^J_{m_1m_2}(z) h^J_{\lambda_1\lambda_2,\lambda_3\lambda_4}(s) ,
\end{align}
where $m_1 = \lambda_1-\lambda_2$, $m_2=\lambda_3-\lambda_4$.

Unitarity is diagonal for helicity partial waves, i.e.
\begin{align}
	\Imspipi h^J_{\lambda_1\lambda_2,\lambda_3\lambda_4}(s) = \eta_i \eta_f \frac{\sigma_\pi(s)}{16\pi S} h_{J,\lambda_1\lambda_2}(s) h_{J,\lambda_3\lambda_4}^*(s) ,
\end{align}
where $S$ is the symmetry factor of the two pions and
\begin{align}
	\eta_i = \left\{ \begin{matrix} -1 & \text{if }\lambda_1 - \lambda_2 = -1 , \\ 1 & \text{otherwise} , \end{matrix} \right. \quad \eta_f = \left\{ \begin{matrix} -1 & \text{if }\lambda_3-\lambda_4 = -1 , \\ 1 & \text{otherwise} \phantom{,} \end{matrix} \right.
\end{align}
account for the sign convention in~\eqref{eq:ggpipiHelicityPW}. We find the relation
\begin{align}
	\Imspipi h^J_{\lambda_1\lambda_2,-\lambda_3-\lambda_4}(s) = \Imspipi h^J_{\lambda_1\lambda_2,\lambda_3\lambda_4}(s) ,
\end{align}
where the ratio of $\eta_f$ factors compensates the sign $(-1)^{\lambda_3+\lambda_4}$ from~\eqref{eq:ggpipiFlippedHelicities}.

The HLbL tensor is written in terms of the redundant BTT Lorentz decomposition as
\begin{align}
	\label{eq:HLbLTensorDecomposition}
	\Pi^{\mu\nu\lambda\sigma} = \sum_{i=1}^{54} T_i^{\mu\nu\lambda\sigma} \Pi_i = \sum_{i=1}^{43} \mathcal{B}_i^{\mu\nu\lambda\sigma} \tilde \Pi_i .
\end{align}
For fixed $t=q_2^2$ and $q_4^2=0$, we have defined the singly-on-shell basis functions $\check\Pi_i$. The helicity amplitudes form a basis of the HLbL tensor, hence
\begin{align}
	\label{eq:ScalarFunctionsInTermsOfHelAmps}
	\Pi_i = \sum_{j=1}^{41} c_{ij} \bar H_j, \quad \tilde\Pi_i = \sum_{j=1}^{41} \tilde c_{ij} \bar H_j , \quad \check\Pi_i = \sum_{j=1}^{41} \check c_{ij} \bar H_j , \quad j = \{\lambda_1,\lambda_2,\lambda_3,\lambda_4\} .
\end{align}
The coefficients $c_{ij}$ contain 13 redundancies, the $\tilde c_{ij}$ still two (in four space-time dimensions). In the relation for $\check\Pi_i$, fixed-$t$ kinematics is implicit and the coefficients $\check c_{ij}$ are free from redundancies.

We define the following ``canonical'' ordering of $j$:
{\small
\begin{align*}
	j \in \{ 1&=\{{+}{+},{+}{+}\}, \quad & 
		2&=\{{+}{+},{+}{0}\}, \quad & 
		3&=\{{+}{+},{+}{-}\}, \quad & 
		4&=\{{+}{+},{0}{+}\}, \quad & 
		5&=\{{+}{+},{0}{0}\}, \\
		6&=\{{+}{+},{0}{-}\}, \quad & 
		7&=\{{+}{+},{-}{+}\}, \quad & 
		8&=\{{+}{+},{-}{0}\}, \quad & 
		9&=\{{+}{+},{-}{-}\}, \quad & 
		10&=\{{+}{0},{+}{+}\}, \\ 
		11&=\{{+}{0},{+}{0}\}, \quad & 
		12&=\{{+}{0},{+}{-}\}, \quad & 
		13&=\{{+}{0},{0}{+}\}, \quad & 
		14&=\{{+}{0},{0}{0}\}, \quad & 
		15&=\{{+}{0},{0}{-}\}, \\
		16&=\{{+}{0},{-}{+}\}, \quad & 
		17&=\{{+}{0},{-}{0}\}, \quad & 
		18&=\{{+}{0},{-}{-}\}, \quad & 
		19&=\{{+}{-},{+}{+}\}, \quad & 
		20&=\{{+}{-},{+}{0}\}, \\
		21&=\{{+}{-},{+}{-}\}, \quad & 
		22&=\{{+}{-},{0}{+}\}, \quad & 
		23&=\{{+}{-},{0}{0}\}, \quad & 
		24&=\{{+}{-},{0}{-}\}, \quad & 
		25&=\{{+}{-},{-}{+}\}, \\
		26&=\{{+}{-},{-}{0}\}, \quad & 
		27&=\{{+}{-},{-}{-}\}, \quad & 
		28&=\{{0}{+},{+}{+}\}, \quad & 
		29&=\{{0}{+},{+}{0}\}, \quad & 
		30&=\{{0}{+},{+}{-}\}, \\
		31&=\{{0}{+},{0}{+}\}, \quad & 
		32&=\{{0}{+},{0}{0}\}, \quad & 
		33&=\{{0}{+},{0}{-}\}, \quad & 
		34&=\{{0}{+},{-}{+}\}, \quad & 
		35&=\{{0}{+},{-}{0}\}, \\
		36&=\{{0}{+},{-}{-}\}, \quad & 
		37&=\{{0}{0},{+}{+}\}, \quad & 
		38&=\{{0}{0},{+}{0}\}, \quad & 
		39&=\{{0}{0},{+}{-}\}, \quad & 
		40&=\{{0}{0},{0}{+}\}, \\
		41&=\{{0}{0},{0}{0}\} \}, 
		\mytag
\end{align*} }%
and the subsets
\begin{align*}
	\label{eq:HLbLHelicitySubsets}
	\{ l_j \}_j &:= \{ 5, 14, 23, 32, 37, 38, 39, 40, 41 \} , \\
	\{ k_j \}_j &:= \{ 1,2,3,4,10,19,28 \} , \\
	\{ \bar k_j \}_j &:= \{ 9,8,7,6,18,27,36 \} , \\
	\{ n_j \}_j &:= \{ 11,12,13,20,21,22,29,30,31 \} , \\
	\{ \bar n_j \}_j &:= \{ 17,16,15,26,25,24,35,34,33 \} .
	\mytag
\end{align*}
The meaning of these subsets is the following: the subset $\{ l_j \}_j$ corresponds to helicity amplitudes with $\bar H_{\bar j} = \pm \bar H_{j}$, where $\bar j := \{ \lambda_1, \lambda_2, -\lambda_3, -\lambda_4\}$. For the subset $\{k_j\}_j$, the Wigner $d$-functions for $j$ and $\bar j$ are identical up to a sign, while for the subset $\{n_j\}_j$ this is not the case.

The imaginary parts of the scalar functions are given by
\begin{align*}
	\label{eq:HLbLScalarFunctionsPWImaginaryParts}
	\begin{split}
		\Imspipi \check \Pi_i &= \sum_{j=1}^{41} \sum_J \check c_{ij} (2J+1) d^J_{m_1^jm_2^j}(z)\Imspipi h^J_{j}(s) \\
			&= \sum_J \bigg[ \begin{aligned}[t]
				& \sum_{j=1}^9 \check c_{il_j} (2J+1) d^J_{l_j}(z) \, \Imspipi h^J_{l_j}(s) \\
				& + \sum_{j=1}^{7} \left( \check c_{ik_j} + \zeta_j \check c_{i\bar k_j}  \right) (2J+1) d^J_{k_j}(z) \, \Imspipi h^J_{{k_j}}(s) \\
				& + \sum_{j=1}^{9} \left( \check c_{i n_j}  d^J_{n_j}(z) + \check c_{i\bar n_j}  d^J_{\bar n_j}(z)  \right) (2J+1) \Imspipi h^J_{{n_j}}(s) \bigg] , \end{aligned}
	\end{split} \mytag
\end{align*}
where the signs
\begin{align}
	\{ \zeta_j \}_j &= \{ +, -, +, -, +, +, + \}
\end{align}
come from the relation
\begin{align}
	\label{eq:WignerDSwapped}
	d^J_{-m_1 -m_2}(z) = (-1)^{m_1-m_2} d^J_{m_1m_2}(z) = d^J_{m_2m_1}(z) .
\end{align}
The explicit Wigner $d$-functions are
\begin{align*}
	\{ d^J_{l_j} \}_j &= \{ d^J_{00}, d^J_{10}, d^J_{20}, - d^J_{10}, d^J_{00}, - d^J_{10}, d^J_{20}, d^J_{10}, d^J_{00} \} , \\
	\{ d^J_{k_j} \}_j &= \{ d^J_{00}, -d^J_{10}, d^J_{20}, d^J_{10}, d^J_{10}, d^J_{20}, - d^J_{10} \} , \\
	\{ d^J_{n_j} \}_j &= \{ d^J_{11}, -d^J_{21}, d^J_{1-1}, d^J_{21}, d^J_{22}, d^J_{2-1}, d^J_{1-1}, -d^J_{2-1}, d^J_{11} \} , \\
	\{ d^J_{\bar n_j} \}_j &= \{ d^J_{1-1}, d^J_{2-1}, d^J_{11}, d^J_{2-1}, d^J_{2-2}, d^J_{21}, d^J_{11}, d^J_{21}, d^J_{1-1} \} ,
	\mytag
\end{align*}
where the signs are due to the use of relation~\eqref{eq:WignerDSwapped}.

In order to identify the coefficients $\check c_{il_j}$ and $(\check c_{ik_j} + \zeta_j \check c_{i\bar k_j})$, it is sufficient to know the contribution to the unitarity relation from the lowest partial waves $h^J_{l_j}$ and $h^J_{k_j}$ (which are either $S$- or $D$-waves). 
However, as the Wigner $d$-functions $d^J_{n_j}$ are different from $d^J_{\bar n_j}$, we need to know the contribution from the two lowest partial waves $h^J_{n_j}$ in order to identify the coefficients $\check c_{in_j}$ and $\check c_{i\bar n_j}$ separately. Therefore, the generalization to arbitrary partial waves is possible as soon as the contributions from $S$-, $D$-, and $G$-waves are determined.

The explicit calculation of the partial-wave unitarity relation involves rather complicated phase-space integrals, see App.~\ref{sec:TensorPhaseSpaceIntegrals}. By calculating the fully-off-shell unitarity relation, projecting onto BTT, and working out the imaginary parts of the functions $\check\Pi_i$, we have verified explicitly that the coefficients $\check c_{ij}$ for the longitudinal polarization $\lambda_4 = 0$ vanish. Therefore, $\check c_{ij}$ is effectively an invertible $27\times27$ matrix. As mentioned above, we have also computed the matrix $\check c_{ij}$ by direct inversion of the basis change from helicity amplitudes to the scalar functions, see App.~\ref{sec:BasisChangeHelAmpsDirectInversion}. The fact that the result agrees with the one from the phase-space calculation provides a very strong cross check, and in addition the full inversion allows one to separate the $\check c_{ik_j}$ and $\check c_{i\bar k_j}$ coefficients.

\subsubsection{Approximate partial-wave sum rules}
\label{sec:PWSumRules}

Before returning to the final result, we comment on the role of the sum rules in the context of a partial-wave expansion.
In Sect.~\ref{sec:PhysicalSumRules}, we have derived a set of 15 sum rules for the $\check\Pi_i$ functions, which, after a basis change, can be written in terms of the 27 singly-on-shell helicity amplitudes for HLbL scattering. By construction, these sum rules only hold true for the full helicity amplitudes. In particular, when expanding the imaginary part of the helicity amplitudes into partial waves and truncating the partial-wave series, there is no reason why the sum rules should still be satisfied exactly: sum-rule violations of a size consistent with higher partial waves are expected, so that the sum rules are fulfilled only approximately. This has some important consequences.

Due to the presence of the sum rules, the formal relation between the master formula input $\hat\Pi_i$ at $q_4 = 0$ and the singly-on-shell basis functions $\check\Pi_i$ is not unique, but can be modified by linear combinations of the sum rules. If the sum rules hold exactly, all these representations are equivalent. Violating the sum rules by a truncation of the partial-wave series implies that a dependence on the precise representation of the $\check\Pi_i$ functions is introduced. Our preferred representation of the $\check\Pi_i$ functions, discussed in Sect.~\ref{sec:Pichecksolution} and App.~\ref{sec:PiCheckFunctions}, leads to a fast convergence of the partial-wave expansion in the test case of the pion box, see Sect.~\ref{sec:PionBoxTests}, but we also checked other variants and convinced ourselves in each case that indeed the sum-rule violations are consistent with a meaningful partial-wave expansion and only slight losses in the rate of convergence.

The dependence on the representation of the $\check\Pi_i$ functions or the violation of the sum rules concerns only the truncated higher partial waves. Hence, we can reverse the argument: with the assumption that sufficiently high partial waves are negligible, the included partial waves have to fulfill the sum rules, which also removes the dependence on the representation. This can be used as a check of or even a constraint on the input for the $\gamma^*\gamma^*\to\pi\pi$ helicity partial waves, in a similar way as sum rules for forward HLbL scattering have been used to derive constraints on transition form factors of higher resonances~\cite{Pascalutsa:2012pr,Pauk:2014rta,Danilkin:2016hnh}.

Out of the 15 sum rules, only a single one involves helicity amplitudes starting with $S$-waves. If we truncate the partial-wave expansion after $S$-waves, this sum rule reads
\begin{align}
	\label{eq:SwaveSumRule}
	0 &= \int_{4M_\pi^2}^\infty ds' \frac{1}{s'-q_3^2} \frac{1}{\lambda_{12}(s')} \Big( 2 \Im h^0_{++,++}(s') - (s'-q_1^2-q_2^2) \Im h^0_{00,++}(s') \Big) + \text{higher waves} ,
\end{align}
where $\lambda_{12}(s) := \lambda(s,q_1^2,q_2^2)$ denotes the K\"all\'en triangle function. Verifying that the corresponding sum rule is approximately fulfilled for the $\gamma^*\gamma^*\to\pi\pi$ amplitudes constructed in Sect.~\ref{sec:Rescattering} provides an important check on the calculation. In fact, it is precisely this sum rule that proves that the $S$-wave result derived here based on the BTT formalism and the one from~\cite{Colangelo:2014dfa} are equivalent. We note that in the limit of forward kinematics the sum rule~\eqref{eq:SwaveSumRule} reduces to the $S$-wave approximation of the sum rule~(27b) in~\cite{Pascalutsa:2012pr}.

\subsubsection{Result for arbitrary partial waves}
\label{sec:ResultsPWDispRel}

The calculations of the previous sections allow one to reconstruct the full result for the dispersion relation for HLbL two-pion contributions to $(g-2)_\mu$. The imaginary part of the functions $\check\Pi_i$, which have to be inserted into the dispersion integrals, are provided by~\eqref{eq:HLbLScalarFunctionsPWImaginaryParts}. Evaluated at $s=q_3^2$, the dispersion relations give the $s$-channel contribution for the fixed-$t$ representation of all 19 $\hat\Pi_i$ functions that contribute to $(g-2)_\mu$. Using the crossing relations~\eqref{eq:CrossingRelationsPiHat} and~\eqref{eq:InternalCrossingSymmetriesPiHat}, we obtain the five other contributions: the $u$-channel contribution for fixed-$t$ as well as both channels in the fixed-$s$ and fixed-$u$ representations. Hence, all six integrals in a dispersion relation for the functions $\hat\Pi_i$ of the form~\eqref{eq:DRrescattering} or~\eqref{eq:FixedstuDRforPionBox} can be calculated.

The crucial ingredient in this calculation is the basis change $\check
c_{ij}$ from scalar functions to helicity amplitudes, which enables the
generalization of the $S$-wave result of~\cite{Colangelo:2014dfa} to
arbitrary partial waves. The matrix $\check c_{ij}$ contains two types of
ostensible kinematic singularities: 
\begin{enumerate}
	\item The kinematic singularities of the singly-on-shell basis
          $\check\Pi_i$ are present, as explained in
          Sect.~\ref{sec:Pichecksolution}. In the dispersion relation,
          their residues vanish due to the sum rules, hence they can be
          subtracted explicitly in the master formula for $(g-2)_\mu$. 
	\item Additional kinematic singularities
          $(-q_2^2)^{-n/2}$, $n=1,\ldots,4$, show up in the
          coefficients $\check c_{ij}$. They are introduced by the basis
          change to helicity amplitudes, i.e.\ they cancel against
          kinematic zeros in the helicity amplitudes, present
          in~\eqref{eq:HLbLScalarFunctionsPWImaginaryParts} in the
          Wigner-$d$ functions for fixed-$t$ kinematics. 
\end{enumerate}
Unfortunately, the matrix $\check c_{ij}$ is too lengthy to be shown here
in full, but is provided as supplemental material in the form of a
\textsc{Mathematica} notebook.\footnote{In this notebook, we make use of \textsc{FeynCalc}~\cite{Mertig:1990an,Shtabovenko:2016sxi}.}

In contrast, the explicit results for the two-pion dispersion relation in
the $S$-wave approximation are very compact: 
\begin{align}
	\begin{split}
		\label{eq:SWaveDR}
		\hat\Pi_4^{J=0} &= \frac{1}{\pi} \int_{4M_\pi^2}^\infty ds' \frac{-2}{\lambda_{12}(s')(s'-q_3^2)^2} \Big( 4s' \Im h^0_{++,++}(s') - (s'+q_1^2-q_2^2)(s'-q_1^2+q_2^2) \Im h^0_{00,++}(s') \Big) , \\
		\hat\Pi_5^{J=0} &= \frac{1}{\pi} \int_{4M_\pi^2}^\infty dt' \frac{-2}{\lambda_{13}(t')(t'-q_2^2)^2} \Big( 4t' \Im h^0_{++,++}(t') - (t'+q_1^2-q_3^2)(t'-q_1^2+q_3^2) \Im h^0_{00,++}(t') \Big) , \\
		\hat\Pi_6^{J=0} &= \frac{1}{\pi} \int_{4M_\pi^2}^\infty du' \frac{-2}{\lambda_{23}(u')(u'-q_1^2)^2} \Big( 4u' \Im h^0_{++,++}(u') - (u'+q_2^2-q_3^2)(u'-q_2^2+q_3^2) \Im h^0_{00,++}(u') \Big) , \\
		\hat\Pi_{11}^{J=0} &= \frac{1}{\pi} \int_{4M_\pi^2}^\infty du' \frac{4}{\lambda_{23}(u')(u'-q_1^2)^2} \Big( 2 \Im h^0_{++,++}(u') - (u'-q_2^2-q_3^2) \Im h^0_{00,++}(u') \Big) , \\
		\hat\Pi_{16}^{J=0} &= \frac{1}{\pi} \int_{4M_\pi^2}^\infty dt' \frac{4}{\lambda_{13}(t')(t'-q_2^2)^2} \Big( 2 \Im h^0_{++,++}(t') - (t'-q_1^2-q_3^2) \Im h^0_{00,++}(t') \Big) , \\
		\hat\Pi_{17}^{J=0} &= \frac{1}{\pi} \int_{4M_\pi^2}^\infty ds' \frac{4}{\lambda_{12}(s')(s'-q_3^2)^2} \Big( 2 \Im h^0_{++,++}(s') - (s'-q_1^2-q_2^2) \Im h^0_{00,++}(s') \Big) ,
	\end{split}
\end{align}
where the dependence of the helicity amplitudes on the virtualities is not written explicitly. This result agrees with~\cite{Stoffer:2014rka}. It slightly differs from the $S$-wave result presented in~\cite{Colangelo:2014dfa}, but, as explained in the previous section, this difference is precisely of the form of the sum rule~\eqref{eq:SwaveSumRule} and thus simply related to a different choice of basis.

The above result is given in a form that corresponds to the dispersion relation~\eqref{eq:DRrescattering}. In order to apply it to the pion box, one has to use~\eqref{eq:FixedstuDRforPionBox}, hence the dispersion integrals in~\eqref{eq:SWaveDR} need to be multiplied by a factor $2/3$. For the proper evaluation of the $\pi\pi$-rescattering corrections, the contribution of the pion box to the partial waves has to be subtracted: we define the operator $\mathcal{S}$, which takes care of the symmetry factor and the subtraction of the pole $\times$ pole term~\cite{Colangelo:2014dfa}. The imaginary part for the $\pi\pi$-rescattering contribution is then given by
\begin{align}
	\Imspipi h^J_{\lambda_1\lambda_2,\lambda_3\lambda_4}(s) = \eta_i \eta_f \frac{\sigma_\pi(s)}{16\pi} \mathcal{S} \Big[ h_{J,\lambda_1\lambda_2}(s) h_{J,\lambda_3\lambda_4}^*(s) \Big] ,
\end{align}
where
\begin{align}
	\begin{split}
		\mathcal{S}\Big[ h_{J,\lambda_1\lambda_2}^\text{c}(s) h_{J,\lambda_3\lambda_4}^{\text{c}*}(s) \Big] &:= h_{J,\lambda_1\lambda_2}^\text{c}(s) h_{J,\lambda_3\lambda_4}^{\text{c}*}(s) - N_{J,\lambda_1\lambda_2}(s) N_{J,\lambda_3\lambda_4}^*(s) , \\
		\mathcal{S}\Big[ h_{J,\lambda_1\lambda_2}^\text{n}(s) h_{J,\lambda_3\lambda_4}^{\text{n}*}(s) \Big] &:= \frac{1}{2} h_{J,\lambda_1\lambda_2}^\text{n}(s) h_{J,\lambda_3\lambda_4}^{\text{n}*}(s).
	\end{split}
\end{align}
The superscripts refer to charged (c) and neutral (n) pions, respectively, and $N_{J,\lambda_i\lambda_j}$ denotes the partial-wave projection of the pure pion-pole term, explicitly given in App.~\ref{sec:PionPolePartialwaves}.

\subsection{Summary of the formalism}
\label{sec:FormalismOverview}

Arguably the most important result of this paper, especially in view of future applications and generalizations, concerns the derivation of the $\check\Pi_i$ functions, which allows us to establish a direct correspondence between singly-on-shell helicity amplitudes and the scalar functions $\bar\Pi_i$ that enter the master formula~\eqref{eq:MasterFormulaPolarCoord} for the HLbL contribution to $(g-2)_\mu$. The key quantities in this construction are the various scalar amplitudes, a glossary of which is provided in
 Table~\ref{tab:FormalismOverview}, including a reference to the equation where they are defined and a short definition and explanation.
They can be roughly divided into four classes: first, the $\Pi_i$ and $\tilde\Pi_i$ are related to the general BTT decomposition of the HLbL tensor, irrespective of any application to $(g-2)_\mu$ or dispersion relations. Second,  the $\hat\Pi_i$ and $\bar\Pi_i$ isolate the functions actually relevant for $(g-2)_\mu$, by forming suitable subsets and taking the appropriate kinematic limit, but are otherwise still completely general. Third and fourth, the $\check\Pi_i$ are constructed as the crucial intermediate step in the derivation of single-variable partial-wave dispersion relations, by eliminating redundancies in the representation and thereby allowing a well-defined transition to helicity amplitudes $\bar H_j$. In combination with partial-wave unitarity, this last step completes the derivation of the dispersion relation for two-pion intermediate states in the HLbL contribution to $(g-2)_\mu$.

\begin{table}[t]
\centering
\begin{tabular}{c c c p{2cm} L{2.75cm} p{7cm}}
\toprule
funcs. & \# & Eq.\ & relevant kinematics & description & explanations \\
\midrule
$\Pi_i$ & 54 & \eqref{eqn:HLbLTensorKinematicFreeStructures} & 4 off-shell & BTT scalar functions & redundant set; free of kinematic singularities and zeros; full crossing symmetry \\
$\tilde\Pi_i$ & 43 & \eqref{eqn:HLbLTensorBasisDecomposition} & 4 off-shell & basis & true off-shell basis away from 4 space-time dimensions; no Tarrach redundancies, but two ambiguities in 4 space-time dimensions; kinematic singularities, see \cite{Colangelo:2015ama}  \\
\midrule
$\hat\Pi_i$ & 54 & \eqref{eq:HLbLTensorPiHatDecomposition} & 4 off-shell & ``basis'' change for $(g-2)_\mu$ & redundant set; free of kinematic singularities and zeros; crossing symmetry for photons 1, 2, and 3 \\
$\hat\Pi_{g_i}$ & 19 & \eqref{eq:PiHatFunctions} & $q_4 = 0$ & contributing to $(g-2)_\mu$ & subset of 19 functions $\hat\Pi_i$ that contribute to $(g-2)_\mu$: $\{g_i\} =  \{1,\ldots, 11, 13, 14, 16, \newline 17, 39, 50, 51, 54 \}$ \\
$\bar\Pi_i$ & 12 & \eqref{eq:PibarFunctions} & $q_4 = 0$ & scalar functions in master \newline formula & correspond to the 19 functions $\hat\Pi_i$ contributing to  $(g-2)_\mu$ modulo crossing symmetry $q_1\leftrightarrow -q_2$ \\
\midrule
$\check\Pi_i$ & 27 & \eqref{eq:PicheckFunctions} & fixed $t=q_2^2$, $q_4^2 = 0$ & singly-on-shell basis & fulfill unsubtracted dispersion relations; kinematic singularities depending on $q_1^2$, $q_2^2$, and $q_3^2$ only; contain in the limit $q_4\to0$ as a subset the 19 functions $\hat\Pi_i$ contributing to $(g-2)_\mu$ \\
\midrule
$\bar H_j$ & 41 & \eqref{eq:HLbLFiniteHelicityAmplitudes} & 4 off-shell & helicity amplitudes & off-shell HLbL helicity amplitudes; complicated kinematic singularities; simple unitarity relation \\
$\bar H_j\Big|_{\lambda_4 \neq 0}$ & 27 & & 3 off-shell & singly-on-shell helicity amplitudes & helicity amplitudes for the case of an external on-shell photon \\
\bottomrule
\end{tabular}
\caption{Scalar functions appearing in the formalism for the two-pion HLbL contribution to $(g-2)_\mu$.}
\label{tab:FormalismOverview}
\end{table}


\section{The pion box: test case and numerical evaluation}
\label{sec:PionBoxTests}

The interest in the pion box is twofold. On the one hand, it gives a unique
meaning to the notion of a pion loop, by virtue of its dispersive
definition as two-pion intermediates with a pion-pole LHC, and is expected
to provide the most important contribution to HLbL scattering beyond the
pseudoscalar poles. Phenomenologically, the pion box is fully determined by
the pion vector form factor, which allows us to pin down its numerical
value to very high precision, as we will show in
Sect.~\ref{sec:sQEDevaluation} including an error analysis for the form
factor input.

On the other hand, the pion box constitutes an invaluable test case for the
partial-wave formalism that we have developed in
Sect.~\ref{sec:HelicityFormalism}. Given a certain representation of the
pion vector form factor, the full pion box is known exactly, see
App.~\ref{sec:FeynmanParametrizationPionBox}. Since the partial-wave
expansion and the single-variable dispersion relations are valid not only
for the rescattering contribution but also for the pion box, provided the
correct prefactor in~\eqref{eq:FixedstuDRforPionBox} according to the
counting of double-spectral regions is taken into account, we can use the
pion box to check whether the partial-wave representation converges to the
full result upon resummation of the partial waves, and we can study the
details of the convergence behavior numerically.

In a similar way, the pion box provides a test case for the sum rules for
the HLbL scalar functions. In Sect.~\ref{sec:sQEDSumRules} we demonstrate
that they are indeed fulfilled, which is a prerequisite for the
unsubtracted single-variable dispersion relations derived in
Sect.~\ref{sec:HelicityFormalism}. In Sect.~\ref{sec:PionBoxPWConvergence},
we investigate the convergence behavior of its partial-wave representation
and discuss the implications for applications beyond the pion box, such as
the $\pi\pi$-rescattering contribution discussed in
Sect.~\ref{sec:Rescattering}.


\subsection{Evaluation of the full pion box}
\label{sec:sQEDevaluation}

For the numerical evaluation of the pion box, the representation in terms
of Feynman-parameter integrals given in
App.~\ref{sec:FeynmanParametrizationPionBox} proves most efficient. This
representation is based on the equivalence of the pion box with the FsQED
amplitude~\cite{Colangelo:2014dfa}, which we proved
in~\cite{Colangelo:2015ama}. It requires the numerical evaluation of
two-dimensional Feynman integrals with the pion vector form factor as the
only input. For a reliable evaluation of the pion-box contribution to
$(g-2)_\mu$, we therefore need a precise representation of the pion vector
form factor in the space-like region.  

Since about $95\%$ of the final pion-box $(g-2)_\mu$ integral originate
from virtualities below $1\GeV$, it is most critical that the low-energy
properties be correctly reproduced. Experimentally, the available
constraints derive from $e^+e^-\to\pi^+\pi^-$ data, which determine the
time-like form
factor~\cite{Achasov:2006vp,Akhmetshin:2006bx,Aubert:2009ad,Ambrosino:2010bv,Babusci:2012rp,Ablikim:2015orh},
and space-like measurements by scattering pions off an electron
target~\cite{Dally:1982zk,Amendolia:1986wj}. We have also checked that our
representation is consistent with extractions of the space-like form factor
from $e^-p\to e^- \pi^+ n$
data~\cite{Horn:2006tm,Tadevosyan:2007yd,Blok:2008jy,Huber:2008id},
although due to the remaining model dependence of extrapolating to the pion
pole we do not use these data in our fits. To obtain a representation that
allows us to simultaneously fit space- and time-like data, and thereby
profit from the high-statistics form factor measurements motivated mainly
by the two-pion contribution to HVP, we adopt the formalism suggested
in~\cite{Leutwyler:2002hm,Colangelo:2003yw} (similar representations have
been used
in~\cite{DeTroconiz:2001rip,deTroconiz:2004yzs,Ananthanarayan:2013zua,Ananthanarayan:2016mns,Hoferichter:2016duk,Hanhart:2016pcd}),
whose essential ingredients will be briefly reviewed in the following.

The form factor is decomposed according to
\beq
F_\pi^V(s)=\Omega^1_1(s) G_{\rho\omega}(s) G_\text{inel}(s).
\eeq
The Omn\`es factor 
\beq
\Omega^1_1(s)=\exp\limits\Bigg\{\frac{s}{\pi}\int_{4\mpi^2}^\infty ds'\frac{\delta^1_1(s')}{s'(s'-s)}\Bigg\}
\eeq
would provide the exact answer if only the elastic $\pi\pi$ channel contributed to the unitarity relation of the form factor. It is fully determined by the $P$-wave phase shift $\delta^1_1$.
Next, $G_{\rho\omega}$ describes the isospin-violating coupling to the $3\pi$ system, which becomes relevant in the vicinity of the $\rho$ peak as reflected by $\rho$--$\omega$ mixing. In practice, a one-parameter ansatz
\beq
G_{\rho\omega}(s)=1+\epsilon_{\rho\omega}\frac{s}{s_\omega-s},\qquad s_\omega=\bigg(M_\omega-i\frac{\Gamma_\omega}{2}\bigg)^2,
\eeq
proves indistinguishable from a dispersively improved version that eliminates the imaginary part below the $3\pi$ threshold~\cite{Leutwyler:2002hm,Colangelo:2003yw}. Finally, $G_\text{inel}$ parameterizes the effect of higher inelastic channels. We use a conformal mapping 
\beq
G_\text{inel}(s)=1+\sum_{i=1}^p c_i \big(z(s)^i-z(0)^i\big),\qquad z(s)=\frac{\sqrt{s_{\pi\omega}-s_1}-\sqrt{s_{\pi\omega}-s}}{\sqrt{s_{\pi\omega}-s_1}+\sqrt{s_{\pi\omega}-s}},
\eeq
where $s_{\omega\pi}=(M_\omega+\mpi)^2$ denotes the threshold where phenomenologically $4\pi$ inelasticities first start to set in and the second parameter is fixed at $s_1=-1\GeV^2$. The $\pi\pi$ phase shift is taken from the extended Roy-equation analysis of~\cite{Caprini:2011ky}, which determines $\delta^1_1$ up to $s_m=(1.15\GeV)^2$ in terms of its values at $s_m$ and $s_A=(0.8\GeV)^2$. 
Our representation thus involves $3+p$ free parameters: the $\rho$--$\omega$ mixing parameter $\epsilon_{\rho\omega}$, the two values of the phase shift at $s_m$ and $s_A$, and $p$ parameters from the conformal expansion of $G_\text{inel}$. This representation ensures that the form factor behaves as $1/s$ asymptotically as long as the phase shift approaches $\pi$, up to logarithms in agreement with the expectation from perturbative QCD~\cite{Lepage:1979zb,Lepage:1980fj,Efremov:1979qk,Efremov:1978rn,Farrar:1979aw}. We impose this asymptotic behavior by smoothly extrapolating $\delta^1_1$ to $\pi$ from the boundary $s_m$ of the applicability of the Roy solution, but checked that introducing effects from $\rho'$, $\rho''$ excitations as suggested in~\cite{Schneider:2012ez} does not impact the space-like form factor. 
The form of $G_\text{inel}$ can be further constrained by requiring that the imaginary part exhibit the expected $P$-wave behavior and respect the Eidelman--\L{}ukaszuk bound~\cite{Eidelman:2003uh}, but again the impact on the space-like form factor proves to be small.

\begin{figure}[t]
\centering
\includegraphics[height=6.35cm]{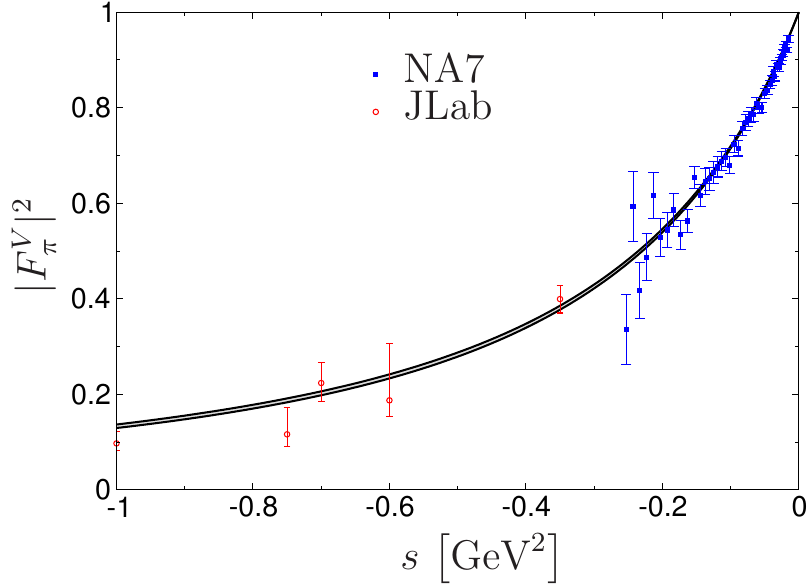}
\includegraphics[height=6.35cm]{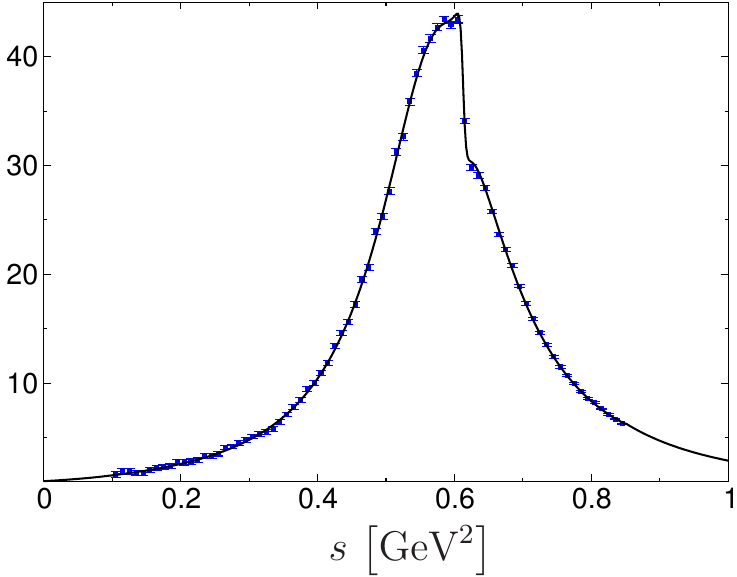}
\caption{Left: space-like pion form factor from our dispersive fit in comparison to data from NA7~\cite{Amendolia:1986wj} and JLab~\cite{Tadevosyan:2007yd,Blok:2008jy,Huber:2008id} (the latter are not included in the fit). The error band represents the variation observed between different time-like data sets. Right: pion form factor in the time-like region from the combined fit to NA7 and~\cite{Ambrosino:2010bv},
chosen here for illustrative purposes only. Fits to the other time-like data sets look very similar and lead to the same numerical results within the accuracy quoted in~\eqref{pion_box}.}
\label{fig:VFF}
\end{figure}

We fit this representation simultaneously to the space-like data from~\cite{Amendolia:1986wj} as well as one of the time-like data sets~\cite{Achasov:2006vp,Akhmetshin:2006bx,Aubert:2009ad,Ambrosino:2010bv,Babusci:2012rp,Ablikim:2015orh} (restricted to data points below $1\GeV$). Moreover, we varied $s_1$, $p=1,2$, and constructed an error band for the uncertainties in $\delta^1_1$ apart from the phase shifts at $s_m$ and $s_A$. We find that the results for the space-like form factor are extremely stable to all these variations, the largest effect being produced by the differences between the time-like data sets. For the accuracy required in HLbL scattering we can therefore simply take the largest variation among them as an uncertainty estimate, without having to perform a careful investigation of the statistical and systematic errors that are crucial when combining the different data sets for HVP. 
The result for the space-like form factor is shown in Fig.~\ref{fig:VFF}, leading to a numerical evaluation\footnote{The multidimensional integrals required for the numerical evaluation of~\eqref{eq:MasterFormulaPolarCoord} are performed using the \textsc{CUBA} library~\cite{Hahn:2004fe}.} for the pion box of
\beq
\label{pion_box}
a_\mu^{\pi\text{-box}}=-15.9(2)\times 10^{-11}.
\eeq


\subsection{Verification of sum rules}

\label{sec:sQEDSumRules}

In Sect.~\ref{sec:SumRulesBTTFunctions}, we have presented sum rules for
the BTT scalar functions that follow from a uniform asymptotic behavior of
the HLbL tensor and ensure the independence from the choice of the tensor
basis. These sum rules prove essential for the derivation of
single-variable dispersion relations that can be used with input on the
$\gamma^*\gamma^*\to\pi\pi$ helicity partial waves. Furthermore, an
important consequence of the BTT sum rules are the physical sum rules in
Sect.~\ref{sec:PhysicalSumRules}, which can be expressed in terms of
helicity amplitudes. 

An important test case for our partial-wave formalism is the pion box: the
fact that we know the full result allows us to test the convergence
behavior of the partial-wave approximation. Before turning to the tests of
the full formalism in Sect.~\ref{sec:PionBoxPWConvergence}, here we check
that the sum rules as a necessary prerequisite for the single-variable
dispersion relations are indeed fulfilled in the case of the pion box. Due
to the equivalence of the pion box with the FsQED
amplitude~\cite{Colangelo:2014dfa,Colangelo:2015ama}, these tests can be
directly performed with sQED. 

Although we have formulated the sum rules in terms of the BTT functions
$\Pi_i$, an explicit calculation must avoid the Tarrach ambiguities present
in this set. In Sect.~\ref{sec:SumRulesBTTFunctions}, we have derived the
sum rules at a certain kinematic point where the ambiguity vanishes. The
most convenient and complete method to check the sum rules uses the basis
coefficient functions $\tilde\Pi_i$, see~\cite{Colangelo:2015ama}. In this
set, the Tarrach redundancy is traded for kinematic singularities. We
remove these singularities by multiplying the $\tilde\Pi_i$ functions with
the denominators of the Tarrach poles, i.e.\ we consider 
\begin{align}
	\begin{split}
		\label{eq:TildeFunctionsTimesPoles}
		q_3 \cdot q_4 \tilde \Pi_7 &= q_3 \cdot q_4 \Pi_7 - q_2 \cdot q_3 q_2 \cdot q_4 \Pi_{31} , \\
		q_3 \cdot q_4 \tilde \Pi_9 &= q_3 \cdot q_4 \Pi_9 + q_1 \cdot q_4 \Pi_{22} , \\
		q_3 \cdot q_4 \tilde \Pi_{19} &= q_3 \cdot q_4 \Pi_{19} + q_1 \cdot q_4 q_2 \cdot q_3 \Pi_{31} , \\
		q_1 \cdot q_2 q_3 \cdot q_4 \tilde \Pi_{21} &= q_1 \cdot q_2 q_3 \cdot q_4 \Pi_{21} - q_1 \cdot q_4 q_2 \cdot q_3 \Pi_{22} , \\
		q_1 \cdot q_2 \tilde \Pi_{36} &= q_1 \cdot q_2 \Pi_{43} + q_1 \cdot q_4 \Pi_{37} .
	\end{split}
\end{align}
The functions $\tilde\Pi_1$, $\ldots$, $\tilde\Pi_6$ are not involved in
sum rules, while the functions $\tilde\Pi_{39}$, $\ldots$, $\tilde\Pi_{43}$
vanish in sQED. All the remaining functions are related to the ones above
by crossing. Apart from $q_1 \cdot q_2 q_3 \cdot q_4 \tilde \Pi_{21}$, the
combinations in~\eqref{eq:TildeFunctionsTimesPoles} have a mass dimension
that suggests an asymptotic behavior $\asymp s^{-1}, t^{-1}, u^{-1}$. The
BTT sum rules can therefore be formulated as the requirement that the
functions in~\eqref{eq:TildeFunctionsTimesPoles} fulfill an unsubtracted
Mandelstam representation. In contrast, in~\cite{Colangelo:2015ama} we only
verified that subtracted Mandelstam representations which
follow from unsubtracted ones for the BTT functions $\Pi_i$ are actually
fulfilled.  

In analogy to~\cite{Colangelo:2015ama}, we extract the sQED double-spectral
densities of these functions from the explicit expression of the loop
calculation in terms of Passarino--Veltman
amplitudes~\cite{Passarino:1978jh,tHooft:1978jhc}: in such a decomposition
into scalar loop functions the double-spectral densities are given by the
coefficients of the $D_0$ functions times the $D_0$ spectral densities. By
inserting the double-spectral densities into an unsubtracted Mandelstam
representation of the form 
\begin{align}
	\frac{1}{\pi^2} \int ds' dt' \frac{\rho_{st}(s',t')}{(s'-s)(t'-t)} + \frac{1}{\pi^2} \int ds' du' \frac{\rho_{su}(s',u')}{(s'-s)(u'-u)} + \frac{1}{\pi^2} \int dt' du' \frac{\rho_{tu}(t',u')}{(t'-t)(u'-u)} ,
\end{align}
we have verified numerically that the functions~\eqref{eq:TildeFunctionsTimesPoles} are reproduced. 
Surprisingly, even $q_1 \cdot q_2 q_3 \cdot q_4 \tilde \Pi_{21}$ fulfills
an unsubtracted Mandelstam representation, which is not expected from the
mass dimension and implies an even higher sum rule in the case of
sQED. Single-variable dispersion relations then follow from the Mandelstam
representation in the appropriate limit, including the explicit cases
discussed in Sect.~\ref{sec:SumRulesBTTFunctions}.  
While the imaginary parts for the single-variable dispersion relations
extracted in this way need to be calculated numerically, it is also
possible to obtain analytic expressions starting from a Feynman-parameter
representation of the BTT functions. The results again confirm the validity
of the sum rules, in agreement with the more general approach via the
Mandelstam representation.   

Although the sum rules for the BTT scalar functions are crucial ingredients
in the derivation of the single-variable dispersion relations, the physical
sum rules~\eqref{eq:SumRulesPicheck} have a more direct significance as
they are formulated in terms of physical quantities for the kinematics of
the $(g-2)_\mu$ single-variable dispersion integrals, and thus allow one to
impose constraints on the $\gamma^*\gamma^*\to\pi\pi$ helicity amplitudes
used as input for a numerical evaluation. We have verified that these sum
rules are fulfilled in the case of the pion box, by extracting the
imaginary parts of the $\check\Pi_i$ functions from the sQED calculation
and calculating the integrals numerically. These sQED tests thereby allow
one to establish the validity of nearly all sum rules---except for the last
one involving $\check\Pi_{17} - \check\Pi_{18} + \check\Pi_{19} =
\tilde\Pi_{40} - \tilde\Pi_{41}$, which vanishes identically in sQED. It
should be stressed that the underlying assumptions follow solely from
demanding a uniform asymptotic behavior of the HLbL tensor, but as the
discussion in App.~\ref{eq:ForwardScattering} shows, similar conclusions
can be drawn from Regge models as well. Together with the explicit checks
in the case of sQED there is therefore compelling evidence for our
assumptions regarding the asymptotic behavior of the HLbL tensor.


\subsection{Convergence of the partial-wave representation}

\label{sec:PionBoxPWConvergence}

In the following, we perform tests of the helicity partial-wave dispersion relations developed in Sect.~\ref{sec:HelicityFormalism} by applying the formalism to the pion box. In this case, a dispersion relation of the form~\eqref{eq:FixedstuDRforPionBox} has to be used in order to account for the fact that only three different double-spectral regions are present. We emphasize that in this test case each single-variable dispersion relation reconstructs the full pion box. Therefore, we can test the three channels separately---each must converge to the full result upon resummation of the partial-wave series.

The input for the $\gamma^*\gamma^*\to\pi\pi$ helicity partial waves in the case of the pion box is given by the partial-wave projection of the pure pion-pole terms, see App.~\ref{sec:PionPolePartialwaves}. In order to simplify the convergence checks, we use a simple vector-meson dominance representation for the pion vector form factor:
\begin{align}
	F_{\pi,\text{VMD}}^V(q^2) = \frac{M_\rho^2}{M_\rho^2 - q^2} .
\end{align}
Such a form factor leads to $a_\mu^{\pi\text{-box, VMD}} = -16.4 \times 10^{-11}$, which is very close to the full result obtained with the dispersive representation of the form factor discussed in Sect.~\ref{sec:sQEDevaluation}. The convergence behavior of the partial-wave expansion is not affected by the details of the form factor implementation.

Since our formalism for single-variable dispersion relations is valid for arbitrary partial waves, we can extend these tests in principle to an arbitrary angular momentum $J$. In practice, our numerical implementation becomes less reliable for large values of $J$, so that we performed the numerical tests up to $J=20$ and estimated the truncation error by extrapolation.

The HLbL contribution to $(g-2)_\mu$ is given as a sum of 12 terms in the master formula~\eqref{eq:MasterFormulaPolarCoord}, which, in principle, are completely independent. However, in the case of the pion box it turns out that especially for the lower partial waves a numerical cancellation occurs that leads to a faster convergence of $a_\mu$ than for the individual terms. Therefore, we define the following vector in the 12-dimensional space of the contributions to the master formula:
\begin{align}
	\avec_\mu^\text{HLbL} &:= \big\{ a_{\mu, i}^\text{HLbL} \big\}_i, \notag\\
	a_{\mu,i}^\text{HLbL} &:= \frac{\alpha^3}{432\pi^2} \int_0^\infty d\tilde\Sigma\, \tilde\Sigma^3 \int_0^1 dr\, r\sqrt{1-r^2} \int_0^{2\pi} d\phi \,T_i(Q_1,Q_2,\tau) \bar\Pi_i(Q_1,Q_2,\tau) ,
\end{align}
so that
\begin{align}
	a_\mu^\text{HLbL} = \sum_{i=1}^{12} a_{\mu, i}^\text{HLbL} .
\end{align}
In order to quantify the convergence behavior, we define the following two quantities: the relative deviation between the full pion-box contribution to $(g-2)_\mu$ and its partial-wave approximation
\begin{align}
	\delta_{J_\text{max}} := 1 - \frac{a_{\mu,J_\text{max}}^{\pi\text{-box, PW}}}{a_\mu^{\pi\text{-box}}} ,
\end{align}
as well as the analogous quantity in the 12-dimensional space of the contributions to the master formula
\begin{align}
	\Delta _{J_\text{max}} := \frac{\norm{\avec_{\mu,J_\text{max}}^{\pi\text{-box, PW}} - \avec_\mu^{\pi\text{-box}}}}{\norm{\avec_\mu^{\pi\text{-box}}}} ,
\end{align}
where $| \cdot |$ denotes the 12-dimensional Euclidean norm. Due to cancellations between the 12 terms in the master formula, $\delta_{J_\text{max}}$ will indicate a faster convergence than $\Delta_{J_\text{max}}$, which is more robust against cancellations.

\begin{table}[t]
	\centering
	\renewcommand{\arraystretch}{1.3}
	\begin{tabular}{c | r r | r r | r r | r r}
		\toprule
		 & \multicolumn{2}{c |}{fixed-$s$} & \multicolumn{2}{c |}{fixed-$t$} & \multicolumn{2}{c |}{fixed-$u$} & \multicolumn{2}{c}{average} \\
		$J_\text{max}$ & $\delta_{J_\text{max}}$ & $\Delta_{J_\text{max}}$ & $\delta_{J_\text{max}}$ & $\Delta_{J_\text{max}}$ & $\delta_{J_\text{max}}$ & $\Delta_{J_\text{max}}$ & $\delta_{J_\text{max}}$ & $\Delta_{J_\text{max}}$ \\
		\midrule
		0	&	$100.0\%$	&	$100.0\%$	&	$-6.2\%$		&	$35.4\%$		&	$-6.2\%$		&	$35.4\%$		&	$29.2\%$		&	$55.4\%$		\\
		2	&	$26.1\%$		&	$50.8\%$		&	$-2.3\%$		&	$5.6\%$		&	$7.3\%$		&	$8.0\%$		&	$10.4\%$		&	$20.9\%$		\\
		4	&	$10.8\%$		&	$28.2\%$		&	$-1.5\%$		&	$2.1\%$		&	$3.6\%$		&	$3.9\%$		&	$4.3\%$		&	$11.0\%$		\\
		6	&	$5.7\%$		&	$16.1\%$		&	$-0.7\%$		&	$1.1\%$		&	$2.1\%$		&	$2.2\%$		&	$2.4\%$		&	$6.2\%$		\\
		8	&	$3.5\%$		&	$9.6\%$		&	$-0.4\%$		&	$0.6\%$		&	$1.3\%$		&	$1.4\%$		&	$1.5\%$		&	$3.7\%$		\\
		10	&	$2.3\%$		&	$5.9\%$		&	$-0.2\%$		&	$0.4\%$		&	$0.9\%$		&	$1.0\%$		&	$1.0\%$		&	$2.4\%$		\\
		12	&	$1.7\%$		&	$3.8\%$		&	$-0.1\%$		&	$0.3\%$		&	$0.7\%$		&	$0.7\%$		&	$0.7\%$		&	$1.6\%$		\\
		14	&	$1.3\%$		&	$2.5\%$		&	$-0.1\%$		&	$0.2\%$		&	$0.5\%$		&	$0.5\%$		&	$0.6\%$		&	$1.1\%$		\\
		16	&	$1.0\%$		&	$1.7\%$		&	$-0.0\%$		&	$0.2\%$		&	$0.4\%$		&	$0.4\%$		&	$0.4\%$		&	$0.7\%$		\\
		18	&	$0.8\%$		&	$1.2\%$		&	$-0.0\%$		&	$0.1\%$		&	$0.3\%$		&	$0.3\%$		&	$0.4\%$		&	$0.5\%$		\\
		20	&	$0.7\%$		&	$0.9\%$		&	$-0.0\%$		&	$0.1\%$		&	$0.3\%$		&	$0.3\%$		&	$0.3\%$		&	$0.4\%$		\\
		\bottomrule
	\end{tabular}
	\caption{Convergence of the partial-wave expansion in the case of the pion box: the three single-variable dispersion relations and their average are compared. See main text for the definition of the relative deviations.}
	\label{tab:PWConvergence}
	\renewcommand{\arraystretch}{1.0}
\end{table}

Table~\ref{tab:PWConvergence} shows the results of a detailed study of the convergence behavior of the partial-wave representation for the test case of the pion box. Both measures $\delta_{J_\text{max}}$ and $\Delta_{J_\text{max}}$ for the deviation from the full pion box result are displayed for fixed-$s$, fixed-$t$, and fixed-$u$ dispersion relations, as well as for the average of the three single-variable dispersion relations~\eqref{eq:FixedstuDRforPionBox}. In this context we use the notion of fixed-$(s,t,u)$ as follows: it defines the dispersion relation for each of the six representatives in~\eqref{eq:PiHatFunctions}, while the remaining scalar functions are obtained via the crossing relation~\eqref{eq:CrossingRelationsPiHat}. In particular, this implies that in the so-called fixed-$s$ evaluation, we do use a fixed-$s$ dispersion relation for $\hat\Pi_1$, but a fixed-$t$ dispersion relation for $\hat\Pi_2$ and a fixed-$u$ dispersion relation for $\hat\Pi_3$.

Next, we comment on the following two observations:
\begin{enumerate}
	\item The $S$-wave approximation shows a particular pattern: the fixed-$s$ representation vanishes, while fixed-$t$ and -$u$ agree.
	\item The fixed-$s$ representation exhibits a slower convergence than the other two dispersion relations.
\end{enumerate}
In order to understand the first point, consider the explicit $S$-wave representation for the $\hat\Pi_i$ functions,~\eqref{eq:SWaveDR}. We note that $S$-waves contribute only to the $s$-channel discontinuities in $\hat\Pi_4$ and $\hat\Pi_{17}$, while the $t$- and $u$-channel discontinuities for these two functions start with $D$-waves (the situation for the other functions follows from crossing symmetry). A discontinuity in the $s$-channel contributes to a fixed-$t$ and fixed-$u$ dispersion relation, while in a fixed-$s$ dispersion relation the integral runs only over $t$- and $u$-channel discontinuities. This means that in the fixed-$s$ representation in Table~\ref{tab:PWConvergence}, no $S$-wave discontinuity is encountered at all, hence in this representation the first non-vanishing contribution is obtained from $D$-waves. Furthermore, because the $S$-wave $s$-channel discontinuity has no angular dependence, it contributes identically to a fixed-$t$ and fixed-$u$ dispersion relation, which makes the fixed-$t$ and fixed-$u$ results in Table~\ref{tab:PWConvergence} agree at $J_\text{max}=0$.

The second point can be understood as follows. For each of the six
representatives in~\eqref{eq:PiHatFunctions} the $s$-channel is special
with respect to the other two. This is due to the fact that the associated
Lorentz structure exhibits an $s$-channel symmetry, either $\Cr{12}{}$,
$\Cr{34}{}$, or both, the only special case being $\hat\Pi_{39}$, which is
totally crossing symmetric in all three channels. For instance, the Lorentz
structure $\hat T_1^{\mu\nu\lambda\sigma}$ is the one that belongs to the
$s$-channel (pseudo-scalar) $\pi^0$-pole contribution to HLbL scattering,
while $\hat T_4^{\mu\nu\lambda\sigma}$ can be related to an $s$-channel
scalar amplitude, which manifests itself as the $S$-wave $s$-channel
$\pi\pi$ contribution. It is therefore not surprising that even in the case
of the pion box, the $s$-channel discontinuity for the
functions~\eqref{eq:PiHatFunctions} is more important than the other two
discontinuities. Since this is the discontinuity that evades the fixed-$s$
dispersion relation, we observe a slower convergence pattern in this case.

We have performed these convergence tests not only with our preferred
representation for the $\check\Pi_i$ functions, but also with different
versions that are modified by terms that vanish due to the sum
rules~\eqref{eq:SumRulesPicheck}. While the exact numbers do differ---as
expected given the fact that the sum rules only hold for the full
amplitudes but not the individual partial waves---the sum rule violations
in the case of the pion box due to the partial-wave approximation are
reasonably small and the overall picture remains the same.

\begin{figure}[t]
\centering
\includegraphics[width=11cm]{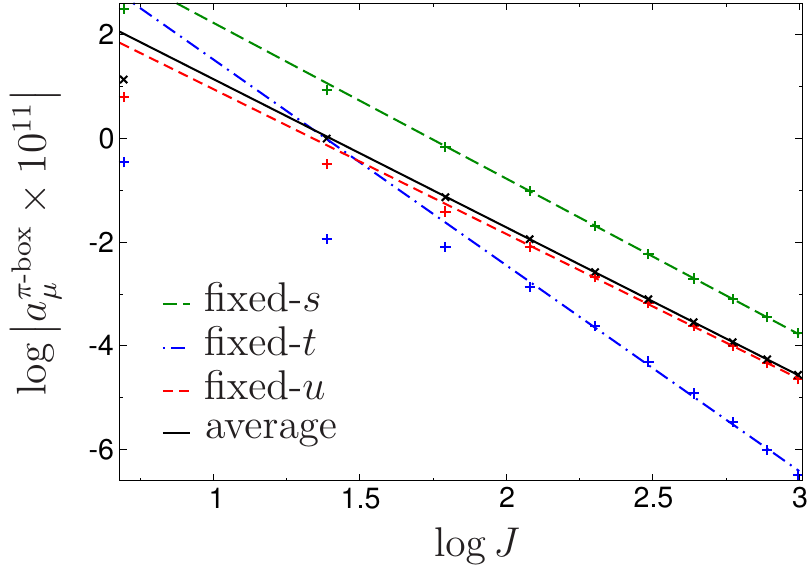}	
\caption{Extrapolation of the partial-wave results for the fixed-$(s,t,u)$  representations as well as their average, see main text for details.}
\label{fig:convergence}
\end{figure}

To fully understand the partial-wave convergence of the pion box we also
studied the remaining deviation from the full result at
$J=20$. Empirically, we observe that the size of the individual terms for
given $J$ is well described by a fit function\footnote{We thank Martin
  J.~Savage for suggesting this ansatz.} 
\beq
\label{fit}
a_{\mu,J}^{\pi\text{-box, PW}}\sim c J^x,
\eeq
which in a double-log plot produces the straight lines in
Fig.~\ref{fig:convergence}. The fact that the first few terms do not fall
on this line indicates that the form~\eqref{fit} is only asymptotic, and
might also be related to the abovementioned cancellations for low $J$ (the
fit therefore excludes the points for $J\leq 6$). The figure shows that the
rate of convergence is actually similar for fixed-$s$ and fixed-$u$, both
of which yield an exponent $x\approx -3$, while the fixed-$t$
representation converges with $x\approx -4$.  
The slower convergence of the fixed-$s$ results seen in
Table~\ref{tab:PWConvergence} is therefore a remnant of the missed $S$-wave
contribution that leads to larger deviations for small $J$, not the overall
rate of convergence. The resummation of the terms with $J>20$ based on the
fit function then removes all remaining discrepancies, providing a strong
check of the partial-wave formalism developed in
Sect.~\ref{sec:HelicityFormalism}. 

Finally, we discuss the consequences for the application of the formalism
to the case of two-pion contributions beyond the pion box, most importantly
the unitarity (or rescattering) correction. The most important difference
is related to the fact that for these applications, instead
of~\eqref{eq:FixedstuDRforPionBox}, the dispersion
relation~\eqref{eq:DRrescattering} applies, where due to the different
double-spectral regions an overall factor $1/2$ instead of $1/3$ is
required. However, this means that for the rescattering contribution the
slower convergence of the fixed-$s$ dispersion relation is of no
significance: let us assume that an important resonant contribution shows
up in a partial wave in the $s$-channel. This resonance will be captured by
the fixed-$t$ and fixed-$u$ dispersion relation (though not by the
fixed-$s$ dispersion relation). Since the full result is given by the sum
of the three dispersion relations weighted by $1/2$, this behavior is
actually expected and the absence of the resonance in the fixed-$s$
representation does not impact the average in the symmetric
representation~\eqref{eq:DRrescattering} (in contrast to the pion box,
where the missed $S$-wave contribution needs to be recovered by the higher
partial waves). Therefore, the average in the case of the pion box in
Table~\ref{tab:PWConvergence} should rather be regarded as a worst-case
scenario---for the convergence behavior in the case of rescattering
contributions the fixed-$t$ and fixed-$u$ dispersion relations are more
representative.

The second important difference concerns the presence of resonances in the
rescattering contribution, a feature that does not occur in the pion box.
We expect the rescattering contribution to be dominated by resonant
effects, whereas the convergence behavior established for the pion box can
be understood as a weighting of the partial waves. In the truncated
partial-wave series, the resonances in the included partial waves are fully
reproduced. The approximate fulfillment of the sum rules indicates then
whether neglected higher partial waves still play an important role, to the
effect that the size of the sum-rule violations allows one to estimate the
accuracy of the calculation.


\section{Application: two-pion rescattering}
\label{sec:Rescattering}

The natural application of the partial-wave formalism developed in the main part of this paper concerns $\pi\pi$ rescattering effects, which can be considered a unitarization
of the pure pion-pole LHC that defines the pion box. To isolate this contribution, it suffices to subtract the pure pion-pole piece in the partial-wave unitarity relation, 
and insert for the remainder phenomenological input for the $\gamma^*\gamma^*\to\pi\pi$ partial waves. 
The construction of such input is by itself challenging, given that direct experimental results, at least for the doubly-virtual case, are not expected in the near future.

In the on-shell case, available data on $\gamma\gamma\to\pi\pi$~\cite{Marsiske:1990hx,Boyer:1990vu,Behrend:1992hy,Mori:2007bu,Uehara:2008ep,Uehara:2009cka}
(in combination with $\gamma\gamma\to\bar K K$~\cite{Althoff:1982df,Althoff:1985yh,Aihara:1986qk,Behrend:1988hw,Albrecht:1989re,Abe:2003vn,Uehara:2013mbo})
are now sufficient to perform a partial-wave analysis~\cite{Dai:2014zta}, but such an approach appears unrealistic to control the dependence on the photon virtualities.
However, approaches that exploit more comprehensively the analytic properties of the amplitude, see~\cite{GarciaMartin:2010cw,Hoferichter:2011wk,Moussallam:2011zg} for on-shell photons, 
can be extended towards the off-shell case with limited data input required to determine parameters,
as demonstrated for the singly-virtual process in~\cite{Moussallam:2013una}. 
The essential features of the generalization towards the doubly-virtual case, i.e.\ the appearance of anomalous thresholds for time-like kinematics~\cite{Hoferichter:2013ama}
and the modifications to tensor basis and kernel functions~\cite{Colangelo:2014dfa,Colangelo:2015ama}, have already been laid out in previous work, but 
the practical implementation involves a number of challenges: due to the strong coupling between the $\pi\pi/\bar K K$ channels in the isospin-$0$ $S$-wave a single-channel analysis is 
limited to rather low energies~\cite{GarciaMartin:2010cw,Moussallam:2011zg,Moussallam:2013una,Dai:2014zta},
assumptions for the LHC and number of subtractions need to be carefully studied to reliably assess the sensitivity to the high-energy input in the dispersive integrals~\cite{Hoferichter:2011wk},
a full analysis of the generalized Roy--Steiner equations~\cite{Hoferichter:2011wk,Colangelo:2014dfa,Colangelo:2015ama} involves solving coupled $S$- and $D$-wave systems of various helicity projections, and
last but not least constraints on the $\gamma^*\gamma^*\to\pi\pi$ amplitudes from asymptotic behavior and the sum rules derived in Sect.~\ref{sec:PhysicalSumRules} need to be incorporated.
A full analysis along these lines will be left for future work. 

To obtain a first estimate of rescattering effects, we concentrate on $S$-waves and consider a further simplified system: first, we use $\pi\pi$ phase shifts from the inverse-amplitude method,
which reproduces the phenomenological phase shifts as well as the $f_0(500)$ properties at low energies, and in addition allows one to separate the $\pi\pi$ rescattering from the $\bar K K$ channel in a well-defined manner, all while providing a reasonable extrapolation for high energies. In addition, we restrict ourselves to a pion-pole LHC in the solution of the Roy--Steiner equations, which has the advantage that the off-shell behavior is still described by the pion vector form factor. 
In the following, we lay out the details of this calculation, and discuss the consequences for rescattering effects in $(g-2)_\mu$. 

\subsection[$\gamma^*\gamma^*\to\pi\pi$ helicity partial waves from the inverse-amplitude method]{\boldmath $\gamma^*\gamma^*\to\pi\pi$ helicity partial waves from the inverse-amplitude method}
\label{sec:IAM}

Unitarization within the inverse-amplitude method (IAM)~\cite{Dobado:1989qm,Dobado:1992ha,Dobado:1996ps,Guerrero:1998ei,GomezNicola:2001as,Nieves:2001de}
is based on the observation that elastic unitarity
\beq
\Im t(s)=\sigma_\pi(s)|t(s)|^2
\eeq
for a $\pi\pi$ partial-wave amplitude $t(s)$ implies
\beq
\Im \frac{1}{t(s)}=-\sigma_\pi(s),
\eeq
which together with the chiral expansion $t(s)=t_2(s)+t_4(s)+\O(p^6)$ and perturbative unitarity
\beq
\Im t_2(s)=0,\qquad \Im t_4(s)=\sigma_\pi(s) |t_2(s)|^2,
\eeq
already concludes the naive derivation of the IAM prescription
\beq
t^\text{IAM}(s)=\frac{1}{\Re \frac{1}{t(s)}-i\sigma_\pi(s)}=\frac{\big(t_2(s)\big)^2}{t_2(s)-t_4(s)}.
\eeq

However, in the single-channel case the IAM approach can be justified much more rigorously based on dispersion relations, where the only approximation involves replacing the LHC by its chiral expansion~\cite{GomezNicola:2007qj}. In this way, one can also remedy the fact that the standard IAM fails to correctly reproduce the Adler zero~\cite{Adler:1964um,Adler:1965ga}, and is thus not fully consistent with chiral symmetry. The modified form of the IAM (mIAM) becomes~\cite{GomezNicola:2007qj}
\beq
t^\text{mIAM}(s)=\frac{\big(t_2(s)\big)^2}{t_2(s)-t_4(s)+A^\text{mIAM}(s)},
\eeq
where the additional term\footnote{For $\pi\pi$ scattering the expression simplifies because $t''_2(s_2)=0$ and $t_2(s)/t'_2(s_2)=s-s_2$.}
\beq
A^\text{mIAM}(s)=\bigg(\frac{t_2(s)}{t'_2(s_2)}\bigg)^2\bigg[\frac{t_4(s_2)}{(s-s_2)^2}-\frac{s_2-s_A}{(s-s_2)(s-s_A)}\bigg(t'_2(s_2)-t'_4(s_2)+\frac{t_4(s_2)t''_2(s_2)}{t'_2(s_2)}\bigg)\bigg]
\eeq
ensures that the Adler zero $s_A=s_2+s_4+\O(p^6)$ occurs at its $\O(p^4)$ position, i.e.
\beq
t_2(s_2)=0,\qquad t_2(s_2+s_4)+t_4(s_2+s_4)=0.
\eeq
This form of the IAM thus correctly describes the low-energy phase shifts as well as resonance properties, and has indeed been used in recent years to determine the quark-mass dependence of $\sigma$ and $\rho$ resonances~\cite{Hanhart:2008mx,Pelaez:2010fj}. For our purposes, the single-channel IAM for $\pi\pi$ scattering conveniently separates the $\pi\pi$ channel from its mixing to $\bar K K$ in the vicinity of the $f_0(980)$ and defines a reasonable continuation to high energies, without compromising the low-energy physics. 

\begin{figure}[t]
\centering
\includegraphics[width=8cm]{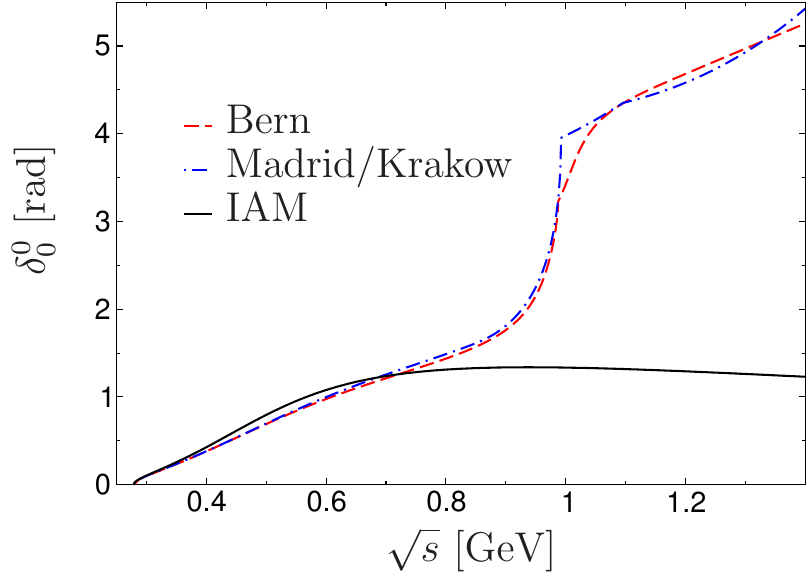}\qquad
\includegraphics[width=8cm]{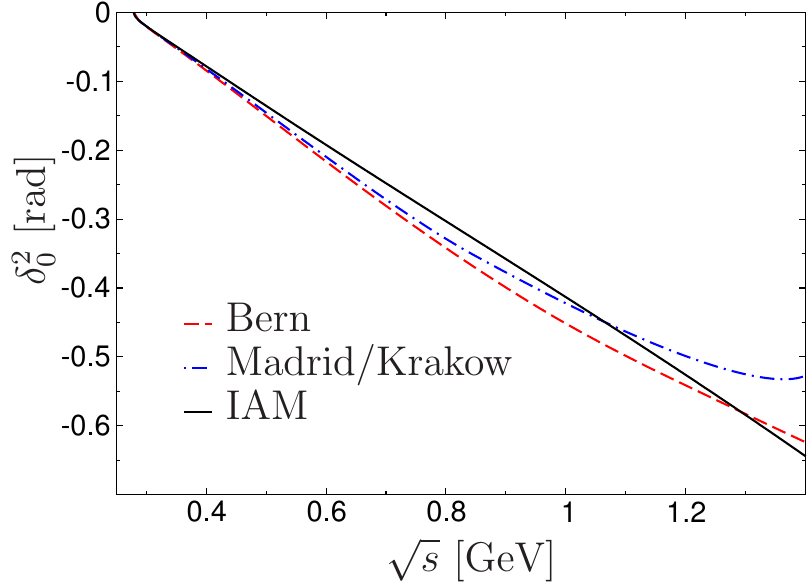}		
\caption{$I=0$ (left) and $I=2$ (right) $\pi\pi$ $S$-wave phase shifts from the IAM (black solid line), in comparison to the Bern (red dashed line)~\cite{Colangelo:2001df,Caprini:2011ky} and Madrid/Krakow (blue dot-dashed line)~\cite{GarciaMartin:2011cn} Roy-equation analyses.}
\label{fig:IAM_delta}
\end{figure}

We use the $1$-loop IAM with low-energy constants as specified in~\cite{Pelaez:2010fj}, which produces the phase shifts shown in Fig.~\ref{fig:IAM_delta}. As expected, there is good agreement throughout, apart from the fact that the IAM $I=0$ phase shift avoids the rise related to the $f_0(980)$ and the coupling to the $\bar KK$ channel. We also checked that the $\sigma$ properties~\cite{Pelaez:2015qba} are reproduced: for the pole position we find $\sqrt{s_\sigma}=(0.443+i 0.217)\GeV$, to be compared to $\sqrt{s_\sigma}=(0.441+i 0.272)\GeV$~\cite{Caprini:2005zr} and similar numbers from other recent dispersive extractions~\cite{GarciaMartin:2011jx,Moussallam:2011zg}. Accordingly, the width comes out a bit too low, as does the residue at the pole $g_{\sigma\pi\pi}$.   
This deviation is consistent with earlier IAM analyses, see e.g.~\cite{Hanhart:2008mx} for the analogous calculation including the mIAM correction, and can certainly be tolerated to obtain an estimate for 
the HLbL rescattering contribution, which, after all, only requires the amplitude on the real axis, not the analytic continuation into the complex plane where the slight discrepancy in the width would matter most.
Similarly, one can check the coupling to two photons $|g_{\sigma\gamma\gamma}/g_{\sigma\pi\pi}|\sim 0.014$, well in line with $|g_{\sigma\gamma\gamma}/g_{\sigma\pi\pi}|=0.014$ and $0.015$ from~\cite{Hoferichter:2011wk} and~\cite{Moussallam:2011zg}, respectively.

With the input for the $\pi\pi$ phase shifts specified, the $\gamma^*\gamma^*\to\pi\pi$ amplitudes follow by solving the generalized Roy--Steiner equations derived in~\cite{Colangelo:2014dfa,Colangelo:2015ama} for doubly-virtual kinematics. For the $S$-waves, these dispersion relations take the form (isospin indices are suppressed for the time being)
\begin{align}
 h_{0,++}(s)&=\Delta_{0,++}(s)+\frac{1}{\pi}\int_{4\mpi^2}^\infty ds'\bigg[\bigg(\frac{1}{s'-s}-\frac{s'-q_1^2-q_2^2}{\lambda_{12}(s')}\bigg)\Im h_{0,++}(s')+\frac{2q_1^2q_2^2}{\lambda_{12}(s')}\Im h_{0,00}(s')\bigg],\notag\\
 h_{0,00}(s)&=\Delta_{0,00}(s)+\frac{1}{\pi}\int_{4\mpi^2}^\infty ds'\bigg[\bigg(\frac{1}{s'-s}-\frac{s'-q_1^2-q_2^2}{\lambda_{12}(s')}\bigg)\Im h_{0,00}(s')+\frac{2}{\lambda_{12}(s')}\Im h_{0,++}(s')\bigg],
\end{align}
with LHC singularities represented by the inhomogeneities $\Delta_{0,++}(s)$ and $\Delta_{0,00}(s)$. These equations can be rewritten as
\begin{align}
	\begin{split}
		h_{0,++}(s)\pm \sqrt{q_1^2q_2^2}\, h_{0,00}(s)&=\Delta_{0,++}(s)\pm \sqrt{q_1^2q_2^2}\,\Delta_{0,00}(s)\\
			&+ \frac{s-\big(\sqrt{q_1^2}\mp\sqrt{q_2^2}\big)^2}{\pi}\int_{4\mpi^2}^\infty ds'\frac{\Im \big[h_{0,++}(s')\pm \sqrt{q_1^2q_2^2}\, h_{0,00}(s')\big]}{(s'-s)\big(s'-(\sqrt{q_1^2}\mp\sqrt{q_2^2})^2\big)}.
	\end{split}
\end{align}
The new combinations still fulfill Watson's theorem~\cite{Watson:1954uc}
\beq
\Im \Big[h_{0,++}(s)\pm \sqrt{q_1^2q_2^2}\, h_{0,00}(s)\Big]=\sin\delta_0(s)e^{-i\delta_0(s)}\Big[h_{0,++}(s)\pm \sqrt{q_1^2q_2^2}\, h_{0,00}(s)\Big]\theta\big(s-4\mpi^2\big),
\eeq
so that the dispersion relation reduces to a standard Muskhelishvili--Omn\`es (MO) problem~\cite{Muskhelishvili:1953,Omnes:1958hv}, whose solution reads
\begin{align}
h_{0,++}(s)\pm \sqrt{q_1^2q_2^2}\, h_{0,00}(s)&=\Delta_{0,++}(s)\pm \sqrt{q_1^2q_2^2}\,\Delta_{0,00}(s)\\
&\hspace{-1cm}+\frac{\Omega_0(s)\Big(s-\big(\sqrt{q_1^2}\mp\sqrt{q_2^2}\big)^2\Big)}{\pi}\int_{4\mpi^2}^\infty ds'\frac{\big[\Delta_{0,++}(s')\pm \sqrt{q_1^2q_2^2}\, \Delta_{0,00}(s')\big]\sin\delta_0(s')}{(s'-s)\big(s'-(\sqrt{q_1^2}\mp\sqrt{q_2^2})^2\big)|\Omega_0(s')|},\notag
\end{align}
with the Omn\`es function
\beq
\Omega_0(s)=\exp\Bigg\{\frac{s}{\pi}\int_{4\mpi^2}^\infty ds'\frac{\delta_0(s')}{s'(s'-s)}\Bigg\}.
\eeq
For convenience, we finally rewrite the result in terms of the original helicity amplitudes according to
\begin{align}
\label{MO_solution}
 h_{0,++}(s)&=\Delta_{0,++}(s)\notag\\
 &+\frac{\Omega_0(s)}{\pi}\int_{4\mpi^2}^\infty ds'\frac{\sin\delta_0(s')}{|\Omega_0(s')|}\bigg[\bigg(\frac{1}{s'-s}-\frac{s'-q_1^2-q_2^2}{\lambda_{12}(s')}\bigg)\Delta_{0,++}(s')+\frac{2q_1^2q_2^2}{\lambda_{12}(s')}\Delta_{0,00}(s')\bigg],\notag\\
 h_{0,00}(s)&=\Delta_{0,00}(s)\notag\\
 &+\frac{\Omega_0(s)}{\pi}\int_{4\mpi^2}^\infty ds'\frac{\sin\delta_0(s')}{|\Omega_0(s')|}\bigg[\bigg(\frac{1}{s'-s}-\frac{s'-q_1^2-q_2^2}{\lambda_{12}(s')}\bigg)\Delta_{0,00}(s')+\frac{2}{\lambda_{12}(s')}\Delta_{0,++}(s')\bigg].
\end{align}
For a pion-pole LHC $\Delta_{0,++}(s)$ and $\Delta_{0,00}(s)$ simply correspond to the partial-wave projection of the Born terms, given in App.~\ref{sec:PionPolePartialwaves}, which shows that the dependence on the virtualities, apart from the modified kernel functions in the MO solution, is still governed by the pion vector form factor. In particular, the corresponding factor $F_\pi^V(q_1^2)F_\pi^V(q_2^2)$ can be moved out of the integrals in~\eqref{MO_solution}, so that one can simply calculate a reduced amplitude, with the dependence on the pion form factors fully factorized. 
Further, in the solution of Roy--Steiner equations, a MO representation similar to~\eqref{MO_solution} is often required for the low-energy region only, in order to match to some known high-energy input, and to this end a finite matching point is introduced~\cite{Hoferichter:2011wk,Buettiker:2003pp,Ditsche:2012fv,Hoferichter:2012wf,Hoferichter:2015hva}. In case the amplitudes are assumed to vanish above the matching point, it effectively acts as a cutoff both in~\eqref{MO_solution} and in the Omn\`es function. We will use this variant of the MO solution to estimate the sensitivity to the high-energy extrapolation of the phase shifts, referring for more details of its implementation to~\cite{Hoferichter:2011wk,Buettiker:2003pp}. 

\begin{figure}[t]
\centering
\includegraphics[width=8cm]{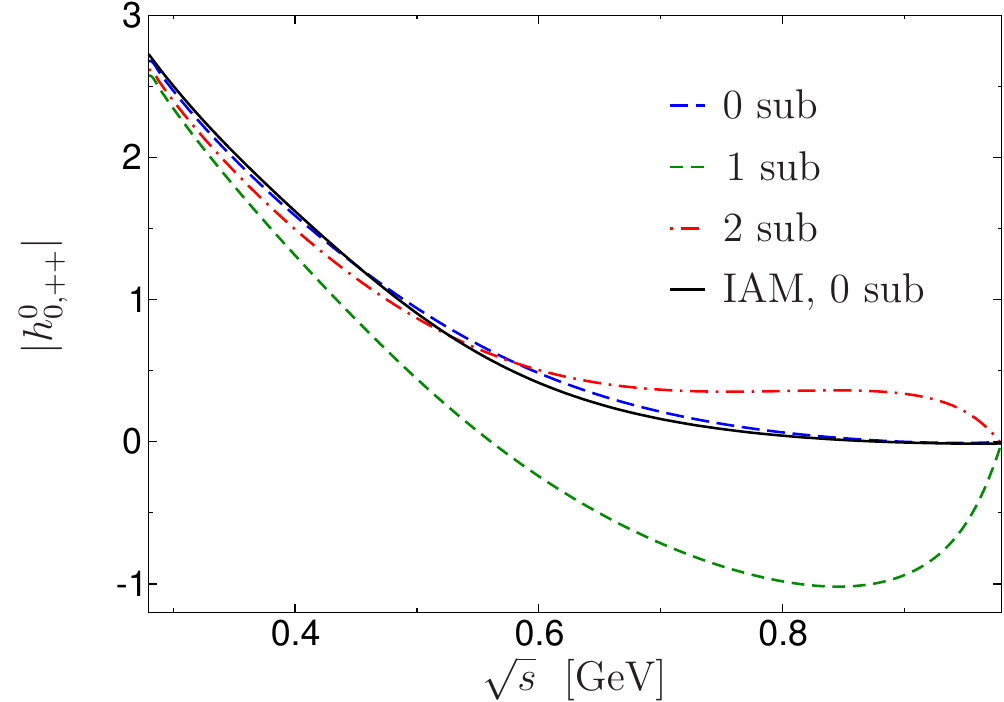}\qquad
\includegraphics[width=8cm]{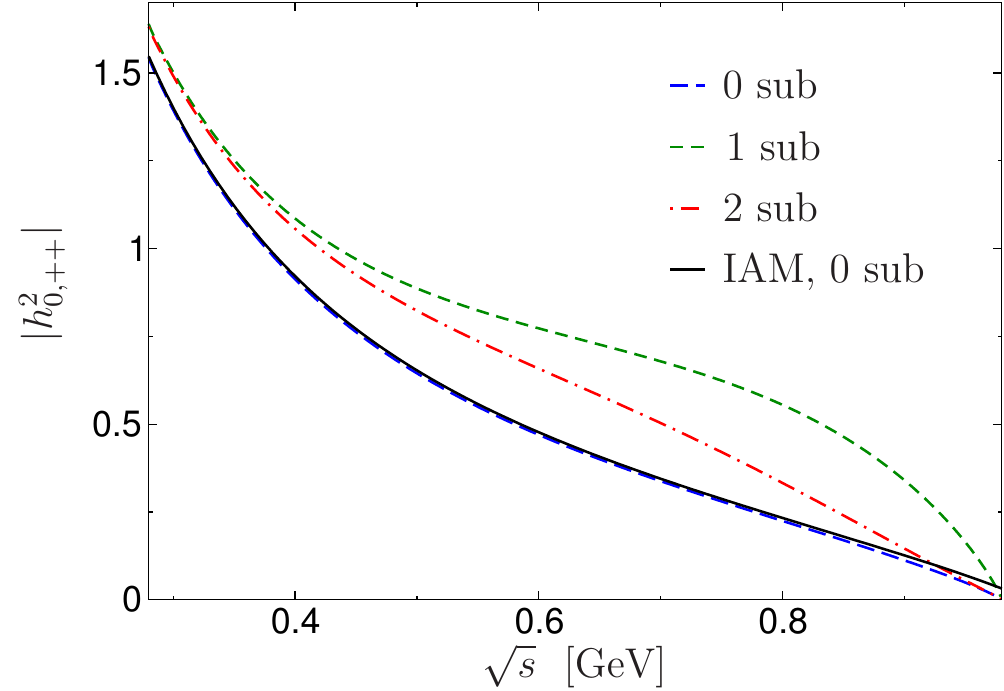}\\[2mm]	
\includegraphics[width=8cm]{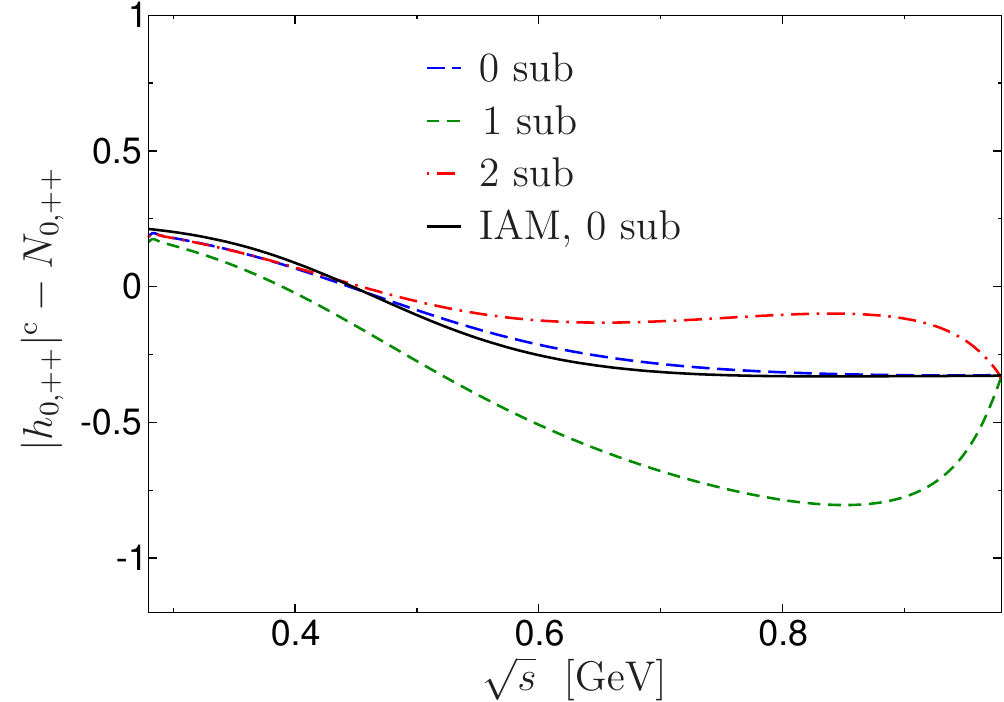}\qquad
\includegraphics[width=8cm]{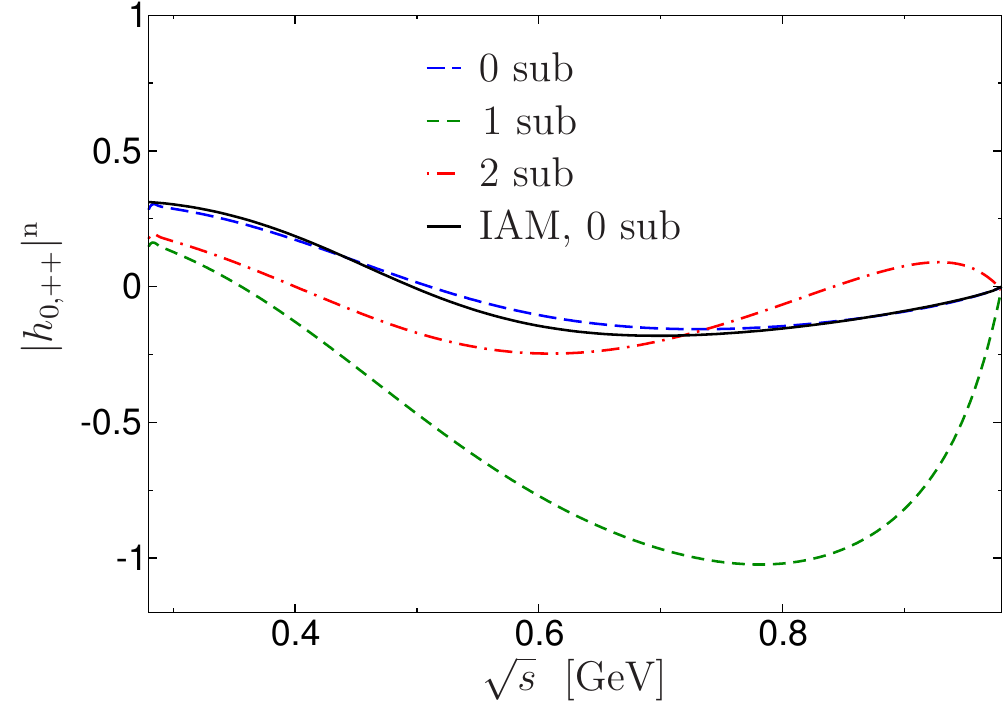}
\caption{Comparison of the $\gamma\gamma\to\pi\pi$ $S$-waves from this work (black solid line) to the different subtraction schemes from~\cite{Hoferichter:2011wk} as indicated. Upper/lower panel: left/right corresponds to $I=0/2$ and charged/neutral channel, respectively, as explained in the main text.}
\label{fig:HPS}
\end{figure}

Finally, the justification why an unsubtracted representation such as~\eqref{MO_solution} is still expected to provide a decent description is two-fold: first, by removing the $\bar K K$ intermediate states the Omn\`es functions are smoothened considerably around the nominal $f_0(980)$ position, which eliminates most of the need for subtractions necessary otherwise in a single-channel description to suppress the corresponding peak in the Omn\`es function. Second, while in general a precision description does require subtractions~\cite{GarciaMartin:2010cw,Hoferichter:2011wk}, we observe in the on-shell case that the
results particularly for the charged channel are reasonably close to the twice-subtracted variants studied in~\cite{Hoferichter:2011wk}, see Fig.~\ref{fig:HPS} for a cutoff $\Lambda=1\GeV$.
The upper panel shows the modulus $|h_{0,++}^I|$ for isospin $I=0$ and $I=2$, which for the unsubtracted IAM emerges remarkably close to the twice-subtracted variant in both cases. However, this agreement is largely driven by the projection of the Born term, while a more realistic picture can be obtained by considering the rotated amplitudes
\beq
|h_{0,++}|^\text{c}=\frac{1}{\sqrt{3}}|h_{0,++}^0|+\frac{1}{\sqrt{6}}|h_{0,++}^2|,\qquad
|h_{0,++}|^\text{n}=\frac{1}{\sqrt{3}}|h_{0,++}^0|-\sqrt{\frac{2}{3}}|h_{0,++}^2|,
\eeq
and subtracting the Born term in the charged channel. In this way, we find that the agreement is still very good for the charged combination, while the neutral channel is less well reproduced based on the pion-pole LHC alone, see lower panel in Fig.~\ref{fig:HPS}. To improve the quantitative agreement, the introduction of subtraction constants becomes unavoidable. These subtraction constants can be identified with pion polarizabilities and were taken from $2$-loop \ChPT{}~\cite{Gasser:2005ud,Gasser:2006qa} in~\cite{Hoferichter:2011wk}. The agreement in the charged channel implies that the corresponding sum rules for the subtraction constants, just based on the pion-pole LHC, are reasonably well fulfilled, while significant corrections are expected in the neutral channel. This interplay with the pion polarizabilities will be discussed in more detail in Sect.~\ref{sec:pion_pol}. For the moment, the fact that the
dominant rescattering correction is generated by the charged-pion intermediate states, with neutral pions first entering at three-loop order in the chiral expansion, ensures that the Roy--Steiner solution~\eqref{MO_solution} captures the phenomenology of unitarity corrections to the pion-pole LHC, i.e.\ the rescattering effects required to unitarize the pion-box contribution.

\subsection[A first numerical estimate of the $\pi\pi$-rescattering contribution to $(g-2)_\mu$]{\boldmath A first numerical estimate of the $\pi\pi$-rescattering contribution to $(g-2)_\mu$}

Based on the amplitudes calculated from~\eqref{MO_solution} we are now in the position to present a first numerical evaluation for the $S$-wave $\pi\pi$ rescattering effects. For simplicity, we use a VMD pion form factor, which proves to be very close to a full phenomenological determination extrapolated from the time-like region~\cite{prep}, see Sect.~\ref{sec:sQEDevaluation}. Restoring isospin indices, symmetry factors, virtualities, and subtracting the corresponding isospin projection of the pion-pole terms $N_{J,\lambda_1\lambda_2}$, the relevant imaginary parts in the HLbL integral become
\begin{align}
\Im h_{++,++}^{0,I}\big(s;q_1^2,q_2^2,q_3^2,0\big)&=\frac{\sigma_\pi(s)}{32\pi}\Big(h^I_{0,++}(s;q_1^2,q_2^2)h^I_{0,++}(s;q_3^2,0)-c_I N_{0,++}(s;q_1^2,q_2^2)N_{0,++}(s;q_3^2,0)\Big),\notag\\
\Im h_{00,++}^{0,I}\big(s;q_1^2,q_2^2,q_3^2,0\big)&=\frac{\sigma_\pi(s)}{32\pi}\Big(h^I_{0,00}(s;q_1^2,q_2^2)h^I_{0,++}(s;q_3^2,0)-c_I N_{0,00}(s;q_1^2,q_2^2)N_{0,++}(s;q_3^2,0)\Big),
\end{align}
with isospin factors $c_0=4/3$, $c_2=2/3$. 

\begin{table}[t]
\centering
\renewcommand{\arraystretch}{1.3}
\begin{tabular}{crrrr}
\toprule
cutoff & $1\GeV$ & $1.5\GeV$ & $2\GeV$ & $\infty$\\
\midrule
$I=0$ & $-9.2$ & $-9.5$ & $-9.3$ & $-8.8$\\
$I=2$ & $2.0$ & $1.3$ & $1.1$ & $0.9$\\
sum & $-7.3$ & $-8.3$ & $-8.3$ & $-7.9$\\
\bottomrule
\end{tabular}
\caption{Results for the $S$-wave rescattering contribution to $(g-2)_\mu$ in units of $10^{-11}$. The cutoff refers to the finite-matching-point analog of~\eqref{MO_solution}.}
\label{tab:rescatt}
\renewcommand{\arraystretch}{1.0}
\end{table}

The numerical results for the $S$-wave contribution then follow from~\eqref{eq:DRrescattering} together with the dispersive representation for the scalar functions derived in Sect.~\ref{sec:HelicityFormalism}.
Since the full integration becomes numerically costly---with the dispersion integral in~\eqref{MO_solution}, the $(g-2)_\mu$ dispersion integral, and three integrals in the master formula~\eqref{eq:MasterFormulaPolarCoord} this would amount to a delicate $5$-dimensional integral, wherein in addition the Omn\`es factor requires the numerical evaluation of yet another integral---we calculate the $\gamma^*\gamma^*\to\pi\pi$ amplitudes on a three-dimensional grid in $(s,q_1^2,q_2^2)$ and then interpolate in the remaining $4$-dimensional $(g-2)_\mu$ integration. Using up to $50$ grid points in each variable the results become insensitive to the interpolation uncertainty, and we obtain the values listed in Table~\ref{tab:rescatt}. As expected based on the size of the phase shifts, the $I=2$ contribution is much smaller than its $I=0$ counterpart, while in both cases the variation with respect to the cutoff amounts to about one unit. Accordingly, this estimate can be interpreted as evidence for a rescattering contribution corresponding to $f_0(500)$ degrees of freedom of about $-9\times 10^{-11}$ in the HLbL contribution to $(g-2)_\mu$. 

\begin{table}[t]
\centering
\renewcommand{\arraystretch}{1.3}
\begin{tabular}{ccrrrr}
\toprule
&cutoff & $1\GeV$ & $1.5\GeV$ & $2\GeV$ & $\infty$\\
\midrule
&$++,++$ & $6.3$ & $6.5$ & $6.4$ & $6.1$\\
$I=0$ &$00,++$ & $-6.8$ & $-7.0$ & $-6.8$ & $-6.4$\\
&sum & $-0.6$ & $-0.4$ & $-0.4$ & $-0.3$\\\midrule
& $++,++$ & $-1.3$ & $-0.9$ & $-0.7$ & $-0.7$\\
$I=2$ & $00,++$ & $1.5$ & $1.0$ & $0.8$ & $0.7$\\
& sum & $0.2$ & $0.1$ & $0.1$ & $0.0$\\
\bottomrule
\end{tabular}
\caption{Contribution to the sum rule~\eqref{eq:SwaveSumRule} from $h^0_{++,++}$ and $h^0_{00,++}$ as well as their sum once integrated over momenta and virtualities in the $(g-2)_\mu$ master formula as explained in the main text, in units of $10^{-11}$.}
\label{tab:sum_rule}
\renewcommand{\arraystretch}{1.0}
\end{table}

Another check on our input for $\gamma^*\gamma^*\to\pi\pi$ follows from the sum rule~\eqref{eq:SwaveSumRule}. In fact, it is precisely this sum rule that ensures that the $S$-wave rescattering contribution as formulated in~\cite{Colangelo:2014dfa} and the one from Sect.~\ref{sec:HelAmpPWE} are strictly equivalent. Furthermore, this observation immediately suggests a way how to condense the full sum rule into a single number: the difference between the two representations amounts to a shift in $\hat\Pi_4$ of the size
\beq
\Delta\hat\Pi_4=\frac{2}{\pi}\int_{4\mpi^2}^\infty ds' \frac{1}{(s'-q_3^2)\lambda_{12}(s')} \Big( 2 \Im h^0_{++,++}(s') - \big(s'-q_1^2-q_2^2\big) \Im h^0_{00,++}(s') \Big),
\eeq
and accordingly in $\hat \Pi_5$ and $\hat\Pi_6$ from crossing, so that the convolution in the $(g-2)_\mu$ integral should be done with the corresponding kernel function. 
Still subtracting the pion-pole terms since the validity of the sum rule in sQED is already known, we find the results for the separate contribution from $h^0_{++,++}$ and $h^0_{00,++}$ as listed in Table~\ref{tab:sum_rule}. The expected cancellation already works at the level of $10\%$ with $S$-waves only, and even better for the larger values of the cutoff. Such a $10\%$ error on the actual rescattering contributions from Table~\ref{tab:rescatt} would yield a very similar uncertainty estimate as the variation observed from the cutoff dependence before.
In total, these results lead us to quote
\beq
\label{amupipi}
a_{\mu,J=0}^{\pi\pi,\pi\text{-pole LHC}}=-8(1)\times 10^{-11}
\eeq
for the $S$-wave rescattering corrections to the pion-pole LHC.

\subsection{Role of the pion polarizabilities}
\label{sec:pion_pol}

The low-energy behavior of the on-shell $\gamma\gamma\to\pi\pi$ amplitudes is strongly constrained by the pion polarizabilities, which therefore encode valuable information on the two-pion rescattering contributions to HLbL. The precise relation can be expressed in terms of the expansion
\beq
\frac{2\alpha}{\mpi s}\hat h_{0,++}(s)=\alpha_1-\beta_1 +\frac{s}{12}(\alpha_2-\beta_2)+\O(s^2)
\eeq
for the Born-term-subtracted on-shell amplitudes $\hat h_{0,++}=h_{0,++}-N_{0,++}$. Here, $\alpha_1-\beta_1$ and $\alpha_2-\beta_2$ refer to dipole and quadrupole polarizabilities, respectively.
The soft-photon zero required as a consequence of Low's theorem~\cite{Low:1958sn} ensures that $\hat h_{0,++}$ indeed vanishes for $s\to 0$.

Accordingly, the representation~\eqref{MO_solution} implies the following sum rules for the pion polarizabilities
\begin{align}
\label{sum_rule_pol}
 \frac{\mpi}{2\alpha}(\alpha_1-\beta_1)&=\bigg[\frac{\Delta_{0,++}(s)-N_{0,++}(s)}{s}\bigg]_{s=0}+\frac{1}{\pi}\int_{4\mpi^2}^\infty ds'\frac{\sin\delta_0(s')\Delta_{0,++}(s')}{|\Omega_0(s')|s'^2},\\
 \frac{\mpi}{24\alpha}(\alpha_2-\beta_2)&=\bigg[\frac{\partial}{\partial s}\frac{\Delta_{0,++}(s)-N_{0,++}(s)}{s}\bigg]_{s=0}+\frac{1}{\pi}\int_{4\mpi^2}^\infty ds'\frac{\sin\delta_0(s')\Delta_{0,++}(s')}{|\Omega_0(s')|s'^2}\bigg(\dot \Omega_0(0)+\frac{1}{s'}\bigg),\notag
\end{align}
where $\dot \Omega_0(0)$ denotes the derivative of the Omn\`es factor at $s=0$ and the first term in each line disappears for a pion-pole LHC.

\begin{table}[t]
\centering
\renewcommand{\arraystretch}{1.3}
\begin{tabular}{crrrrr}
\toprule
 & $1\GeV$ & $1.5\GeV$ & $2\GeV$ & $\infty$ & \ChPT\\
\midrule
$(\alpha_1-\beta_1)^{\pi^\pm} \ \big[10^{-4}\,\text{fm}^3\big]$ & $5.4$ & $5.8$ & $5.8$ & $5.7$ & $5.7(1.0)$\\
$(\alpha_1-\beta_1)^{\pi^0} \ \big[10^{-4}\,\text{fm}^3\big]$ & $11.2$ & $9.7$ & $9.3$ & $8.9$ & $-1.9(2)$\\
$(\alpha_2-\beta_2)^{\pi^\pm} \ \big[10^{-4}\,\text{fm}^5\big]$ & $19.9$ & $20.1$ & $20.0$ & $19.9$ & $16.2\ [21.6]$\\
$(\alpha_2-\beta_2)^{\pi^0} \ \big[10^{-4}\,\text{fm}^5\big]$ & $28.4$ & $27.1$ & $26.7$ & $26.3$ & $37.6(3.3)$\\
\bottomrule
\end{tabular}
\caption{Pion polarizabilities from the sum rules~\eqref{sum_rule_pol} for a pion-pole LHC and different values of the cutoff $\Lambda$, in comparison to the chiral two-loop prediction from~\cite{Gasser:2005ud,Gasser:2006qa}. The two numbers in the case of the charged-pion quadrupole polarizability refer to two different sets of low-energy constants.}
\label{tab:pol}
\renewcommand{\arraystretch}{1.0}
\end{table}

The numerical evaluation for $\Delta_{0,++}=N_{0,++}$, see Table~\ref{tab:pol}, confirms the observation from Sect.~\ref{sec:IAM} that the charged-pion amplitude is better reproduced than its neutral-pion analog. In fact, the charged-pion dipole polarizability comes out in perfect agreement with  \ChPT{}~\cite{Gasser:2006qa}, as well as with the recent measurement by COMPASS $(\alpha_1-\beta_1)^{\pi^\pm}=4.0(1.2)_\text{stat}(1.4)_\text{syst}\times 10^{-4}\,\text{fm}^3$~\cite{Adolph:2014kgj}. The quadrupole polarizability is more sensitive to poorly-determined low-energy constants, but the sum-rule value lies within the range quoted in~\cite{Gasser:2006qa} and is also close to $(\alpha_2-\beta_2)^{\pi^\pm}=15.3(3.7)\times 10^{-4}\,\text{fm}^5$ obtained in~\cite{Hoferichter:2011wk} by combining the more stable chiral prediction for the neutral-pion quadrupole polarizability with a finite-matching-point sum rule for $I=2$.  

In contrast, both neutral-pion polarizabilities differ by about $10$ units each from the full result, a deficiency that signals the impact of higher contributions to the LHC, as we will demonstrate in the following. The next such contribution is generated by the exchange of vector-meson resonances $V=\rho,\omega$, whose impact can be roughly estimated within a narrow-width approximation. Starting from a vector--pion--photon coupling of the form
\beq
\mathcal{L}_{V\pi\gamma}=e C_V\epsilon^{\mu\nu\lambda\sigma}F_{\mu\nu}\partial_\lambda \pi V_\sigma,
\eeq
with coupling constant related to the partial width according to
\beq
\Gamma_{V\to\pi\gamma}=\alpha C_V^2\frac{(\mv^2-\mpi^2)^3}{6\mv^3},
\eeq
we obtain~\cite{Ko:1989yd,GarciaMartin:2010cw}
\beq
\label{LHC_V}
\Delta_{0,++}^V(s)=2C_V^2\bigg[-\frac{\mv^2}{\sigma_\pi(s)}\log\frac{x_V(s)+1}{x_V(s)-1}+s\bigg],\qquad x_V(s)=\frac{s+2(\mv^2-\mpi^2)}{s \sigma_\pi(s)}.
\eeq
Unfortunately, the polynomial piece $\propto s$ is ambiguous and would even appear with a different sign in an antisymmetric-tensor description of the vector-meson fields~\cite{Ecker:1988te,GarciaMartin:2010cw}. It is for this reason that in a full Roy--Steiner approach only the imaginary parts are employed, while the low-energy parameters enter via subtraction constants. However, in order to predict the numerical values of the polarizabilities in terms of the lowest contributions to the LHC in $\gamma\gamma\to\pi\pi$ we do need the full amplitude in~\eqref{LHC_V}. Parameterizing the ambiguity according to $s\to \xi_V s$, we find
\beq
\frac{\mpi}{2\alpha}(\alpha_1-\beta_1)_V=2C_V^2\bigg[\xi_V-\frac{\mv^2}{\mv^2-\mpi^2}\bigg],\qquad \frac{\mpi}{24\alpha}(\alpha_2-\beta_2)_V=C_V^2\frac{\mv^2(3\mv^2-\mpi^2)}{3(\mv^2-\mpi^2)^3}.
\eeq
Adding $\rho,\omega$ contributions using masses and partial widths from~\cite{Olive:2016xmw}, the quadrupole polarizabilities are shifted by $(\alpha_2-\beta_2)_V^{\pi^{\pm}}=0.9\times 10^{-4}\,\text{fm}^5$ and $(\alpha_2-\beta_2)_V^{\pi^{0}}=10.3\times 10^{-4}\,\text{fm}^5$, which explains how vector-meson contributions can restore agreement with \ChPT{} for the neutral pion without spoiling the charged channel. In fact, the hierarchy can be attributed almost exclusively to the large $\omega\to\pi^0\gamma$ branching fraction
\beq
\frac{\Gamma_\omega\BR[\omega\to\pi^0\gamma]+\Gamma_\rho\BR[\rho^0\to\pi^0\gamma]}{\Gamma_\rho\BR[\rho^\pm\to\pi^\pm\gamma]}\sim 12,
\eeq
which ensures that the same mechanism applies for the dipole polarizability as well.

In any case, such corrections are not contained in our estimate~\eqref{amupipi}, but at least at the on-shell point the impact is expected to be moderate due to the fact that the charged-pion intermediate states are most important. In particular, the physics related to the low-energy constants $\bar l_6-\bar l_5$, which appear at two-loop level in the chiral expansion for the HLbL tensor~\cite{Engel:2012xb}, only contribute to the charged-pion polarizability (a more detailed comparison to ChPT is provided in App.~\ref{app:ChPT}). Our calculation therefore demonstrates in a model-independent way that such next-to-leading-order corrections are moderate in size, in agreement with~\cite{Bijnens:2016hgx}, but in contradiction to the large corrections suggested in~\cite{Engel:2013kda}. This conclusively settles the role of the charged-pion dipole polarizability in the HLbL contribution to $(g-2)_\mu$.


\newpage
\section{Conclusions}
\label{sec:Conclusion}

In this paper we presented an in-depth derivation of the general formalism
required for the analysis of two-pion-intermediate-state contributions to
HLbL scattering in $(g-2)_\mu$. As a first step we gained a detailed
understanding of the properties of the HLbL tensor, including its
decomposition into scalar functions, projection onto helicity amplitudes,
and the relation between the different sets we needed to introduce in the
course of our derivation, see Table~\ref{tab:FormalismOverview}.  Some of
the more subtle issues that arose in this derivation are related to the
fact that, in order to write down dispersion relations for the HLbL tensor,
we had to start with a redundant set of functions. At first sight, the
relation between the latter and the physically observable helicity
amplitudes seems to suffer from ambiguities. To show that this
arbitrariness is only apparent we invoked a set of sum rules, which follow
from a simple assumption on the asymptotic behavior of the HLbL
tensor. These sum rules allowed us to construct a basis for kinematics with
one single on-shell photon (singly-on-shell) that satisfies unsubtracted
dispersion relations. In addition they lead to physically relevant sum
rules that constrain the helicity amplitudes for
$\gamma^*\gamma^*\to\pi\pi$.  After working out the basis change from the
singly-on-shell basis to helicity amplitudes, we combined this general
formalism with a partial-wave expansion to address two-pion-rescattering
contributions.

In a second step we thoroughly tested our formalism using the example of
the pion box, whose full result is known thanks to an exact relation to the
scalar QED pion loop we established earlier. In particular, we demonstrated
that the sum rules that follow from our assumptions on
the asymptotic behavior of the HLbL tensor are fulfilled. Moreover we
studied whether the partial-wave expansion of the pion box converges to the
full answer after resummation, and demonstrated that it does so
sufficiently quickly. Given that the pion-box contribution can be expressed
exactly in terms of the pion vector form factor---much as the HVP
contribution of two pion intermediate states is completely determined by
this form factor---we showed that by fitting a dispersive representation of
the pion vector form factor to a combination of space- and time-like data,
the space-like form factor required for the HLbL application can be
constrained to a very high precision, leading to
$a_\mu^{\pi\text{-box}}=-15.9(2)\times 10^{-11}$ for the pion-box
contribution.

The main motivation for developing a partial-wave framework is to be able
to calculate rescattering corrections, since only in a partial-wave basis
for helicity amplitudes do unitarity relations become diagonal.
Accordingly, as a first application of the formalism developed
here we studied the unitarization of the pion box, a correction whose
evaluation requires the use of partial-wave amplitudes. Concentrating on
$S$-wave $\pi\pi$-rescattering effects, we presented a first numerical
estimate, which, together with the pion-box evaluation, combines to
\beq
\label{final}
a_\mu^{\pi\text{-box}} + a_{\mu,J=0}^{\pi\pi,\pi\text{-pole
    LHC}}=-24(1)\times 10^{-11}
\eeq
for the leading two-pion contributions to $(g-2)_\mu$. The improvement in
accuracy with respect to previous model-dependent analyses is striking.
It derives: (i) from our model-independent approach based on dispersion
relations that allows us to express this contribution, in a rigorous way, 
in terms of hadronic observables, and (ii) from the fact that all
quantities needed in this calculation (the pion vector form factor and the
$\pi\pi$ $S$-wave phase shifts) are very well known. Remaining two-pion
contributions that have not been addressed yet are likely to lead to larger
uncertainties, but given that the error quoted in~\eqref{final} lies an
order of magnitude below the experimental accuracy goal, we are confident
that the final estimate for the total HLbL contribution should be
sufficiently accurate to make these measurements of $(g-2)_\mu$ a sensitive
test of the Standard Model.

Many of the technical advances described here are not specific to the
two-pion intermediate state but completely general and thus lay the
groundwork for a full phenomenological analysis of HLbL scattering. Armed
with these, we are now poised to study other contributions and apply
further refinements to the numerical analysis of the two-pion channel and
beyond.

\section*{Acknowledgements}
\addcontentsline{toc}{section}{Acknowledgements}

We thank B.~Kubis, A.~Manohar, M.~J.~Ramsey-Musolf, and M.~J.~Savage for useful discussions.
Financial support by
the DFG (SFB/TR 16, ``Subnuclear Structure of Matter,'' SFB/TR 110, ``Symmetries and the Emergence of Structure in QCD''),   
the DOE (Grant No.\ DE-FG02-00ER41132 and DE-SC0009919), 
the National Science Foundation (Grant No.\ NSF PHY-1125915),
and the Swiss National Science Foundation
is gratefully acknowledged.
M.P.\ is supported by a Marie Curie Intra-European Fellowship of the
European Community's 7th Framework Programme under contract number
PIEF-GA-2013-622527
and P.S.\ by a grant of the Swiss National Science Foundation (Project No.\
P300P2\_167751).

\begin{appendices}


\section{\boldmath Transformed tensor decomposition for the contribution to $(g-2)_\mu$}

\label{sec:BasisChangeGm2}

For the calculation of $(g-2)_\mu$, we make a linear transformation of the BTT tensor decomposition:
\begin{align}
	\Pi^{\mu\nu\lambda\sigma} &= \sum_{i=1}^{54} T_i^{\mu\nu\lambda\sigma} \Pi_i = \sum_{i=1}^{54} \hat T_i^{\mu\nu\lambda\sigma} \hat \Pi_i .
\end{align}
Only 19 of the new structures $\hat T_i^{\mu\nu\lambda\sigma}$ contribute to $(g-2)_\mu$, which is the minimal number of independent contributions in the $(g-2)_\mu$ kinematic limit. The symmetry under $q_1 \leftrightarrow -q_2$ reduces this to 12 terms in the master formula.

\subsection{Tensor structures}

Here, we give the tensor structures $\hat T_i^{\mu\nu\lambda\sigma}$ explicitly in terms of the BTT structures~\cite{Colangelo:2015ama}. The 19 structures contributing to $(g-2)_\mu$ are defined in~\eqref{eq:ThatStructures}. The remaining 35 structures, which do not contribute to $(g-2)_\mu$, are defined by
\begin{align}
	\begin{split}
		\hat T_i^{\mu\nu\lambda\sigma} &= T_i^{\mu\nu\lambda\sigma}, \quad i=12, 15, 18, 23, 24, 27, 28, 29, 30, 32, 35, 36, 37, 38, 41, 44, 45, 48, 49, 52, 53 , \\
		\hat T_{19}^{\mu\nu\lambda\sigma} &= q_1 \cdot q_3 T_{4}^{\mu\nu\lambda\sigma} + T_{7}^{\mu\nu\lambda\sigma} + T_{19}^{\mu\nu\lambda\sigma} , \\
		\hat T_{31}^{\mu\nu\lambda\sigma} &= - q_1 \cdot q_3 q_2 \cdot q_3 T_{4}^{\mu\nu\lambda\sigma} - q_2 \cdot q_3 T_{7}^{\mu\nu\lambda\sigma} - q_1 \cdot q_3 T_{8}^{\mu\nu\lambda\sigma} + T_{31}^{\mu\nu\lambda\sigma} , \\
		\hat T_{40}^{\mu\nu\lambda\sigma} &= T_{40}^{\mu\nu\lambda\sigma} -  T_{39}^{\mu\nu\lambda\sigma} , \\
		\hat T_{42}^{\mu\nu\lambda\sigma} &= -q_1 \cdot q_3 ( T_{2}^{\mu\nu\lambda\sigma} + T_{4}^{\mu\nu\lambda\sigma} + T_{6}^{\mu\nu\lambda\sigma} ) - T_{11}^{\mu\nu\lambda\sigma} + T_{16}^{\mu\nu\lambda\sigma} - T_{17}^{\mu\nu\lambda\sigma} \\
			&\quad + T_{42}^{\mu\nu\lambda\sigma} - T_{51}^{\mu\nu\lambda\sigma} - T_{54}^{\mu\nu\lambda\sigma} ,
	\end{split}
\end{align}
together with the crossed structures
\begin{align}
	\begin{split}
		\hat T_{20}^{\mu\nu\lambda\sigma} &= \mathcal{C}_{12}[ \hat T_{19}^{\mu\nu\lambda\sigma} ] , \quad \hat T_{21}^{\mu\nu\lambda\sigma} = \mathcal{C}_{23}[ \hat T_{19}^{\mu\nu\lambda\sigma} ] , \quad \hat T_{22}^{\mu\nu\lambda\sigma} = \mathcal{C}_{23}[ \mathcal{C}_{12}[ \hat T_{19}^{\mu\nu\lambda\sigma} ] ] , \\
		\hat T_{25}^{\mu\nu\lambda\sigma} &= \mathcal{C}_{12}[ \mathcal{C}_{23}[ \hat T_{19}^{\mu\nu\lambda\sigma} ] ] , \quad \hat T_{26}^{\mu\nu\lambda\sigma} = \mathcal{C}_{13}[ \hat T_{19}^{\mu\nu\lambda\sigma} ] , \\
		\hat T_{33}^{\mu\nu\lambda\sigma} &= \mathcal{C}_{23}[ \hat T_{31}^{\mu\nu\lambda\sigma} ] , \quad \hat T_{34}^{\mu\nu\lambda\sigma} = \mathcal{C}_{13}[ \hat T_{31}^{\mu\nu\lambda\sigma} ] , \\
		\hat T_{46}^{\mu\nu\lambda\sigma} &= \mathcal{C}_{13}[ \hat T_{40}^{\mu\nu\lambda\sigma} ] , \\
		\hat T_{43}^{\mu\nu\lambda\sigma} &= \mathcal{C}_{12}[ \hat T_{42}^{\mu\nu\lambda\sigma} ] , \quad \hat T_{47}^{\mu\nu\lambda\sigma} = \mathcal{C}_{23}[ \hat T_{42}^{\mu\nu\lambda\sigma} ] .
	\end{split}
\end{align}

\subsection{Scalar functions}

In terms of the BTT functions $\Pi_i$, the transformed scalar functions $\hat \Pi_i$ that contribute to $(g-2)_\mu$ are defined in~\eqref{eq:PiHatFunctions} and~\eqref{eq:CrossingRelationsPiHat}. The ones that do not contribute to $(g-2)_\mu$ are given by:
\begin{align}
	\begin{split}
		\hat\Pi_i &= \Pi_i , \quad i = 12, 15, 18, \ldots, 38, 41, \ldots, 45, 47, 48, 49, 52, 53 , \\
		\hat\Pi_{40} &= \frac{1}{3} \left( -\Pi_{39} + 2 \Pi_{40} - \Pi_{46} \right) , \quad \hat\Pi_{46} = \mathcal{C}_{13}[ \hat\Pi_{40} ] .
	\end{split}
\end{align}


\section{\boldmath New kernel functions for the master formula}

\label{sec:MaFoKernels}

Compared to~\cite{Colangelo:2015ama}, we choose a different basis for the Lorentz structures contributing to $(g-2)_\mu$ in order to preserve crossing symmetry between all three off-shell photons. This modifies slightly the kernel functions in the master formula~\eqref{eq:MasterFormula3Dim}.

The kernel functions $T_1$, $\ldots$, $T_9$ are identical to the ones in \cite{Colangelo:2015ama}, while $T_{10} = \frac{1}{2} T_{10}^\text{\cite{Colangelo:2015ama}}$. For completeness, here we provide the full set of the new kernels, superseding Sect.~E.2 in~\cite{Colangelo:2015ama}:
\begin{align*}
		T_1 &= \frac{Q_1^2 \tau  \left(\sigma_1^E-1\right) \left(\sigma_1^E+5\right)+Q_2^2 \tau  \left(\sigma_2^E-1\right) \left(\sigma_2^E+5\right)+4 Q_1 Q_2 \left(\sigma_1^E+\sigma_2^E-2\right)-8 \tau  m_{\mu }^2}{2 Q_1 Q_2 Q_3^2 m_{\mu }^2} \\
			&\quad + X \left(\frac{8 \left(\tau ^2-1\right)}{Q_3^2}-\frac{4}{m_{\mu }^2}\right) , \\
		T_2 &= \frac{Q_1 \left(\sigma _1^E-1\right) \left(Q_1 \tau  \left(\sigma _1^E+1\right)+4 Q_2 \left(\tau ^2-1\right)\right)-4 \tau  m_{\mu}^2}{Q_1 Q_2 Q_3^2 m_{\mu }^2} +  X \frac{8 \left(\tau ^2-1\right) \left(2 m_{\mu }^2-Q_2^2\right)}{Q_3^2 m_{\mu }^2} , \\
		T_3 &= \frac{1}{Q_3^2} \begin{aligned}[t]
			& \bigg( -\frac{2 \left(\sigma _1^E+\sigma _2^E-2\right)}{m_{\mu }^2}-\frac{Q_1 \tau  \left(\sigma _1^E-1\right) \left(\sigma_1^E+7\right)}{2 Q_2 m_{\mu }^2} + \frac{8 \tau }{Q_1Q_2} \\
			& -\frac{Q_2 \tau  \left(\sigma _2^E-1\right) \left(\sigma _2^E+7\right)}{2 Q_1 m_{\mu }^2} + \frac{Q_1^2 \left(1-\sigma _1^E\right)}{Q_2^2 m_{\mu }^2} + \frac{Q_2^2 \left(1-\sigma _2^E\right)}{Q_1^2 m_{\mu}^2}+\frac{2}{Q_1^2}+\frac{2}{Q_2^2}\bigg) \end{aligned} \\
			&\quad + X \left(\frac{4}{m_{\mu }^2}-\frac{8 \tau }{Q_1 Q_2}\right) , \\
		T_4 &= \frac{1}{Q_3^2} \begin{aligned}[t]
			& \Bigg(\frac{4 \left(\tau ^2 \left(\sigma _1^E-1\right)+\sigma _2^E-1\right)}{m_{\mu }^2}-\frac{Q_1 \tau  \left(\sigma _1^E-5\right) \left(\sigma _1^E-1\right)}{Q_2 m_{\mu }^2} + \frac{4 \tau }{Q_1Q_2} \\
			& - \frac{Q_2 \tau  \left(\sigma _2^E-3\right) \left(\sigma_2^E-1\right)}{Q_1m_{\mu }^2} +\frac{2 Q_2^2 \left(\sigma _2^E-1\right)}{Q_1^2m_{\mu }^2}-\frac{4}{Q_1^2} \\
   			& +X \left(-\frac{8 Q_2^2 \tau^2}{m_{\mu }^2}-\frac{16 Q_2 Q_1 \tau }{m_{\mu }^2}-\frac{8 Q_1^2}{m_{\mu }^2}+\frac{16 Q_2 \tau }{Q_1}+16\right)\Bigg) , \end{aligned} \\
		T_5 &= \frac{1}{Q_3^2} \begin{aligned}[t]
			& \Bigg( Q_1^2 \left(\frac{\tau ^2 \left(\sigma _1^E-1\right) \left(\sigma _1^E+3\right)+4 \left(\sigma _1^E+\sigma _2^E-2\right)}{2 m_{\mu }^2} -\frac{4}{Q_2^2}\right) -\frac{Q_2^2 \tau ^2 \left(\sigma _2^E-5\right) \left(\sigma _2^E-1\right)}{2 m_{\mu}^2} \\
			& +\frac{Q_1^3 \tau  \left(\sigma _1^E-1\right) \left(\sigma _1^E+5\right)}{Q_2 m_{\mu }^2} + Q_1 \left(\frac{Q_2 \tau  \left(\sigma_1^E+5 \sigma _2^E-6\right)}{m_{\mu }^2}-\frac{12 \tau }{Q_2}\right)+\frac{2 Q_1^4 \left(\sigma _1^E-1\right)}{Q_2^2 m_{\mu }^2} \\
			&  - 4 \tau ^2 +X \begin{aligned}[t]
				& \Bigg(Q_1 \left(8 Q_2 \left(\tau ^3+\tau \right)-\frac{2 Q_2^3 \tau }{m_{\mu }^2}\right)+Q_1^2 \bigg(32 \tau ^2-\frac{4 Q_2^2 \left(\tau ^2+1\right)}{m_{\mu }^2}\bigg) \\
				& +Q_1^3 \left(\frac{16 \tau }{Q_2}-\frac{10 Q_2 \tau }{m_{\mu }^2}\right)-\frac{4 Q_1^4}{m_{\mu }^2}\Bigg) \Bigg) , \end{aligned} \end{aligned} \\
   		T_6 &= \frac{1}{Q_3^2} \begin{aligned}[t]
			& \Bigg(\frac{Q_1^2 \left(\tau ^2 \left(\left(\sigma _1^E-22\right) \sigma _1^E-8 \sigma _2^E+29\right)+2 \left(-5 \sigma _1^E+\sigma_2^E+4\right)\right)}{2 m_{\mu }^2} \\
			& +Q_1 \left(\frac{Q_2 \tau  \left(2 \tau ^2 \left(\left(\sigma _2^E-3\right)^2-4 \sigma_1^E\right)-26 \sigma _1^E+\sigma _2^E \left(\sigma _2^E-12\right)+37\right)}{2 m_{\mu }^2}-\frac{4 \tau }{Q_2}\right) \\
			& +\frac{Q_2^2 \left(\tau ^2 \left(-8 \sigma _1^E+\sigma _2^E \left(5 \sigma _2^E-26\right)+29\right)-4 \left(\sigma _1^E+2 \sigma_2^E-3\right)\right)}{2 m_{\mu }^2}+\frac{Q_1^3 \tau  \left(\sigma _1^E-9\right) \left(\sigma _1^E-1\right)}{2 Q_2 m_{\mu}^2} \\
			& +\frac{Q_2^3 \tau  \left(\sigma _2^E-9\right) \left(\sigma _2^E-1\right)}{Q_1m_{\mu }^2}+\frac{8 Q_2 \tau }{Q_1}+\frac{2 Q_2^4 \left(1- \sigma _2^E\right)}{Q_1^2 m_{\mu }^2}+\frac{4 Q_2^2}{Q_1^2} \\
			& +X \begin{aligned}[t]
				& \Bigg(\frac{Q_2 Q_1^3 \left(8 \tau ^3+22 \tau \right)}{m_{\mu}^2}+\frac{Q_1^4 \left(8 \tau ^2-2\right)}{m_{\mu }^2}+Q_1^2 \left(\frac{Q_2^2 \left(36 \tau ^2+18\right)}{m_{\mu }^2}-8 \left(\tau^2+1\right)\right) \\
				& +\frac{Q_2^4 \left(8 \tau ^2+4\right)}{m_{\mu }^2}+Q_1 \left(\frac{Q_2^3 \left(8 \tau ^3+34 \tau \right)}{m_{\mu}^2}-8 Q_2 \tau  \left(\tau ^2+5\right)\right) \\
				& -16 Q_2^2 \left(2 \tau ^2+1\right)-\frac{16 Q_2^3 \tau }{Q_1}\Bigg)\Bigg) , \end{aligned} \end{aligned} \\
   		T_7 &= \frac{1}{Q_3^2} \begin{aligned}[t]
			& \Bigg( \frac{Q_1^2 \left(2 \left(\sigma _1^E+\sigma _2^E-2\right)-\tau ^2 \left(\left(\sigma _1^E+10\right) \sigma _1^E+8 \sigma_2^E-19\right)\right)}{2 m_{\mu }^2} \\
			& +Q_1 \left(\frac{Q_2 \tau  \left(2 \tau ^2 \left(\sigma _2^E-5\right) \left(\sigma_2^E-1\right)-2 \sigma _1^E+\sigma _2^E \left(\sigma _2^E+4\right)-3\right)}{2 m_{\mu }^2}-\frac{4 \tau }{Q_2}\right) \\
			& +\frac{Q_2^2 \tau ^2 \left(\sigma _2^E-5\right) \left(\sigma _2^E-1\right)}{2 m_{\mu }^2}+\frac{Q_1^3 \tau  \left(\sigma _1^E-9\right) \left(\sigma _1^E-1\right)}{2 Q_2 m_{\mu }^2} + 4 \tau ^2 \\
			& +X \begin{aligned}[t]
				& \Bigg(\frac{Q_2 Q_1^3 \left(8 \tau ^3+6 \tau \right)}{m_{\mu }^2}+Q_1 \left(\frac{2 Q_2^3 \tau }{m_{\mu }^2}-8 Q_2 \left(\tau ^3+\tau \right)\right) \\
				& +\frac{Q_1^4 \left(8 \tau ^2-2\right)}{m_{\mu }^2}+Q_1^2 \left(\frac{2 Q_2^2 \left(6 \tau ^2-1\right)}{m_{\mu }^2}-8 \left(\tau ^2+1\right)\right)\Bigg) \Bigg) , \end{aligned} \end{aligned} \\
   		T_8 &= \frac{1}{Q_3^2} \begin{aligned}[t]
			& \Bigg( Q_1^2 \left(\frac{4}{Q_2^2}-\frac{2 \left(2 \tau ^2+1\right) \left(\sigma _1^E+\sigma _2^E-2\right)}{m_{\mu }^2}\right)+Q_1 \left(\frac{4 \tau }{Q_2}-\frac{4 Q_2 \tau  \left(\tau ^2+1\right) \left(\sigma _2^E-1\right)}{m_{\mu }^2}\right) \\
			& -\frac{6 Q_1^3 \tau \left(\sigma _1^E-1\right)}{Q_2 m_{\mu }^2}+\frac{Q_1^4 \left(2-2 \sigma _1^E\right)}{Q_2^2 m_{\mu }^2} \\
			& +X \left(\frac{Q_1^4 \left(8 \tau ^2+4\right)}{m_{\mu }^2}+Q_1^3 \left(\frac{8 Q_2 \tau  \left(\tau ^2+2\right)}{m_{\mu }^2}-\frac{16 \tau }{Q_2}\right)+Q_1^2 \left(\frac{Q_2^2 \left(8 \tau ^2+4\right)}{m_{\mu }^2}-16 \tau ^2\right)\right)\Bigg) , \end{aligned} \\ 
   		T_9 &=  Q_3^2 \left(\frac{\sigma _1^E-1}{Q_2^2 m_{\mu }^2}+\frac{\sigma _2^E-1}{Q_1^2 m_{\mu }^2}-\frac{2}{Q_1^2Q_2^2} \right)+X \left(-\frac{2 Q_3^2}{m_{\mu }^2}+\frac{8 Q_2 \tau }{Q_1}+\frac{8 Q_1 \tau }{Q_2}+8 \left(\tau ^2+1\right)\right) , \\
   		T_{10} &= \frac{1}{2Q_3^2} \begin{aligned}[t]
			& \Bigg( -\frac{Q_1^2 \left(\tau ^2 \left(\sigma _1^E-1\right) \left(\sigma _1^E+3\right)+2 \left(\sigma _1^E+\sigma_2^E-2\right)\right)}{m_{\mu }^2} -\frac{Q_2^3 \tau  \left(\sigma _2^E-1\right) \left(\sigma_2^E+3\right)}{Q_1m_{\mu }^2} \\
			& -\frac{Q_2^2 \left(\tau ^2 \left(\sigma _2^E-1\right) \left(\sigma _2^E+3\right)+2 \left(\sigma_1^E+\sigma _2^E-2\right)\right)}{m_{\mu }^2}-\frac{Q_1^3 \tau  \left(\sigma _1^E-1\right) \left(\sigma _1^E+3\right)}{Q_2 m_{\mu}^2} \\
			& +Q_1 \left(\frac{8 \tau }{Q_2}-\frac{Q_2 \tau  \left(\left(\sigma _1^E+4\right) \sigma _1^E+\sigma _2^E \left(\sigma_2^E+4\right)-10\right)}{m_{\mu }^2}\right)+\frac{8 Q_2 \tau}{Q_1} \\
			& +8 \tau ^2 +X \left(-16 Q_1^2 \left(\tau ^2-1\right)-16 Q_2 Q_1 \tau  \left(\tau ^2-1\right)-16 Q_2^2 \left(\tau ^2-1\right)\right) \Bigg) \end{aligned} \\
   			&\quad +\frac{X}{2} \left(\frac{4 Q_2 Q_1 \tau }{m_{\mu }^2}+\frac{4 Q_1^2}{m_{\mu }^2}+\frac{4 Q_2^2}{m_{\mu }^2}\right) , \\
		T_{11} &= \frac{1}{2m_\mu^2 Q_1 Q_2^2 Q_3^2} \begin{aligned}[t]
			& \Bigg(
				Q_2^5 \tau  \left(-6 \sigma _2^E+ {\sigma_2^E}^2 + 5\right) + 8 Q_1^5 \left(-\sigma _1^E+2 Q_2^2 \left(\tau ^2+1\right) X+1\right) \\
				& + 4 Q_2 Q_1^4 \tau  \left(-7 \sigma _1^E + 2 Q_2^2 \left(2 \tau ^2+9\right) X+7\right) \\
				& + 4 Q_2^2 Q_1^3 \left(2 \tau ^2 \left(-3 \sigma_1^E - \sigma _2^E + 8 Q_2^2 X+4\right)-2 \left(\sigma _1^E+\sigma _2^E-2\right)+5 Q_2^2 X\right) \\
				& + Q_2^3 Q_1^2 \tau  \left(8 \tau ^2 \left(-\sigma _1^E - \sigma _2^E + 2 Q_2^2 X+2\right)-6 \sigma _1^E - {\sigma _1^E}^2 - 28 \sigma _2^E + 16 Q_2^2 X+35\right) \\
				& + 2 Q_2^4 Q_1 \left(\tau ^2 \left(-10 \sigma _2^E+ {\sigma _2^E}^2 +9\right) - \sigma _1^E-3 \sigma _2^E+2 Q_2^2 X+4\right) \\
				& - 8 m_{\mu }^2 \begin{aligned}[t]
					&\Big( -Q_2^3 \tau +2 Q_1^3 \left(2 Q_2^2 \left(4 \tau ^2 X+X\right)-1\right)+Q_2 Q_1^2 \tau  \left(4 Q_2^2 \left(\tau ^2+3\right) X-5\right) \\
					& +Q_2^2 Q_1 \left(2 \tau ^2 \left(Q_2^2 X-1\right)+2 Q_2^2 X-1\right)+8 Q_2 Q_1^4 \tau  X\Big)
			\Bigg) , \end{aligned}  \end{aligned} \\
		T_{12} &= \frac{1}{4 m_\mu^2 Q_1 Q_2 Q_3^2} \begin{aligned}[t]
			& \Bigg( Q_2^2 \tau  \left(-Q_3^2 {\sigma _2^E}^2 + Q_2^2 \left(6 \sigma _2^E-5\right)-8 m_{\mu }^2\right) \\
			& - 2 Q_2 Q_1^3 \left(\tau ^2 \left(2 \sigma _1^E+8 X m_{\mu }^2-1\right)-3 \sigma _1^E+\sigma _2^E+8 X m_{\mu }^2+2\right) \\
			& + Q_1^2 \tau  \left(-2 Q_2^2 \left(4 \tau^2-5\right) \left(\sigma _1^E-\sigma _2^E\right)+Q_3^2 {\sigma _1^E}^2 + 8 m_{\mu }^2+8 Q_2^4 \left(2 \tau ^2-3\right) X\right) \\
			& + 2 Q_2^3 Q_1 \left(\tau ^2 \left(2 \sigma _2^E+8 X m_{\mu }^2-1\right)+\sigma _1^E - 3 \sigma _2^E+8 X m_{\mu }^2-2 Q_2^2 X+2\right) \\
			& + Q_1^4 \tau  \left(-6 \sigma _1^E - 8 Q_2^2 \left(2 \tau ^2-3\right) X+5\right)+4 Q_2 Q_1^5 X \Bigg) , \end{aligned} \mytag
\end{align*}
where
\begin{align}
	\begin{split}
		X &= \frac{1}{Q_1 Q_2 x} \atan\left( \frac{z x}{1 - z \tau} \right) , \quad x = \sqrt{1 - \tau^2} , \\
		z &= \frac{Q_1 Q_2}{4m_\mu^2} (1-\sigma^E_1)(1-\sigma^E_2) , \quad \sigma^E_i = \sqrt{ 1 + \frac{4 m_\mu^2}{Q_i^2} } , \\
		Q_3^2 &= Q_1^2 + 2 Q_1 Q_2 \tau + Q_2^2 .
	\end{split}
\end{align}


\section{Feynman-parameter representation of the pion box}

\label{sec:FeynmanParametrizationPionBox}

In the limit $q_4\to0$, the pion-box contribution to the scalar functions that appear in the master formula can be written as a two-dimensional Feynman parameter integral:
\begin{align}
	\hat \Pi_i^{\pi\text{-box}}(q_1^2,q_2^2,q_3^2) = F_\pi^V(q_1^2) F_\pi^V(q_2^2) F_\pi^V(q_3^2) \frac{1}{16\pi^2} \int_0^1 dx \int_0^{1-x} dy \, I_i(x,y) ,
\end{align}
where
\begin{align*}
	I_1(x,y) &= \frac{8 xy (1-2x) (1-2y)}{\Delta_{123} \Delta_{23}}  , \\
	I_4(x,y) &= \frac{4(1-x-y)(1-2x-2y)\Delta_{21}}{\Delta_{321}^2}\left( \frac{(1-2x-2y)^2}{\Delta_{321}} - \frac{1-x(3-2x) - y(3-2y)}{\Delta_{21}} \right) \\
		&\quad + \frac{16xy(1-2x)(1-2y)}{\Delta_{321} \Delta_{21}}, \\
	I_7(x,y) &= - \frac{8 x y (1-x-y) (1-2x)^2(1-2y)}{\Delta_{123}^3} , \\
	I_{17}(x,y) &= \frac{16xy^2(1-2x)(1-2y)}{\Delta_{123} \Delta_{23}} \left( \frac{1-x-y}{\Delta_{123}} + \frac{1-y}{\Delta_{23}} \right) , \\
	I_{39}(x,y) &= \frac{8xy(1-x-y)(1-2x)(1-2y)(1-2x-2y)}{\Delta_{123}^3} , \\
	I_{54}(x,y) &= - \frac{8xy(1-x-y)(1-2x)(1-2y)(x-y)}{\Delta_{321} \Delta_{21}} \left( \frac{1}{\Delta_{321}} + \frac{1}{\Delta_{21}} \right) , 
	\mytag
\end{align*}
and
\begin{align}
	\begin{split}
		\Delta_{ijk} &= M_\pi^2 - x y q_i^2 - x (1-x-y) q_j^2 - y(1-x-y) q_k^2 , \\
		\Delta_{ij} &= M_\pi^2 - x (1-x) q_i^2 - y (1-y) q_j^2 .
	\end{split}
\end{align}
The remaining functions entering the master formula can be obtained with the crossing relations~\eqref{eq:CrossingRelationsPiHat}.


\section{Scalar functions for the two-pion dispersion relations}

\label{sec:PiCheckFunctions}

Here, we give the explicit solution for the scalar functions $\check\Pi_i$, which fulfill unsubtracted single-variable dispersion relations and only depend on physical helicity amplitudes. First, we define the following linear combinations of BTT functions:
\begin{align}
	\begin{split}
		\Pi_A &:= \Pi_{47} + \Pi_{49} - \Pi_{51} , \\
		\Pi_B &:= \Pi_{41} - \Pi_{42} + \Pi_{45} , \\
		\Pi_C &:= \Pi_{38} + \Pi_{47} - \Pi_{51} + \Pi_{52} , \\
		\Pi_D &:= \Pi_{27} - \Pi_{28} - 2 \Pi_{49} + 2 \Pi_{52} - 2 \Pi_{53} , \\
		\Pi_E &:= \Pi_{41} - \Pi_{45} , \\
		\Pi_F &:= \Pi_{42} - \Pi_{45} - \Pi_{53} , \\
		\Pi_G &:= \Pi_{15} + \Pi_{16} - \Pi_{24} - \Pi_{27} - \Pi_{28} + \Pi_{38} + \Pi_{47} - \Pi_{52} + \Pi_{54} , \\
		\Pi_H &:= \Pi_{12} + \Pi_{41} + \Pi_{53} , \\
		\Pi_I &:= 4 \Pi_{16}+\Pi_{23}-4 \Pi_{27}-\Pi_{30}+\Pi_{37}+\Pi_{38}+\Pi_{41}+2 \Pi_{42}+\Pi_{43}-\Pi_{45}+\Pi_{47}+\Pi_{49} \\
			&\quad +3 \Pi_{51}-7 \Pi_{52}+6 \Pi_{53}+3 \Pi_{54} , \\
		\Pi_J &:= \Pi_{23}+\Pi_{30}-\Pi_{37}-\Pi_{38}-\Pi_{41}+2 \Pi_{42}-\Pi_{43}-\Pi_{45}-\Pi_{47}-\Pi_{49}+\Pi_{51}-\Pi_{52}+\Pi_{54} , \\
		\Pi_K &:= \Pi_{50} - \Pi_{51} + \Pi_{54} ,
	\end{split}
\end{align}
as well as $\Pi_i^c := \Cr{13}{\Pi_i}$.

The 19 functions that contribute to $(g-2)_\mu$ can be written in the form (for $q_4^2=0$ and $t=q_2^2$)
\begin{align}
	\tag{\ref{eq:PicheckFunctions}}
	\check\Pi_i &= \hat\Pi_{g_i} + (s-q_3^2) \bar\Delta_i + (s-q_3^2)^2 \bar{\bar{\Delta}}_i ,
\end{align}
where $\{ g_i \}  = \{1,\ldots, 11, 13, 14, 16, 17, 39, 50, 51, 54\}$,
\begin{align}
	\begin{split}
		\bar\Delta_1 &= -\frac{1}{2} \Pi_A + \frac{q_2^2 q_{123}}{\lambda_{123}} \Pi_B + \frac{q_{123}^2}{2\lambda_{123}} \Pi_C - \frac{q_2^2 q_{312}}{\lambda_{123}} \Pi_C^c , \\
		\bar\Delta_2 &= -\frac{\Sigma q_{312}(q_1^2-q_3^2)}{2\lambda_{123}(q_1^2+q_3^2)} \Pi_B - \frac{\Sigma q_{312}(2q_1^2+q_{312})}{4\lambda_{123}(q_1^2+q_3^2)} \Pi_C + \frac{\Sigma q_{312}(2q_3^2+q_{312})}{4\lambda_{123}(q_1^2+q_3^2)} \Pi_C^c \\
			&\quad - \frac{q_{312}}{4(q_1^2+q_3^2)} \Pi_D - \frac{\Sigma}{4(q_1^2+q_3^2)} \Pi_E + \frac{\Sigma q_{312}(q_2^2+2q_{312})}{4 \lambda_{123} (q_1^2+q_3^2)} \Pi_K , \\
		\bar\Delta_3 &= -\Cr{13}{\bar\Delta_1} , \\
		\bar\Delta_4 &= \frac{1}{2} \Pi_A^c + \frac{q_{231}(q_1^2-q_2^2)}{\lambda_{123}} \Pi_B - \frac{q_{312}(q_1^2-q_2^2)}{\lambda_{123}} \Pi_C + \frac{2q_3^2(q_1^2-q_2^2)}{\lambda_{123}} \Pi_C^c - \frac{1}{2} \Pi_F - \frac{q_2^2 q_{312}}{\lambda_{123}} \Pi_K , \\
		\bar\Delta_5 &= -\frac{1}{2} \Pi_A + \frac{1}{2} \Pi_A^c + \frac{2 q_2^2(q_1^2-q_3^2)}{\lambda_{123}} \Pi_B + \left( \frac{1}{2} + \frac{q_2^2(q_{312} + 2q_1^2)}{\lambda_{123}} \right) \Pi_C - \left( \frac{1}{2} + \frac{q_2^2(q_{312} + 2q_3^2)}{\lambda_{123}} \right) \Pi_C^c \\
			&\quad + \frac{q_{312}(2q_{312} + q_2^2)\Sigma}{4\lambda_{123}(q_1^2+q_3^2)} \Pi_K , \\
		\bar\Delta_6 &= -\Cr{13}{\bar\Delta_4} , \\
		\bar\Delta_{11} &= - \bar\Delta_{15} = - \bar\Delta_{17} =  -\frac{2 q_2^2}{\lambda_{123}} \Pi_B - \frac{q_{123}}{\lambda_{123}} \Pi_C - \frac{q_{231}}{\lambda_{123}} \Pi_C^c , \\
		\bar\Delta_{14} &= -\frac{\Sigma (q_1^2-q_3^2)}{\lambda_{123}(q_1^2+q_3^2)} \Pi_B - \frac{\Sigma(q_{312}+2q_1^2)}{2\lambda_{123}(q_1^2+q_3^2)} \Pi_C + \frac{\Sigma(q_{312}+2q_3^2)}{2\lambda_{123}(q_1^2+q_3^2)} \Pi_C^c - \frac{1}{2(q_1^2+q_3^2)} \left( \Pi_D - \Pi_E \right) , \\
		\bar\Delta_{18} &=  \frac{q_{123}}{\lambda_{123}} \Pi_B + \frac{2q_1^2}{\lambda_{123}} \Pi_C - \frac{q_{312}}{\lambda_{123}} \Pi_C^c , \\
		\bar\Delta_{19} &= -\Cr{13}{\bar\Delta_{18}} ,
	\end{split}
\end{align}
and
\begin{align}
	\bar{\bar\Delta}_2 &= \frac{q_2^2}{4(q_1^2+q_3^2)} \Pi_{36} , \quad \bar{\bar\Delta}_{14} = -\frac{1}{2(q_1^2+q_3^2)} \Pi_{36} .
\end{align}
All other $\bar\Delta_i$ and $\bar{\bar\Delta}_i$ are zero. We use the abbreviations $q_{ijk} := q_i^2 + q_j^2 - q_k^2$, $\Sigma = q_1^2 + q_2^2 + q_3^2$, and $\lambda_{123} := \lambda(q_1^2, q_2^2, q_3^2)$ for the K\"all{\'e}n function.

We define five additional scalar functions $\check\Pi_i$ that appear in sum rules:
\begin{align}
	\begin{split}
		\check\Pi_{20} &:= q_{312} \Pi_G - q_{231} \Pi_H + \frac{q_{312}(s-q_1^2-q_2^2)}{2} \Pi_{35} + \frac{q_{123} q_{231}}{2} \Pi_{36} , \\
		\check\Pi_{21} &:= \Cr{13}{\check\Pi_{20}} , \\
		\check\Pi_{22} &:= \lambda_{123} \Pi_{23} - 2 q_3^2 q_{123} \Pi_B - q_{312}^2 \Pi_C + 2 q_3^2 q_{312} \Pi_C^c + \lambda_{123} \Pi_F^c , \\
		\check\Pi_{23} &:= \Cr{13}{\check\Pi_{23}} , \\
		\check\Pi_{24} &:= q_1^2 \Pi_I + q_2^2 \Pi_J + q_3^2 \Pi_I^c - 2(q_1^2-q_3^2)(s-q_3^2) \Pi_{36} .
	\end{split}
\end{align}
The singly-on-shell basis consists of 27 elements. The three functions $\check\Pi_{25}$, $\check\Pi_{26}$, and $\check\Pi_{27}$ are not given explicitly as they have no significance in the connection with $(g-2)_\mu$.


\section{Basis change and sum rules}

\subsection{Unphysical polarizations}

\label{sec:UnphysicalPolarizations}

In the following, we explain why unphysical polarizations are not trivially absent in any representation. In short, although unphysical polarizations cannot contribute to any observable, the absence of such unphysical contributions is manifest only if the basis is well chosen. Otherwise, their apparent contribution vanishes only due to the presence of sum rules for the scalar functions.

Suppose we have a decomposition of the HLbL tensor into a ``physical'' and an ``unphysical'' piece,
\begin{align}
	\label{eq:HLbLSeparationUnphysicalHelicities}
	\Pi^{\mu\nu\lambda\sigma} = \Pi^{\mu\nu\lambda\sigma}_\text{phys} + \Pi^{\mu\nu\lambda\sigma}_\text{unph}  = \sum_i T_{i,\text{phys}}^{\mu\nu\lambda\sigma} \Pi_i^\text{phys} +  \sum_i T_{i,\text{unph}}^{\mu\nu\lambda\sigma} \Pi_i^\text{unph},
\end{align}
where the scalar functions $\Pi_i^\text{phys}$ are linear combinations of helicity amplitudes with only transverse polarizations of the external photon. The scalar functions $\Pi_i^\text{unph}$ contain also contributions from the longitudinal polarization. Because these scalar functions cannot contribute to an observable, the unphysical tensor structures have to fulfill
\begin{align}
	T_{i,\text{unph}}^{\mu\nu\lambda\sigma} \propto q_4^\sigma, q_4^2 .
\end{align}
Such structures do not contribute to $(g-2)_\mu$, because the derivative with respect to $q_4^\rho$ either vanishes for $q_4\to0$ or is symmetric in $\rho\leftrightarrow\sigma$.

Next, we apply the following transformation, which mixes the physical and unphysical part:
\begin{align}
	T_{a,\text{phys}}^{\mu\nu\lambda\sigma} \Pi_a^\text{phys} + T_{b,\text{unph}}^{\mu\nu\lambda\sigma} \Pi_b^\text{unph} = T_{a,\text{phys}}^{\mu\nu\lambda\sigma} \left( \Pi_a^\text{phys} + \alpha \Pi_b^\text{unph} \right) + \left( T_{b,\text{unph}}^{\mu\nu\lambda\sigma} - \alpha T_{a,\text{phys}}^{\mu\nu\lambda\sigma} \right) \Pi_b^\text{unph} .
\end{align}
Because not all tensor structures have the same mass dimension, the
coefficient $\alpha$ can be dimensionful, e.g.~$\alpha = q_3 \cdot q_4$ if
the mass dimension of $T_{b,\text{unph}}^{\mu\nu\lambda\sigma}$ is larger
by two units than the one of $T_{a,\text{phys}}^{\mu\nu\lambda\sigma}$, all
while avoiding kinematic singularities. The new structure $\left(
  T_{b,\text{unph}}^{\mu\nu\lambda\sigma} - \alpha
  T_{a,\text{phys}}^{\mu\nu\lambda\sigma} \right)$ still cannot contribute
to $(g-2)_\mu$ if $\alpha \propto q_4$. However, we have introduced a new
combination of unphysical and physical helicity amplitudes into the scalar
coefficient functions of $T_{a,\text{phys}}^{\mu\nu\lambda\sigma}$. If we
make such a transformation in the discontinuity appearing in an $s$-channel
dispersion integral, the factor $\alpha = q_3 \cdot q_4$ becomes in the
$(g-2)_\mu$ limit 
\begin{align}
	q_3 \cdot q_4 \to -\frac{1}{2} (s'-q_3^2) ,
\end{align}
where we have replaced the Mandelstam variable $s$ by the integration variable of the dispersion integral $s'$. This factor cancels with the Cauchy kernel $1/(s'-q_3^2)$, producing an apparent polynomial contribution that depends on both physical and unphysical helicity amplitudes. As shown in Sect.~\ref{sec:SumRulesBTTFunctions} this polynomial contribution actually vanishes due to sum rules, but in practice it can be tedious to identify the combination of physical and unphysical helicity amplitudes that corresponds to this vanishing polynomial, and, worse, in a partial-wave expansion these sum rules are only fulfilled after resumming all partial waves. Since the above example implies that setting by hand only the unphysical polarizations to zero leads to a wrong result, a practical implementation requires a basis where this contribution is manifestly absent from the beginning. The construction of this basis is performed in Sect.~\ref{sec:SinglyOnShellBasis}.

\subsection{Comparison to forward-scattering sum rules}

\label{eq:ForwardScattering}

In \cite{Pascalutsa:2012pr}, sum rules have been derived for the case of forward HLbL scattering. In the following, we compare them to our fixed-$t$ sum rules derived in Sect.~\ref{sec:PhysicalSumRules}. To this end, we consider the case of general forward kinematics, i.e.
\begin{align}
	\label{eq:ForwardKinematics}
	q_3 = - q_1, \quad q_4 = q_2 ,
\end{align}
which implies for the Lorentz invariants
\begin{align}
	t = 0 , \quad u = 2 q_1^2 + 2 q_2^2 - s , \quad q_3^2 = q_1^2, \quad q_4^2 = q_2^2.
\end{align}
The common limit of forward and singly-on-shell fixed-$t$ kinematics is obtained for $q_2^2 \to 0$.

It is convenient to define the variable~\cite{Budnev:1971sz}
\begin{align}
	\nu := q_1 \cdot q_2 = \frac{1}{4}(s-u) .
\end{align}

In the case of forward scattering, only eight independent helicity amplitudes exist~\cite{Budnev:1971sz}. Consistently, starting with the BTT decomposition~\eqref{eqn:HLbLTensorKinematicFreeStructures} and taking the limit of forward kinematics, only eight independent Lorentz structures survive. Interestingly, the two ambiguities in four space-time dimensions~\cite{Eichmann:2014ooa} disappear, but even for forward kinematics one redundancy of Tarrach's type remains~\cite{Tarrach:1975tu}. Therefore, the forward HLbL tensor can be written as
\begin{align}
	\Pi^{\mu\nu\lambda\sigma}_{\FW} &= \sum_{i=1}^9 T_{i,\FW}^{\mu\nu\lambda\sigma} \Pi_i^{\FW} ,
\end{align}
where the tensor structures are given by
\begin{align*}
	\begin{alignedat}{2}
		T_{1,\FW}^{\mu\nu\lambda\sigma} &= \frac{1}{2} \left( T_1^{\mu\nu\lambda\sigma} + T_3^{\mu\nu\lambda\sigma} \right) , \quad & 
		T_{2,\FW}^{\mu\nu\lambda\sigma} &= T_5^{\mu\nu\lambda\sigma} , \\
		T_{3,\FW}^{\mu\nu\lambda\sigma} &= \frac{1}{2} \left( T_4^{\mu\nu\lambda\sigma} + T_6^{\mu\nu\lambda\sigma} \right) , \quad &
		T_{4,\FW}^{\mu\nu\lambda\sigma} &= \frac{1}{2} \left( T_9^{\mu\nu\lambda\sigma} + T_{10}^{\mu\nu\lambda\sigma} \right) , \\
		T_{5,\FW}^{\mu\nu\lambda\sigma} &= \frac{1}{2} \left( T_{15}^{\mu\nu\lambda\sigma} + T_{16}^{\mu\nu\lambda\sigma} \right) , \quad &
		T_{6,\FW}^{\mu\nu\lambda\sigma} &= \frac{1}{4} \left( T_{49}^{\mu\nu\lambda\sigma} + T_{51}^{\mu\nu\lambda\sigma} + T_{52}^{\mu\nu\lambda\sigma} + T_{54}^{\mu\nu\lambda\sigma} \right) , \\
		T_{7,\FW}^{\mu\nu\lambda\sigma} &= \frac{1}{2} \left( T_4^{\mu\nu\lambda\sigma} - T_6^{\mu\nu\lambda\sigma} \right) , \quad &
		T_{8,\FW}^{\mu\nu\lambda\sigma} &= \frac{1}{2} \left( T_1^{\mu\nu\lambda\sigma} - T_3^{\mu\nu\lambda\sigma} - T_4^{\mu\nu\lambda\sigma} + T_6^{\mu\nu\lambda\sigma} \right) , \\
		T_{9,\FW}^{\mu\nu\lambda\sigma} &= \frac{1}{4} \left( T_{19}^{\mu\nu\lambda\sigma} - T_{24}^{\mu\nu\lambda\sigma} - T_{26}^{\mu\nu\lambda\sigma} + T_{29}^{\mu\nu\lambda\sigma} \right) , \hspace{-2cm} &
	\end{alignedat} \mytag
\end{align*}
with BTT structures on the right-hand side of the equations evaluated in the limit~\eqref{eq:ForwardKinematics}. The redundancy reads
\begin{align}\label{eq:Red67FW}
	\nu \, T_{6,\FW}^{\mu\nu\lambda\sigma} + q_1^2 q_2^2 \, T_{7,\FW}^{\mu\nu\lambda\sigma}  = 0 .
\end{align}
In terms of the BTT functions, the forward scalar functions are given by
\begin{align}
	\begin{split}
		\label{eq:ForwardScalarFunctions}
		\Pi_1^{\FW} &= \Pi_{1}+\Pi_{3}-\nu \big(\Pi_{49}-\Pi_{51}-\Pi_{52}+\Pi_{54} \big) , \\
		\Pi_2^{\FW} &= \Pi_{5} - \nu \big( \Pi_{49}- \Pi_{51} - \Pi_{52} + \Pi_{54}\big) , \\
		\Pi_3^{\FW} &= \Pi_{4}+\Pi_{6} + q_1^2 \big(\Pi_{7}+\Pi_{11}+\Pi_{13}+\Pi_{17} \big) + q_2^2 \big(\Pi_{8}+\Pi_{12}+\Pi_{14}+\Pi_{18}\big) \\
			&\quad - q_1^2 q_2^2 \big( \Pi_{31} + \Pi_{32} + \Pi_{34} + \Pi_{35}\big)+\nu \big(\Pi_{20}-\Pi_{23}-\Pi_{25}+\Pi_{30}-\Pi_{49}+\Pi_{51}+\Pi_{52}-\Pi_{54}\big) , \\
		\Pi_4^{\FW} &= \Pi_{9}+\Pi_{10}-\Pi_{21}-\Pi_{22} , \\
		\Pi_5^{\FW} &= \Pi_{15}+\Pi_{16}-\Pi_{27}-\Pi_{28} , \\
		\Pi_6^{\FW} &= - \frac{1}{2} \big( \Pi_{19} + \Pi_{24} + \Pi_{26} + \Pi_{29} \big) - \frac{\nu}{2} \big( \Pi_{31} + \Pi_{32} - \Pi_{34} - \Pi_{35}\big) \\
			&\quad - \Pi_{37}-\Pi_{38}-\Pi_{40}-\Pi_{43}-\Pi_{44}-\Pi_{46}-\Pi_{47}-\Pi_{48}+\Pi_{49}+\Pi_{51}+\Pi_{52}+\Pi_{54} , \\
		\Pi_7^{\FW} &= \Pi_{1}-\Pi_{3}+\Pi_{4}-\Pi_{6}+\nu \big(\Pi_{20}+\Pi_{23}+\Pi_{25}+\Pi_{30}\big) + q_1^2  \big(\Pi_{7}-\Pi_{11}-\Pi_{13}+\Pi_{17} -2 \Pi_{50}\big) \\
			&\quad + q_2^2 \big(\Pi_{8}-\Pi_{12}-\Pi_{14}+\Pi_{18}-2 \Pi_{53}\big) + q_1^2 q_2^2 \big(-\Pi_{31}-\Pi_{32}+\Pi_{34}+\Pi_{35}\big) , \\
		\Pi_8^{\FW} &=  \Pi_{1}-\Pi_{3}  - \frac{\nu}{2} \big(\Pi_{19} + \Pi_{24} + \Pi_{26} + \Pi_{29} - 2 \big(\Pi_{37}+\Pi_{38}+\Pi_{40}+\Pi_{43}+\Pi_{44}+\Pi_{46}+\Pi_{47}+\Pi_{48}\big)\big) \\
			&\quad - \frac{\nu^2}{2} \big( \Pi_{31} + \Pi_{32} - \Pi_{34} - \Pi_{35}\big) , \\
		\Pi_9^{\FW} &= \Pi_{19}-\Pi_{24}-\Pi_{26}+\Pi_{29}+2 \big( \Pi_{49} - \Pi_{51} - \Pi_{52} + \Pi_{54} \big) + \nu \big(\Pi_{31}+\Pi_{32}+\Pi_{34}+\Pi_{35}\big) .
	\end{split}
\end{align}
The functions $\Pi_i^{\FW}$ are even in $\nu$ for $i=1, \ldots, 6$ and odd
for $i=7,8,9$, which corresponds to the crossing symmetries $\Cr{13}{}$ or
$\Cr{24}{}$. We further have $\Cr{12}{\Cr{34}{\Pi_{4}^{\FW}}} =
\Pi_{5}^{\FW}$, while the other seven functions are invariant under this
transformation. According to our assumption for the asymptotic
behavior~\eqref{eq:BTTAsymptoticBehaviour}, all the functions $\Pi_i^{\FW}$
fulfill an unsubtracted dispersion relation. Note, however, that due to the
redundancy~\eqref{eq:Red67FW} $\Pi_{6,7}^{\FW}$ enter in observables only
in the linear combination 
\begin{align}
	q_1^2 q_2^2 \, \Pi_6^{\FW} - \nu \, \Pi_7^{\FW} ,
\end{align}
which requires a once-subtracted dispersion relation. The subtraction
constant vanishes in the quasi-real limit of one of the photons. 

With our assumption for the asymptotic behavior, we find three physical sum rules:
\begin{align}
	\int d\nu \, \Im \Pi_i^{\FW}(\nu) = 0 , \qquad i = 4, 5, 9 .
\end{align}
Due to the symmetry in $\nu$, the first two are trivially fulfilled: the
integrals over the left- and right-hand cuts cancel. This leaves a single
sum rule involving $\Pi_9^{\FW}$. 

Next, we consider the basis change to helicity amplitudes. The eight
forward-scattering amplitudes are given
by~\cite{Budnev:1971sz,Pascalutsa:2012pr} 
\begin{align}
	\begin{alignedat}{3}
		H_1^{\FW} &:= H_{++,++} + H_{+-,+-} , \quad & H_2^{\FW} &:= H_{++,--} , \quad & H_3^{\FW} &:= H_{00,00} , \\
		H_4^{\FW} &:= H_{+0,+0} , \quad & H_5^{\FW} &:= H_{0+,0+} , \quad & H_6^{\FW} &:= H_{++,00} + H_{+0,0-} , \\
		H_7^{\FW} &:= H_{++,++} - H_{+-,+-} , \quad & H_8^{\FW} &:= H_{++,00} - H_{+0,0-} ,
	\end{alignedat}
\end{align}
where the first six are even, the last two are odd in $\nu$. With our conventions for the polarization vectors, they are related to the scalar functions~\eqref{eq:ForwardScalarFunctions} by
\begin{align}
	\begin{split}
		H_1^{\FW} &= -(\nu^2 - q_1^2q_2^2) \Pi_1^{\FW} - 2 q_1^2 q_2^2 \Pi_2^{\FW} - \nu^2 \Pi_3^{\FW} - 2 \nu^2 q_1^2 \Pi_4^{\FW} - 2 \nu^2 q_2^2 \Pi_5^{\FW} - \nu q_1^2 q_2^2 \Pi_9^{\FW} , \\
		H_2^{\FW} &= (\nu^2 - q_1^2q_2^2) \Pi_1^{\FW} - \nu^2 \Pi_3^{\FW} - \nu q_1^2 q_2^2 \Pi_9^{\FW} , \\
		H_3^{\FW} &= - \Pi_2^{\FW} - \Pi_3^{\FW} - q_1^2 \Pi_4^{\FW} - q_2^2 \Pi_5^{\FW} - \nu \Pi_9^{\FW} , \\
		H_4^{\FW} &= -q_1^2 \Pi_2^{\FW} - (q_1^2)^2 \Pi_4^{\FW} - \nu^2 \Pi_5^{\FW} , \\
		H_5^{\FW} &= -q_2^2 \Pi_2^{\FW} - \nu^2 \Pi_4^{\FW} - (q_2^2)^2 \Pi_5^{\FW} , \\
		H_6^{\FW} &= q_1^2q_2^2 \Pi_6^{\FW} - \nu \Pi_7^{\FW} + \nu \Pi_8^{\FW} , \\
		H_7^{\FW} &= \nu( q_1^2 q_2^2 \Pi_6^{\FW} - \nu \Pi_7^{\FW} ) + q_1^2 q_2^2 \Pi_8^{\FW} , \\
		H_8^{\FW} &= - \nu \Pi_3^{\FW} - \frac{1}{2} ( \nu^2 + q_1^2 q_2^2 ) \Pi_9^{\FW} .
	\end{split}
\end{align}
In terms of the helicity amplitudes the sum rule reads
\begin{align*}
	 \int_{\nu_0}^\infty d\nu \frac{1}{(\nu^2 - q_1^2q_2^2)^2} \begin{aligned}[t]
	 	&\bigg( \nu \, \Im \Big[ H_1^{\FW}(\nu) + H_2^{\FW}(\nu) + 2 q_1^2 q_2^2 H_3^{\FW}(\nu) - 2 q_2^2 H_4^{\FW}(\nu) - 2 q_1^2 H_5^{\FW}(\nu) \Big] \\
		& - 2(\nu^2 + q_1^2 q_2^2) \Im H_8^{\FW}(\nu) \bigg) = 0 , \end{aligned} \mytag
\end{align*}
where $\nu_0$ denotes the threshold in $\nu$. Taking the quasi-real limit $q_2^2\to0$ of this equation and accounting for the different conventions for the polarization vectors, we reproduce the sum rule~(27b) of~\cite{Pascalutsa:2012pr}. In addition, two more sum rules (superconvergence relations) were derived in~\cite{Pascalutsa:2012pr}. They originate in different assumptions about the asymptotic behavior based on the Regge model of~\cite{Budnev:1971sz}. In Table~\ref{tab:Asymptotics}, we compare the assumptions on the asymptotic behavior of the helicity amplitudes: concerning the number of subtractions needed in a dispersion relation for the helicity amplitudes, this leads in most cases to identical results.\footnote{Note that even (odd) subtractions vanish for a function that is odd (even) in $\nu$. This implies that for $H_{2,4,5,8}^\FW$, the subtraction schemes are identical although the exact assumptions for the asymptotic behavior slightly differ.} For $H_3^{\FW}$, our assumption is more restrictive.  In fact, a similar behavior was used in~\cite{Pascalutsa:2012pr} to derive an additional sum rule for a low-energy constant in the effective photon Lagrangian, stressing that this sum rule cannot be justified based on the Regge model of~\cite{Budnev:1971sz}. In our approach this sum rule emerges naturally by demanding a uniform asymptotic behavior of the HLbL tensor, which in turn determines the asymptotics of the BTT functions and thereby of the helicity amplitudes.  
For $H_{6,7}^{\FW}$, the assumption in~\cite{Pascalutsa:2012pr} is more restrictive and leads to two additional sum rules, Eqs.~(27a) and (27c) in~\cite{Pascalutsa:2012pr}.

\begin{table}[t]
	\centering
	\renewcommand{\arraystretch}{1.3}
	\begin{tabular}{lll}
		\toprule
		 & \quad this work &  \qquad Ref.~\cite{Pascalutsa:2012pr} \\
		\midrule
		$H_1^{\FW}$ 		& $\qquad\asymp \nu^1$ 		& $\qquad\asymp \nu^{\alpha_P(0)}$  \\
		$H_2^{\FW}$ 		& $\qquad\asymp \nu^1$ 		& $\qquad\asymp \nu^{\alpha_\pi(0)}$  \\
		$H_3^{\FW}$ 		& $\qquad\asymp \nu^{-1}$ 	& $\qquad\asymp \nu^{\alpha_P(0)}$  \\
		$H_4^{\FW}$ 		& $\qquad\asymp \nu^0$ 		& $\qquad\asymp \nu^{\alpha_P(0)}$  \\
		$H_5^{\FW}$ 		& $\qquad\asymp \nu^0$ 		& $\qquad\asymp \nu^{\alpha_P(0)}$  \\
		$H_6^{\FW}$ 		& $\qquad\asymp \nu^0$ 		& $\qquad\asymp \nu^{\alpha_\pi(0)-1}$  \\
		$H_7^{\FW}$ 		& $\qquad\asymp \nu^1$ 		& $\qquad\asymp \nu^{\alpha_\pi(0)}$  \\
		$H_8^{\FW}$ 		& $\qquad\asymp \nu^0$ 		& $\qquad\asymp \nu^{\alpha_\pi(0)-1}$  \\
		\bottomrule
	\end{tabular}
	\caption{Comparison of the assumptions about the asymptotic behavior of the helicity amplitudes. In~\cite{Pascalutsa:2012pr}, $\alpha_P(0) \approx 1.08$ and $\alpha_\pi(0) \approx - 0.014$ was assumed.}
	\label{tab:Asymptotics}
	\renewcommand{\arraystretch}{1.0}
\end{table}

We note that the constraints from gauge invariance that were determined in~\cite{Pascalutsa:2012pr} based on an effective photon Lagrangian are all implemented in the BTT decomposition of the HLbL tensor and can be read off directly from the relations between the helicity amplitudes and the BTT scalar functions.
Finally, with the above description of forward scattering in terms of BTT functions, we can easily establish the link to our sum rules derived for singly-on-shell fixed-$t$ kinematics. By setting $q_3^2 = q_1^2$ and taking the limit $q_2^2\to0$ in both situations, we reach the common kinematic configuration, i.e.\ the case of singly-on-shell forward scattering. We can then easily find the embedding of the forward sum rule into the sum rules for the $\check\Pi_i$ functions:
\begin{align}
	\lim_{q_2^2\to0} \Pi_9^{\FW} = - 2 \lim_{\substack{q_2^2 \to 0, \\ q_3^2 = q_1^2}} \left( \check\Pi_7 + \check\Pi_{11} - \check\Pi_{12} - \check\Pi_{15} + 2 \check\Pi_{18} - 2 \check\Pi_{19} \right) ,
\end{align}
where the right-hand side is a combination of functions fulfilling the sum rules~\eqref{eq:SumRulesPicheck}. We also note that in the $S$-wave approximation, the sum rule (27b) of~\cite{Pascalutsa:2012pr} reduces to the forward limit of~\eqref{eq:SwaveSumRule}.


\section{Basis change to helicity amplitudes}

\label{sec:BasisChangeHelAmps}

\subsection{Calculation of tensor phase-space integrals}

\label{sec:TensorPhaseSpaceIntegrals}

If we consider only $S$-waves in $\gamma^*\gamma^*\to\pi\pi$, the phase-space integral in the $\pi\pi$ unitarity relation for HLbL is trivial and the unitarity relation factorizes. We have calculated the $D$-wave unitarity relation in~\cite{Colangelo:2014dfa} for an external on-shell photon and in~\cite{Stoffer:2014rka} for the fully off-shell case by using a tensor decomposition. In this approach, the unitarity relation requires the calculation of tensor integrals with additional factors of the $\gamma^*\gamma^*\to\pi\pi$ scattering angles, which are replaced by scalar products of external and internal (loop) momenta. Then the unitarity relation can be written as contractions of external momenta with tensor integrals that depend only on a single momentum and can be solved by a standard tensor decomposition.

Unfortunately, with this method the expressions become very large, which makes the computation already at the level of $D$-waves extremely inefficient.
In order to avoid the excessive amount of contractions in the calculation of the phase-space integral, here we present an alternative way to calculate the tensor integrals directly by taking derivatives of scalar integrals. This allows us to calculate even the $G$-wave unitarity relation, as required for the present application in Sect.~\ref{sec:HelAmpPWE}. In the case of $D$-waves, we have checked that both methods give the same result.

We first consider the scalar integrals with additional Legendre polynomials of the scattering angles:
\begin{align}
	\label{eq:ScalarIntegralsAngles}
	I_0^{nm} &:= \int \frac{d^3p_1}{(2\pi)^32p_1^0} \frac{d^3p_2}{(2\pi)^32p_2^0} (2\pi)^4 \delta^{(4)}\big(Q - p_1 - p_2\big) P_n(z^\prime) P_m(z^\dprime) , 
\end{align}
where $Q:=q_1+q_2 = q_4 - q_3$ and $z^\prime$, $z^\dprime$ denote the scattering angles
\begin{align}
	z^\prime = \frac{q_1^2 - q_2^2 - 2(q_1-q_2)\cdot p_1}{\sigma_\pi(s) \lambda_{12}^{1/2}(s)} , \quad z^\dprime = \frac{q_3^2 - q_4^2 + 2(q_3+q_4)\cdot p_1}{\sigma_\pi(s) \lambda_{34}^{1/2}(s)}
\end{align}
with $\lambda_{12}(s)=\lambda(s,q_1^2,q_2^2)$, $\lambda_{34}(s)=\lambda(s,q_3^2,q_4^2)$.
The HLbL scattering angle is defined as
\begin{align}
	z = \frac{(q_1^2-q_2^2)(q_3^2-q_4^2) + s(t-u)}{\lambda_{12}^{1/2}(s)\lambda_{34}^{1/2}(s)} .
\end{align}
The angles fulfill
\begin{align}
	\cos\theta^\dprime = \cos\theta^\prime \cos\theta + \sin\theta^\prime \sin\theta \cos\phi^\prime ,
\end{align}
where $z=\cos\theta$, $z^\prime=\cos\theta^\prime$, $z^\dprime=\cos\theta^\dprime$, and $\phi^\prime$ is the azimuthal angle of $\vec p_1$ in the centre-of-mass frame. The phase-space integral can be understood as an integral over the variables $\theta^\prime$ and $\phi^\prime$.

As a first step, direct calculation leads to
\begin{align}
	\begin{split}
		\label{eq:LegendreAdditionTheorem}
		I_0^{nm} &= \frac{1}{16\pi^2} \int_0^\infty dp \frac{p^2}{M_\pi^2 + p^2} \delta(Q^0 - 2\sqrt{M_\pi^2 + p^2})  \int d\Omega \; P_n(z^\prime) P_m(z^\dprime) \\
			&= \frac{1}{8\pi} \sigma_\pi(s) \delta_{nm} \frac{P_n(z)}{2n+1} ,
	\end{split}
\end{align}
where we have used the addition theorem for the Legendre polynomials.
Next, we define $P:=q_1-q_2$ and $R := q_3 + q_4$ and write the angles as
\begin{align}
	\begin{split}
		z &= \frac{Q^2(P\cdot R)-(P\cdot Q)(R\cdot Q)}{((P\cdot Q)^2-P^2Q^2)^{1/2} ((R\cdot Q)^2-R^2Q^2)^{1/2}} , \\
		z^\prime &= \frac{ P\cdot Q - 2 P \cdot p_1}{\sigma_\pi(Q^2) ((P\cdot Q)^2-P^2Q^2)^{1/2}}, \quad  z^\dprime = \frac{2 R \cdot p_1 - R \cdot Q}{\sigma_\pi(Q^2) ((R\cdot Q)^2-R^2Q^2)^{1/2}} .
	\end{split}
\end{align}
Taking the derivatives of the angles with respect to $P_\mu$ and $R_\mu$ gives
\begin{align*}
	\frac{\p z}{\p P_\mu} &= \frac{Q^2}{\lambda_{12}^{3/2}(Q^2) \lambda_{34}^{1/2}(Q^2)} \begin{aligned}[t]
		&\bigg( Q^\mu \big( P^2 (Q\cdot R) - (P\cdot Q)(P\cdot R) \big) \\
		&+ P^\mu \big( Q^2 (P\cdot R) - (Q\cdot P)(Q\cdot R) \big) \\
		&+ R^\mu \big( (P\cdot Q)^2 - P^2 Q^2 \big)  \bigg) , \end{aligned} \\
	\frac{\p z}{\p R_\mu} &= \frac{Q^2}{\lambda_{12}^{1/2}(Q^2) \lambda_{34}^{3/2}(Q^2)} \begin{aligned}[t]
		&\bigg( Q^\mu \big( R^2 (P\cdot Q) - (P\cdot R)(Q\cdot R) \big) \\
		&+ P^\mu \big( (R\cdot Q)^2 - R^2 Q^2 \big) \\
		&+ R^\mu \big( Q^2 (P\cdot R) - (Q\cdot P)(Q\cdot R) \big) \bigg) , \end{aligned} \\
	\frac{\p z^\prime}{\p P_\mu} &= \frac{Q^\mu - 2 p_1^\mu}{\sigma_\pi(Q^2) \lambda_{12}^{1/2}(Q^2)} + z^\prime \frac{Q^2 P^\mu - (P\cdot Q) Q^\mu}{\lambda_{12}(Q^2)} , \\
	\frac{\p z^\dprime}{\p R_\mu} &= \frac{2 p_1^\mu - Q^\mu}{\sigma_\pi(Q^2) \lambda_{34}^{1/2}(Q^2)} + z^\dprime \frac{Q^2 R^\mu - (R\cdot Q) Q^\mu}{\lambda_{34}(Q^2)} , \\
	\frac{\p z^\prime}{\p R_\mu} &= \frac{\p z^\dprime}{\p P_\mu} = 0 .
	\mytag
\end{align*}
Observing that a loop momentum with an open Lorentz index, $p_1^\mu$, can be written in terms of the derivative of a $\gamma^*\gamma^*\to\pi\pi$ angle with respect to $P_\mu$ or $R_\mu$ and functions of angles and external momenta only, we can write all tensor integrals in terms of derivatives of scalar integrals, since the phase-space integral does not depend on $P$ or $R$. With this method no additional contractions of Lorentz indices are necessary and the complexity of the calculation is reduced significantly. This enables the calculation of the $G$-wave unitarity relation.

Explicitly, we define tensor integrals involving factors of the scattering angles according to:
\begin{align}
	I_{i,nm}^{\mu_1\ldots\mu_i} := \int \frac{d^3p_1}{(2\pi)^32p_1^0} \frac{d^3p_2}{(2\pi)^32p_2^0} (2\pi)^4 \delta^{(4)}\big(Q - p_1 - p_2\big) p_1^{\mu_1} \cdots p_1^{\mu_i} \, {z^\prime}^n {z^\dprime}^m .
\end{align}
For the $G$-wave unitarity relation, we need to know the integrals $I_{i,nm}^{\mu_1\ldots\mu_i}$ with $i + n + m \le 8$ and $i \le 4$. The scalar integrals with $i=0$ can be calculated easily using~\eqref{eq:LegendreAdditionTheorem}:
{\small
\begin{align*}
	I_{0,00} &= I_0 , \\
	I_{0,20} &= I_{0,02} = \frac{1}{3} I_0 , & I_{0,11} &= \frac{z}{3} I_0 , \\
	I_{0,40} &= I_{0,04} = \frac{1}{5} I_0 , & I_{0,31} &= I_{0,13} = \frac{z}{5} I_0 , & I_{0,22} &= \frac{1+2z^2}{15} I_0 , \\
	I_{0,60} &= I_{0,06} = \frac{1}{7} I_0 , & I_{0,51} &= I_{0,15} = \frac{z}{7} I_0 , & I_{0,42} &= I_{0,24} = \frac{1+4z^2}{35} I_0 , & I_{0,33} &= \frac{z(3+2z^2)}{35} I_0 , \\
	I_{0,80} &= I_{0,08} = \frac{1}{9} I_0 , & I_{0,71} &= I_{0,17} = \frac{z}{9} I_0 , & I_{0,62} &= I_{0,26} = \frac{1+6z^2}{63} I_0 , & I_{0,53} &= I_{0,35} = \frac{z(3+4z^2)}{63} I_0 , \\
	I_{0,44} &= \frac{3 + 24 z^2 + 8 z^4}{315} I_0 ,
	\mytag
\end{align*} }%
where
\begin{align}
	I_0 &= \frac{1}{8\pi} \sigma_\pi(s) ,
\end{align}
while all $I_{0,nm}$ with $n+m$ odd vanish.

Next, we calculate the remaining integrals with $i=1,\ldots,4$ successively using the derivative trick. The integrals with $n = m = 0$ are pure tensor integrals and can be cross-checked with the results from the tensor decomposition method.

In order to compute the integrals $I_{1,nm}^\mu$, we consider the following derivative:
\begin{align}
	\frac{\p}{\p P_\mu} \left( {z^\prime}^{n+1} {z^\dprime}^m \right) = (n+1) {z^\prime}^n {z^\dprime}^m \frac{\p z^\prime}{\p P_\mu} &= (n+1) {z^\prime}^n {z^\dprime}^m \left( \frac{Q^\mu - 2 p_1^\mu}{\sigma_\pi(Q^2) \lambda_{12}^{1/2}(Q^2)} + z^\prime \frac{Q^2 P^\mu - (P\cdot Q) Q^\mu}{\lambda_{12}(Q^2)} \right) .
\end{align}
Since the phase-space integral does not depend on $P$ or $R$, we can commute it with the derivative and find
\begin{align}
	I_{1,nm}^\mu &= \frac{1}{2} Q^\mu I_{0,nm} + \frac{\sigma_\pi(s)\lambda_{12}^{1/2}(s)}{2} \left[ \frac{s P^\mu - (q_1^2-q_2^2) Q^\mu}{\lambda_{12}(s)} I_{0,n+1\, m} - \frac{1}{n+1} \frac{\p}{\p P_\mu} I_{0,n+1\,m} \right] .
\end{align}
Similarly, the tensor integrals $I_{2,nm}^{\mu\nu}$ can be calculated by considering the double derivative
\begin{align}
	\frac{\p^2}{\p P_\mu \p P_\nu} \left( {z^\prime}^{n+2} {z^\dprime}^m \right) .
\end{align}
Finally, by taking multiple derivatives the tensor integrals $I_{3,nm}^{\mu\nu\lambda}$ and  $I_{4,nm}^{\mu\nu\lambda\sigma}$ can be calculated.

\subsection{Direct matrix inversion}
\label{sec:BasisChangeHelAmpsDirectInversion}

The expressions for the helicity amplitudes in terms of the scalar
coefficient functions in the tensor decomposition are easily obtained by
contracting the HLbL tensor with the polarization vectors. Expressing the
scalar functions in terms of the helicity amplitudes requires the inversion
of these relations. If we consider the singly-on-shell case, this amounts
to the inversion of a $27\times 27$ matrix. The direct analytic inversion
of a general matrix of this size is not possible, but in this case it can
be reconstructed along the following lines. 

Let us define the basis change from the singly-on-shell helicity amplitudes to scalar functions as
\begin{align}
	\bar H_j \Big|_{\lambda_4 \neq 0} = \sum_{i=1}^{27} \eta_{ji} \check\Pi_i ,
\end{align}
where $\eta$ is a $27\times27$ matrix. Its inverse is effectively the matrix $\check c$ in~\eqref{eq:ScalarFunctionsInTermsOfHelAmps} (restricted to $\lambda_4\neq0$) that we need to determine in order to obtain the imaginary parts of the scalar functions through unitarity.

First, we note that the basis of helicity amplitudes suffers from the presence of kinematic singularities, which makes the expressions for $\eta$ more involved. These singularities can be removed by applying the general recipe of~\cite{Martin:1970} for the construction of amplitudes free of kinematic singularities: first, the singularities at the boundary of the physical region can be removed by
\begin{align}
	\hat H_j := \Big(\frac{1+z}{2}\Big)^{-\frac{1}{2}|m_1+m_2|} \Big( \frac{1-z}{2}\Big)^{-\frac{1}{2}|m_1-m_2|} \bar H_j ,
\end{align}
where $m_1 = \lambda_1-\lambda_2$, $m_2=\lambda_3-\lambda_4$, and $z$ is the cosine of the scattering angle. In our case of fixed-$t$ singly-on-shell kinematics, we have $z=\frac{s-q_1^2+q_2^2}{\lambda_{12}^{1/2}(s)}$. Next, the parity-conserving amplitudes
\begin{align}
	\hat H_j \pm \hat H_{\bar j}
\end{align}
are formed (see Sect.~\ref{sec:UnitarityHelAmps} for the notation). Finally, the remaining singularities can be removed by multiplying with the appropriate powers of $\sqrt{s}$, $\lambda_{12}^{1/2}(s)$, and $\lambda_{34}^{1/2}(s)$, see~\cite{Martin:1970}. We note that the Martin--Spearman amplitudes constructed in this way are free of kinematic singularities, but have an asymptotic behavior that is much worse than the one of the BTT scalar functions.

Since all square-root singularities have been removed, the basis change from the scalar functions $\check\Pi_i$ to the Martin--Spearman amplitudes is now meromorphic in $s$, $q_1^2$, $q_2^2$, and $q_3^2$. We determine all matrix entries with partly numerical methods as follows.

Numerically, the inversion of the $27\times27$ matrix is straightforward. The denominators of the meromorphic matrix entries can be guessed from the pole structure of the numerical inversion: they are products of simple polynomials such as $\lambda_{123}$, $\lambda_{12}(s)$, $(q_1^2-q_2^2+q_3^2)$ etc. We calculate numerically the matrix inversion as a function of each of the Lorentz invariants in turn, keeping the other three invariants fixed. A plot of the matrix entries as a function of the varying variable reveals the poles and therefore the exact form of the denominators. This simple but tedious task has to be performed for all $27\times27$ entries.
The remaining numerators are then polynomials of the form
\begin{align}
	\sum_{i+j+k+l=n} a_{ijkl} s^i (q_1^2)^j (q_2^2)^k (q_3^2)^l ,
\end{align}
where the mass dimension of the numerator is $2n$ and known beforehand. In most cases, $n$ is a small number, although for very few entries we encounter a maximal value of $n=9$, which results in a polynomial with $220$ terms. We perform the numerical inversion on a grid consisting of $9^4$ points in the four-dimensional space of $s$, $q_1^2$, $q_2^2$, and $q_3^2$ and determine the integer coefficients $a_{ijkl}$ for each of the numerators of the $27\times27$ matrix entries by a fit. In contrast to the determination of the denominators by hand, this fit of the numerators can be easily automatized.

Combining the results with the (simple) basis change from helicity to Martin--Spearman amplitudes then leads to the full analytic expression for the inverted basis change $\check c$. In particular, it is straightforward to check analytically that the matrix $\check c$ determined partly with numerical methods is indeed the exact inverse of $\eta$. The result is provided as supplementary material in the form of a \textsc{Mathematica} notebook.


\section{\boldmath Partial-wave expansion of the $\gamma^*\gamma^*\to\pi\pi$ pion-pole contribution}

\label{sec:PionPolePartialwaves}

In order to test the partial-wave formalism, we expand the pion-pole contribution to $\gamma^*\gamma^*\to\pi\pi$ into partial waves. The scalar functions are given by~\cite{Colangelo:2015ama} (with isospin conventions from~\cite{Colangelo:2014dfa}):
\begin{align}
	\begin{split}
		A_1^\pi &= F_\pi^V(q_1^2) F_\pi^V(q_2^2) \left( \frac{1}{t-M_\pi^2} + \frac{1}{u-M_\pi^2}\right) , \\
		A_4^\pi &= F_\pi^V(q_1^2) F_\pi^V(q_2^2) \frac{2}{s - q_1^2 - q_2^2} \left( \frac{1}{t-M_\pi^2} + \frac{1}{u-M_\pi^2}\right) , \\
		A_2^\pi &= A_3^\pi = A_5^\pi = 0 .
	\end{split}
\end{align}
The helicity amplitudes become:
\begin{align*}
	\label{eq:ggpipiHelicityAmplitudesBorn}
	\bar H_{++}^\pi = \bar H_{--}^\pi &= F_\pi^V(q_1^2) F_\pi^V(q_2^2) \left( \frac{1}{t-M_\pi^2} + \frac{1}{u-M_\pi^2}\right) \Bigg( -\frac{1}{2}(s-q_1^2-q_2^2) \\
		& \quad + \frac{1}{4} (s-4M_\pi^2)\left( (s - q_1^2 - q_2^2) + \left( \frac{(q_1^2-q_2^2)^2}{s} - (q_1^2 + q_2^2)\right) z^2 \right) \frac{2}{s - q_1^2 - q_2^2} \Bigg) , \\
	\bar H_{+-}^\pi = \bar H_{-+}^\pi &= - F_\pi^V(q_1^2) F_\pi^V(q_2^2) \frac{1}{2} (s-4M_\pi^2) (1-z^2) \bigg( \frac{1}{t-M_\pi^2} + \frac{1}{u-M_\pi^2} \bigg) , \\
	\bar H_{+0}^\pi = - \bar H_{-0}^\pi &= - F_\pi^V(q_1^2) F_\pi^V(q_2^2) \frac{1}{2} \sqrt{\frac{2}{s}} (s-4M_\pi^2) z \sqrt{1-z^2} \frac{s+q_1^2-q_2^2}{s - q_1^2 - q_2^2} \left( \frac{1}{t-M_\pi^2} + \frac{1}{u-M_\pi^2}\right) , \\
	\bar H_{0+}^\pi = - \bar H_{0-}^\pi &=  - F_\pi^V(q_1^2) F_\pi^V(q_2^2)\frac{1}{2} \sqrt{\frac{2}{s}} (s-4M_\pi^2) z \sqrt{1-z^2} \frac{s-q_1^2+q_2^2}{s - q_1^2 - q_2^2} \left( \frac{1}{t-M_\pi^2} + \frac{1}{u-M_\pi^2}\right) , \\
	\bar H_{00}^\pi &= -F_\pi^V(q_1^2) F_\pi^V(q_2^2) \bigg( 1 - \frac{2(s-4M_\pi^2) z^2}{s - q_1^2 - q_2^2} \bigg) \left( \frac{1}{t-M_\pi^2} + \frac{1}{u-M_\pi^2}\right) . \mytag
\end{align*}
We calculate the partial-wave expansion thereof:\footnote{We use a different convention than in~\cite{Colangelo:2015ama} and do not (anti-)symmetrize the partial waves with respect to $q_1^2 \leftrightarrow q_2^2$.}
\begin{align}
	N_{J,\lambda_1\lambda_2}(s) := \frac{1}{2} \int_{-1}^1 dz \; d_{m0}^J(z) \bar H_{\lambda_1\lambda_2}^\pi(s,t(s,z),u(s,z)) ,
\end{align}
where $m=|\lambda_1-\lambda_2|$. With the relation
\begin{align}
	\begin{split}
		\frac{1}{t-M_\pi^2} &= - \frac{2}{\sigma_\pi(s) \lambda_{12}^{1/2}(s)} \frac{1}{x-z} , \\
		\frac{1}{u-M_\pi^2} &= - \frac{2}{\sigma_\pi(s) \lambda_{12}^{1/2}(s)} \frac{1}{x+z} ,
	\end{split}
\end{align}
where
\begin{align}
	x = \frac{s-q_1^2-q_2^2}{\sigma_\pi(s) \lambda_{12}^{1/2}(s)} ,
\end{align}
we can calculate the pion-pole contribution to the helicity partial waves in terms of the Legendre functions of the second kind, defined by
\begin{align}
	Q_J(x) = \frac{1}{2} \int_{-1}^1 \frac{P_J(z)}{x-z} dz .
\end{align}
They satisfy the relations~\cite{Sharma:2006}
\begin{align}
	\begin{split}
		Q_J(x) P_{J-2}(x) &= P_J(x) Q_{J-2}(x) - \frac{2J-1}{J(J-1)} x , \\
		(J+1) Q_{J+1}(x) &= (2J+1) x Q_J(x) - J Q_{J-1}(x) ,
	\end{split}
\end{align}
which, together with the recursion relation for the Legendre polynomials
\begin{align}
	(J+1) P_{J+1}(x) = (2J+1) x P_J(x) - J P_{J-1}(x)
\end{align}
leads to the following expressions for the pion-pole helicity partial waves:
\begin{align}
	\begin{split}
		N_{J,++}(s) &= F_\pi^V(q_1^2) F_\pi^V(q_2^2)  \left\{ \frac{8}{\sigma_\pi(s) \lambda_{12}^{1/2}(s)}\left(\frac{s q_1^2 q_2^2}{\lambda_{12}(s)} + M_\pi^2\right) Q_J(x) +2 \delta_{J0} 
		\frac{(q_1^2-q_2^2)^2-s (q_1^2+q_2^2)}{\lambda_{12}(s)} \right\} , \\
		N_{J,+-}(s) &= F_\pi^V(q_1^2) F_\pi^V(q_2^2) \frac{2s \sigma_\pi(s)}{\lambda_{12}^{1/2}(s)} J \sqrt{\frac{(J-2)!}{(J+2)!}} \Big\{ 2x Q_{J-1}(x) - \left( (J+1)-x^2(J-1) \right) Q_J(x) \Big\} , \\
		N_{J,+0}(s) &= F_\pi^V(q_1^2) F_\pi^V(q_2^2) \frac{2\sqrt{2s}\sigma_\pi(s)}{\lambda_{12}^{1/2}(s)} \frac{s+q_1^2-q_2^2}{s-q_1^2-q_2^2} \sqrt{\frac{J}{J+1}} x \Big\{ x Q_J(x) -Q_{J-1}(x) \Big\} , \\
		N_{J,0+}(s) &= F_\pi^V(q_1^2) F_\pi^V(q_2^2) \frac{2\sqrt{2s}\sigma_\pi(s)}{\lambda_{12}^{1/2}(s)} \frac{s-q_1^2+q_2^2}{s-q_1^2-q_2^2} \sqrt{\frac{J}{J+1}} x \Big\{ x Q_J(x) - Q_{J-1}(x) \Big\} , \\
		N_{J,00}(s) &= F_\pi^V(q_1^2) F_\pi^V(q_2^2) \frac{4}{\lambda_{12}(s)} \left\{ \frac{(q_1^2-q_2^2)^2-s^2}{\sigma_\pi(s) \lambda_{12}^{1/2}(s)} Q_J(x) + 2s \, \delta_{J0} \right\} .
	\end{split}
\end{align}


\section{Pion polarizability and $\boldsymbol{\gamma\gamma\to\pi\pi}$ in \ChPT{}}
\label{app:ChPT}

The one-loop amplitude for $\gamma\gamma\to\pi\pi$ takes the form~\cite{Bijnens:1987dc,Donoghue:1988eea}
\begin{align}
 h_{0,++}^\text{c}(s)\big|_\text{\ChPT{}}&=N_{0,++}(s)+\frac{\bar l_6-\bar l_5}{48\pi^2\Fpi^2}s-\frac{s}{16\pi^2\Fpi^2}\big(1+2\mpi^2C_0(s)\big),\notag\\
 h_{0,++}^\text{n}(s)\big|_\text{\ChPT{}}&=-\frac{s-\mpi^2}{8\pi^2\Fpi^2}\big(1+2\mpi^2C_0(s)\big),
\end{align}
where we have suppressed the arguments for the virtualities, $\bar l_6-\bar l_5$ refers to a combination of $SU(2)$ low-energy constants~\cite{Gasser:1983yg}, and the loop function is given by
\beq
C_0(s)=\int_0^1\frac{dx}{s x}\log\bigg[1-x(1-x)\frac{s}{\mpi^2}\bigg]. 
\eeq
Unitarity is only fulfilled perturbatively, so that at the one-loop level
\begin{align}
 \Im h_{0,++}^\text{c}(s)\big|_\text{\ChPT{}}&=\frac{\sigma_\pi(s)}{3}N_{0,++}(s)\big(2t^0_0(s)+t^2_0(s)\big)=\frac{\mpi^2}{8\pi\Fpi^2}\log\frac{1+\sigma_\pi(s)}{1-\sigma_\pi(s)},\notag\\
 \Im h_{0,++}^\text{n}(s)\big|_\text{\ChPT{}}&=\frac{2\sigma_\pi(s)}{3}N_{0,++}(s)\big(t^0_0(s)-t^2_0(s)\big)=\frac{s-\mpi^2}{4\pi\Fpi^2}\log\frac{1+\sigma_\pi(s)}{1-\sigma_\pi(s)},
\end{align}
with tree-level $\pi\pi$ partial waves $t^I_J(s)$. Due to the pathological high-energy behavior of these imaginary parts, the chiral amplitudes do not fulfill an unsubtracted dispersion relation, but only a subtracted variant of the form
\begin{align}
 h_{0,++}^\text{c}(s)\big|_\text{\ChPT{}}&=N_{0,++}(s)+\frac{\bar l_6-\bar l_5}{48\pi^2\Fpi^2}s+\frac{s^2}{\pi}\int_{4\mpi^2}^\infty ds'\frac{\Im h_{0,++}^\text{c}(s')\big|_\text{\ChPT{}}}{s'^2(s'-s)},\notag\\
 h_{0,++}^\text{n}(s)\big|_\text{\ChPT{}}&=-\frac{s}{96\pi^2\Fpi^2}+\frac{s^2\mpi^2}{\pi}\int_{4\mpi^2}^\infty d s'\frac{\Im h_{0,++}^\text{n}(s')\big|_\text{\ChPT{}}}{s'^3(s'-s)},
\end{align}
to be contrasted with 
\beq
 h_{0,++}(s)=\Delta_{0,++}(s)+\frac{s}{\pi}\int_{4\mpi^2}^\infty ds'\frac{\Im h_{0,++}(s')}{s'(s'-s)}
\eeq
for the full amplitudes provided that the imaginary parts fall off sufficiently fast. If the MO inhomogeneity is approximated by the Born term that is indeed the case, which, by comparison to the chiral amplitudes, allows one to predict the derivatives at $s=0$ and thereby the pion polarizabilities within this approximation. At the one-loop level this implies a sum rule for $\bar l_6-\bar l_5$, whose numerical evaluation $\bar l_6-\bar l_5=2.7\ldots 2.9$ for the same range of cutoffs as in Sect.~\ref{sec:pion_pol} indeed comes out very close to the phenomenological value $\bar l_6-\bar l_5=3.0(0.3)$~\cite{Bijnens:1996wm,Geng:2003mt,Gasser:2006qa}. As discussed in Sect.~\ref{sec:pion_pol}, only the charged-pion polarizability is reproduced in this way, indicating that higher contributions to the LHC are required in the case of the neutral pion.

In \ChPT{} the value of $\bar l_6-\bar l_5$ can be empirically understood in terms of resonance saturation, explicitly one has~\cite{Bijnens:2014lea,Ecker:1988te,Ecker:1989yg}
\beq
\bar l_6-\bar l_5\big|_\text{sat}=48\pi^2\frac{F_A^2}{M_A^2}\sim 24\pi^2\frac{\Fpi^2}{M_\rho^2}=3.4,
\eeq
where $F_A$ and $M_A$ refer to decay constant and mass of axial resonances to be related to pion decay constant and vector masses by short-distance constraints, see~\cite{Bijnens:2014lea}.
The fact that resonance saturation indeed reproduces the empirical value of $\bar l_6-\bar l_5$ rather accurately has motivated the construction of models based on explicit $a_1$ resonances to incorporate the corresponding effects related to the charged-pion polarizability into HLbL scattering~\cite{Engel:2013kda,Bijnens:2016hgx}. Our calculation makes use of an alternative strategy that exploits a sum rule for the relevant low-energy parameters, largely saturating the phenomenological value. In particular, in this framework the impact of higher contributions such as the $a_1$ to the LHC on the polarizability itself is expected to be small---in fact, the exchange of vector mesons would contribute first---so that significant corrections would require a weighting in the $g-2$ integral that emphasizes kinematics away from $s=0$, where the polarizabilities are defined. Such contributions cannot a priori be excluded, all the more since the comparison with the physical polarizability only determines the on-shell properties, but not the dependence on the photon virtualities.

\end{appendices}

\renewcommand\bibname{References}
\renewcommand{\bibfont}{\raggedright}
\bibliographystyle{utphysmod}
\phantomsection
\addcontentsline{toc}{section}{References}
\bibliography{Literature}

\end{document}